\documentclass[twocolumn, twocolappendix]{aastex63}

\usepackage{graphicx}	
\usepackage{amsmath}	
\usepackage{amssymb}	
\usepackage{rotating}

\received{tbd}
\revised{tbd}
\accepted{tbd}
\submitjournal{AJ}

\shorttitle{MYSST - II. N44 PMS Stars}
\shortauthors{Ksoll et al.}

\begin{document}

\title{Measuring Young Stars in Space and Time - II. The Pre-Main-Sequence Stellar Content of N44}


\correspondingauthor{Victor F. Ksoll}
\email{v.ksoll@stud.uni-heidelberg.de}

\author[0000-0002-0294-799X]{Victor F.\ Ksoll}
\affiliation{Universit\"{a}t Heidelberg, Zentrum f\"{u}r Astronomie, Institut f\"{u}r Theoretische Astrophysik,\\ Albert-Ueberle-Str. 2, 69120 Heidelberg, Germany}
\affiliation{Universit\"{a}t Heidelberg, Interdisziplin\"{a}res Zentrum f\"{u}r Wissenschaftliches Rechnen,\\ Im Neuenheimer Feld 205, 69120 Heidelberg, Germany}

\author[0000-0002-2763-0075]{Dimitrios Gouliermis}
\affiliation{Universit\"{a}t Heidelberg, Zentrum f\"{u}r Astronomie, Institut f\"{u}r Theoretische Astrophysik,\\ Albert-Ueberle-Str. 2, 69120 Heidelberg, Germany}
\affiliation{Max Planck Institute for Astronomy, K\"{o}nigstuhl\,17, 69117 Heidelberg, Germany}

\author[0000-0003-2954-7643]{Elena Sabbi}
\affiliation{Space Telescope Science Institute, 3700 San Martin Drive, Baltimore, MD 21218, USA}

\author{Jenna E. Ryon}
\affiliation{Space Telescope Science Institute, 3700 San Martin Drive, Baltimore, MD 21218, USA}

\author[0000-0002-9573-3199]{Massimo Robberto}
\affiliation{Space Telescope Science Institute, 3700 San Martin Drive, Baltimore, MD 21218, USA}
\affiliation{Johns Hopkins University, 3400 N. Charles Street, Baltimore, MD 21218, USA}

\author[0000-0002-5581-2896]{Mario Gennaro}
\affiliation{Space Telescope Science Institute, 3700 San Martin Drive, Baltimore, MD 21218, USA}

\author[0000-0002-0560-3172]{Ralf S.\ Klessen}
\affiliation{Universit\"{a}t Heidelberg, Zentrum f\"{u}r Astronomie, Institut f\"{u}r Theoretische Astrophysik,\\  Albert-Ueberle-Str. 2, 69120 Heidelberg, Germany}
\affiliation{Universit\"{a}t Heidelberg, Interdisziplin\"{a}res Zentrum f\"{u}r Wissenschaftliches Rechnen,\\ Im Neuenheimer Feld 205, 69120 Heidelberg, Germany}

\author[0000-0001-6036-1287]{Ullrich Koethe}
\affiliation{Universit\"{a}t Heidelberg, Heidelberg Collaboratory for Image Processing, Visual Learning Lab,\\  Berliner Str. 43, 69120 Heidelberg, Germany}

\author[0000-0001-7906-3829]{Guido de Marchi}
\affiliation{European Space Research and Technology Centre, Keplerlaan 1, 2200 AG Noordwijk, Netherlands}

\author[0000-0002-3925-9365]{C.-H. Rosie Chen}
\affiliation{Max-Planck-Institut für Radioastronomie, Auf dem Hügel 69, D-53121 Bonn, Germany}

\author{Michele Cignoni}
\affiliation{Department of Physics - University of Pisa, Largo B. Pontecorvo, 3 Pisa, 56127, Italy }
\affiliation{INFN, Largo B. Pontecorvo 3, 56127, Pisa, Italy}
\affiliation{INAF-Osservatorio di Astrofisica e Scienza dello Spazio, Via Gobetti 93/3, 40129, Bologna, Italy} 

\author{Andrew E.\ Dolphin}
\affiliation{Raytheon, 1151 E. Hermans Road, Tucson, AZ 85706, USA}
\affiliation{Steward Observatory, University of Arizona, 933 North Cherry Avenue, Tucson, AZ 85721, USA}
)



\begin{abstract}

The Hubble Space Telescope (HST) survey Measuring Young Stars in Space and Time (MYSST) entails some of the deepest photometric observations of extragalactic star formation, capturing even the lowest mass stars of the active star-forming complex N44 in the Large Magellanic Cloud. We employ the new MYSST stellar catalog to identify and characterize the content of young pre-main-sequence (PMS) stars across N44 and analyze the PMS clustering structure. To distinguish PMS stars from more evolved line of sight contaminants, a non-trivial task due to several effects that alter photometry, we utilize a machine learning classification approach. This consists of training a support vector machine (SVM) and a random forest (RF) on a carefully selected subset of the MYSST data and categorize all observed stars as PMS or non-PMS. Combining SVM and RF predictions to retrieve the most robust set of PMS sources, we find $\sim26,700$ candidates with a PMS probability above 95\% across N44. Employing a clustering approach based on a nearest neighbor surface density estimate, we identify 16 prominent PMS structures at $1\,\sigma$ significance above the mean density with sub-clusters persisting up to and beyond $3\,\sigma$ significance. The most active star-forming center, located at the western edge of N44's bubble, is a subcluster with an effective radius of $\sim 5.6$\,pc entailing more than 1,100 PMS candidates. Furthermore, we confirm that almost all identified clusters coincide with known H II regions and are close to or harbor massive young O stars or YSOs previously discovered by MUSE and Spitzer observations.

\end{abstract}



\section{Introduction}
    Star formation is one of the most fundamental processes in our universe, bringing light to the galaxies and ultimately providing the environments for the nucleosynthesis of all heavier elements. The primary birth places of stars in galaxies are giant molecular clouds, enormous reservoirs of atomic and molecular hydrogen, harboring the necessary material to create stars \cite[for a review, see e.g.][and references therein]{Klessen16}. Within these clouds stars tend to form in clusters and, in some instances, create large star-forming complexes with multiple stellar populations of different ages, where the feedback of the massive, but short-lived, constituents can repeatedly trigger new star-forming events \citep{Lee2007, Elmegreen2011}. These young and bright objects are the signposts of massive star-forming clusters \citep{Zinnecker07,PortegiesZwart2010}, but as studies of the stellar initial mass function (IMF) indicate \cite[see][]{Kroupa02,Chabrier03}, intermediate and low mass objects actually contribute a significant fraction to a cluster's total stellar mass. Contrary to their massive blue siblings these low mass \textit{pre}-main-squence (PMS) stars, still in the Kelvin-Helmholtz contraction phase \citep{Stahler_Palla2005}, require
    increasingly longer time to reach the main sequence as their masses gets smaller, down to the hydrogen burning limit \citep[about 0.072 M$_\odot$,][]{Schulz2012}. In the first few Myrs PMS stars may still be forming, accreting gas from their immediate surroundings and circumstellar disks \citep{Hartmann16}. Low mass PMS objects trace the history of (recent) star formation beyond the few Myr probed by the ephemeral most massive stars. Therefore, our understanding of star formation may greatly benefit from the study and observation of young PMS objects and the stellar clusters within which they are born. 
    
    Large photometric surveys of nearby systems are one of the main astronomical methods to perform in-depth studies of remote stellar clusters and identify star-forming regions. For more than three decades one of the most successful tools for such photometric surveys has been the \textit{Hubble Space Telescope} (HST), providing observations with exceptional spatial resolution and to great depth. In the past the HST has proven especially capable of detecting faint PMS sources in the Magellanic Clouds, the dwarf companion galaxies to our Milky Way \citep{Gouliermis2006, Gouliermis2012b, Nota2006, Sabbi2007, DaRio2010, DaRio2012, Sabbi2016}. Aside from harboring the only extragalactic PMS sources we can spatially resolve, the Magellanic Clouds are characterized by a relatively high star-forming activity, observable at lower extinction, since they are not obscured by the dusty Galactic disc. Therefore the Clouds provide very attractive targets for the study and observations of large ensembles of PMS stars \citep{Gouliermis2012}. 
    
    \begin{figure*}
        \centering
        \includegraphics[width = \linewidth]{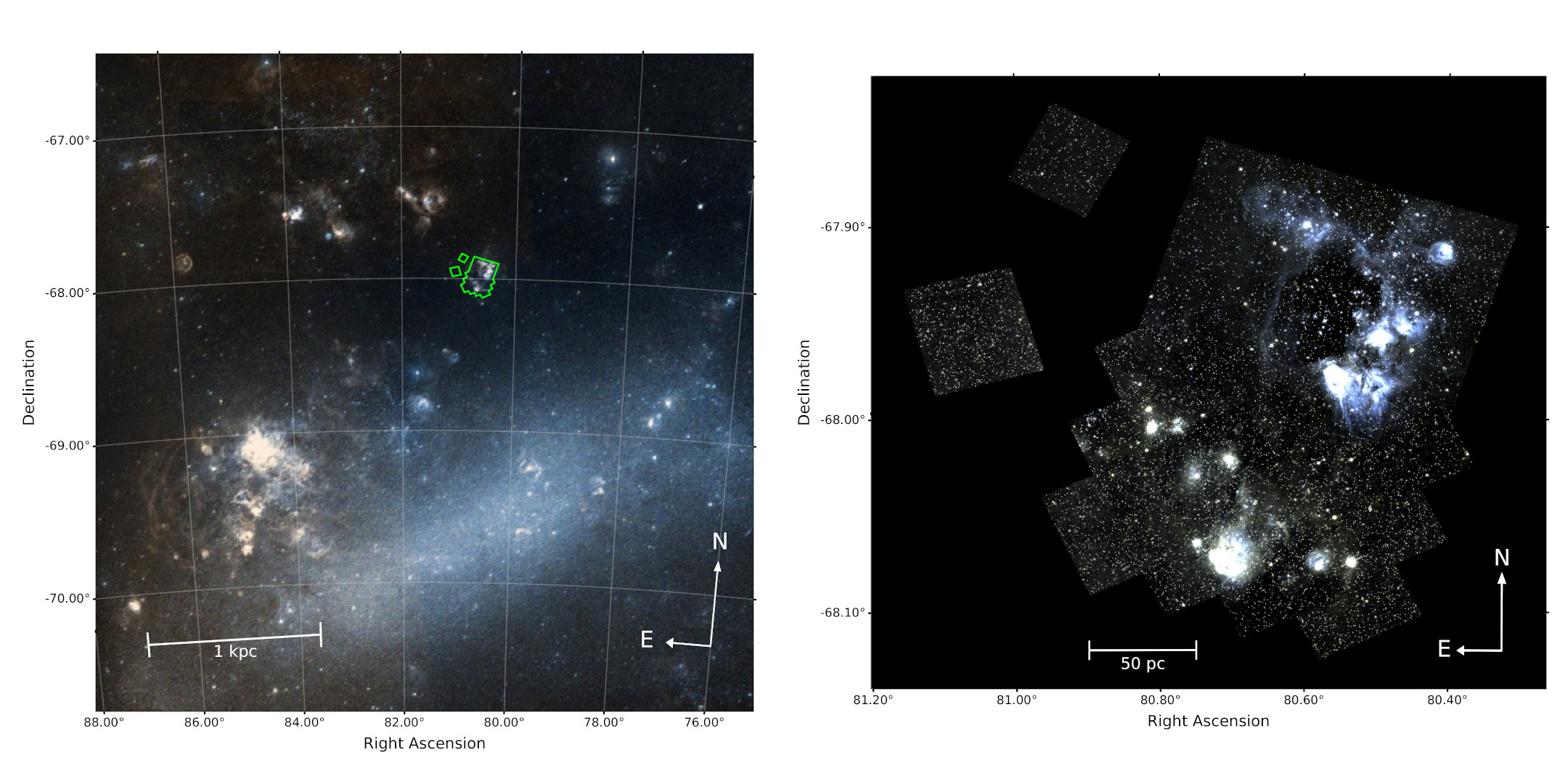}
        \caption{Color composite image from the Digitized Sky Survey \citep[DSS,][]{Lasker1996} of the wider LMC neighborhood of N44 (left). The green contour highlights the main FoV and the observed reference fields of the MYSST survey. Right: Two color composite image of N44 from the MYSST survey, presenting the observations in the F555W and F814W filters in blue and green, respectively. Image from \cite{Ksoll2020b} used with permission.}
        \label{fig:DSS_MYSST_images}
    \end{figure*}
    
    One such complex is the active star-forming region N44 \citep[LH$\alpha$ 120--N44;][]{Henize1956}, located in the Large Magellanic Cloud (LMC). It consists of a giant complex of H II regions, one of the most luminous across the entire LMC after 30 Doradus and N11 \citep{KennicutHodge1986, Pellegrini2012ApJ}, entailing an enormous central super bubble and several compact H II regions along its ridge \citep{Pellegrini2012ApJ, McLeod2019}. The youthfulness of the stars within these ionized gas reservoirs is highlighted by three OB associations \citep[LH47, 48 and 49;][]{LuckeHodge1970} and a plethora of more than 30 massive, short-lived O type stars that have been identified in N44 by spectroscopic studies \citep{McLeod2019, Will1997, Oey1995, Conti1986, Rousseau1978}. N44 also exhibits evidence for multiple star-forming events and feedback triggered star formation, as previous studies have found a $\sim 5$\,Myr difference in age between the stellar populations within and at the rim of N44's bubble \citep{Oey1995}, as well as the presence of a supernova remnant, SNR  0523-679 \citep{Chu1993}, in the vicinity of the bubble \citep{Jaskot2011}. In addition, there is active, ongoing star formation in N44, as \cite{Chen2009} find 59 massive young stellar objects (YSOs) within N44 from observations with the Spitzer Space Telescope. Combining Spitzer data from the SAGE \citep[Surveying the Agents of a Galaxy's Evolution,][]{Meixner2006} legacy program with optical photometry from the Magellanic Clouds Photometric Survey \citep[MCPS,][]{Zaritsky1997} and near-infrared photometry from the InfraRed Survey Facility \citep[IRSF,][]{Kato2007} this list is extended by another 139 YSOs\citep[18 in common with][matched to within 1~arcsec]{Chen2009} by \cite{Carlson2012}. In a recent study, \cite{Zivkov2018} have used near infrared observations from the VISTA Survey of the Magellanic Clouds \citep[VMC,][]{Cioni2011} to estimate the number of PMS sources in N44. Identifying regions containing PMS sources from density excesses in $K_s/(Y-K_s)$ Hess diagrams in comparison to the underlying fields, they find a lower limit to the number of PMS stars in N44 of $1000\pm38$.
    
    N44's complexity is captured by the deep HST imaging of the "Measuring Young Stars in Space and Time" (MYSST) survey, which obtained photometry in two broadband filters for more than 400,000 sources across the extent of N44 \citep[Paper I]{Ksoll2020b}. The rich color magnitude diagram (CMD) of the MYSST survey has not only revealed the presence of significant differential reddening within N44, but also entails many populations of different ages in the observed area. Consequently a significant overlap between the old lower main-sequence (LMS) or red giant branch (RGB), and the PMS population occurs in the CMD making it particularly difficult to distinguish the young N44 cluster constituents from the field contaminants in this large data set without additional information about the excess in emission lines that accompany the PMS phase \citep[e.g.][]{DeMarchi2010}. 
    
    To disentangle the PMS population from the older stars in a statistically sound manner using only broad-band photometry requires sophisticated algorithms, like e.g.~the machine learning (ML) approaches we have demonstrated in a previous study \citep{Ksoll2018}. 
    In the recent years, there have been many examples of established machine learning approaches successfully applied to astronomical problems involving regression, classification, and clustering tasks \citep[see e.g.][for reviews of recent applications]{Baron2019, Fluke2020}.
    
    In this paper we present the identification of the youngest PMS candidates in N44 using the photometric catalog from the HST survey MYSST (Paper I). Our approach, established in \cite{Ksoll2018}, consists of a machine learning based classification of the PMS and Non-PMS constituents of the survey. This study is structured as follows. In Section \ref{sec:Data} we provide a brief summary of the MYSST photometric catalog. In Section \ref{sec:TrainingSet} we begin by describing the construction of the necessary training set for our ML classification approach from a subset of the observational data. This entails the careful selection of a region within N44 that contains distinct PMS and lower main sequence (LMS) populations, as well as the addition of examples of field red giant branch contaminants from suitable areas. Then we present the training and test performance of our models. In Section \ref{sec:Results} we discuss the classification results of our approach while in Section \ref{sec:Clustering} we analyze the spatial clustering structure of the identified PMS candidate stars. The final Section \ref{sec:Summary} provides a summary and considerations on future developments.

\section{Data}
    \label{sec:Data}
        The MYSST program observed the star-forming complex N44, located in the Large Magellanic Cloud, with a deep, high spatial resolution HST survey (Paper I). Its field of view (FoV) of $12.2 \times 14.7\,\mathrm{arcmin^2}$, corresponding to about $180\,\mathrm{pc} \times 215\,\mathrm{pc}$ at the LMC distance \citep[$(m-M)_0 = 18.55 \pm 0.05$;][]{Panagia1991, DeMarchi2016}, entails N44's characteristic super bubble and the region south of it. Figure \ref{fig:DSS_MYSST_images} shows the MYSST FoV in the greater LMC neighborhood of N44 (left) and the MYSST two color composite image (right). The survey was conducted in two broadband filters, F555W and F814W, with the Advanced Camera for Surveys (ACS) and Wide Field Camera 3 (WFC3) instruments of the HST. Reaching down to about 29\,mag in F555W and 28\,mag in F814W the MYSST survey is one of the deepest photometric studies of extragalactic stars, probing even the lowest mass populations of N44. The F555W detection limit implies the capture of e.g.~unreddened 1 Myr pre-main-sequence stars with masses as low as $0.09\,M_\sun$ (see completeness discussion in Paper I) at the distance of the LMC. In this paper we use the MYSST photometric catalog presented in Paper I, consisting of 461,684 sources across the observed FoV of N44 and two smaller LMC reference fields. This catalog only entails objects up to 14\,mag in F555W and 13\,mag in F814W as brighter sources were lost due to saturation. Consequently, the available data is likely missing some of the most massive stars, i.e.~early O stars, of the region.
        
        N44 is also subject to a substantial amount of differential reddening. In Paper I we establish reddening properties for the MYSST survey by fitting the slope of the extinction-elongated red clump using the RANSAC algorithm. Furthermore, we derive individual stellar extinctions using upper main-sequence (UMS) stars as extinction probes and assigning a distance weighted average extinction of the nearest UMS stars to all other sources. This extinction estimate entails some caveats. First, we assume the UMS stars to be on the zero-age-main-sequence (ZAMS) to measure their extinction. For the quickly evolving massive O stars this might not necessarily be the case anymore, even if they are still young. In fact, \cite{Oey1995} estimate the O star population in N44's bubble to be about 10 Myr old while O stars in the bubble rim are 5 Myr younger. However, we find that the error for using the ZAMS instead of e.g.~a 10 Myr isochrone is only on the order of 0.04~mag for our selection of UMS sources. In any case this ZAMS assumption for the UMS sources means that the estimated reddening is at worst only an upper limit of the true extinction for older UMS stars. Second, while it has been found that using the reddening of UMS neighbors returns reasonable values for constituents of young star forming regions \citep{DeMarchi2016}, such as N44, there is no guarantee that the UMS extinction is representative for field sources, leading to occasional over- or underestimates. 
   
\section{Training Set}
    \label{sec:TrainingSet}
    \begin{figure*}
        \centering
        \includegraphics[width = \linewidth]{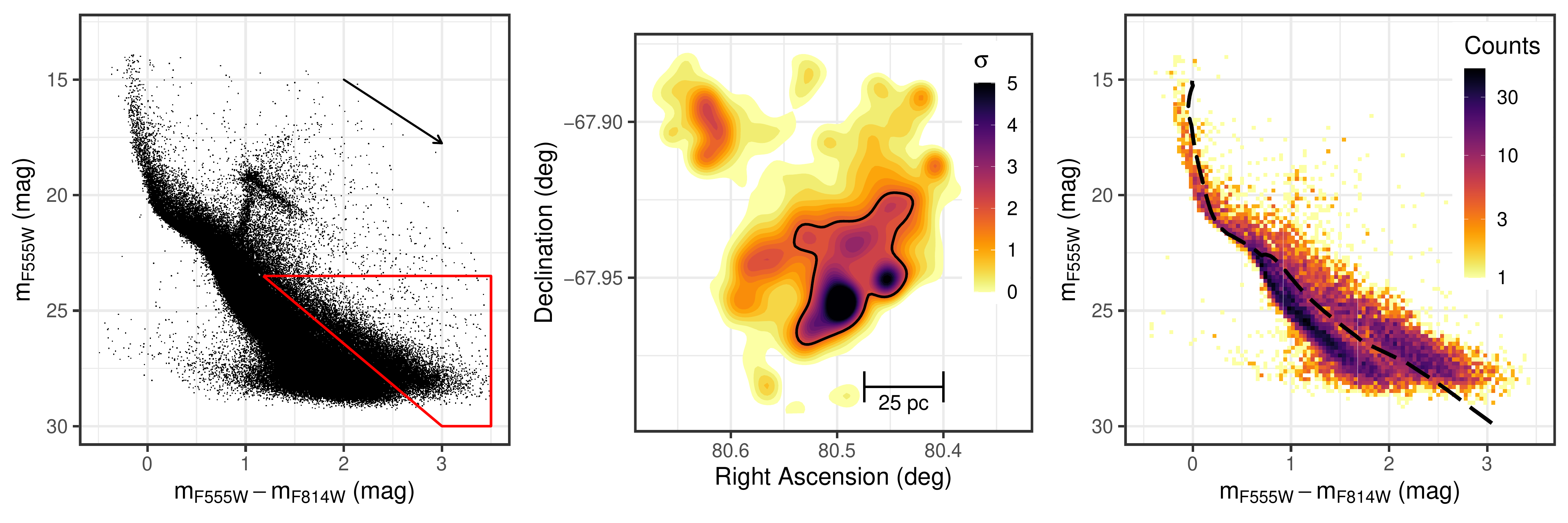}
        \caption{Optical CMD of the northern half of the MYSST main FoV, centered on N44's super bubble (left). The red polygon indicates a rough selection of PMS candidate stars used to identify a training set for our ML approach. The black arrow indicates the direction of the reddening vector of N44, as derived in Paper I. Center: Contour density plot of a kernel density estimate of the rough PMS candidates located in the northern part of the observed field view. The density levels are shown in units of $\sigma$ above the mean estimated density. The contour highlighted by the solid black line indicates the region selected as a base for the training set. Right: Hess diagram of the black outlined region in the center panel. This density diagrams highlights the presence of two distinct populations of stars in this FoV, namely a clear main-sequence and pre-main-sequence. For comparison the black dashed line indicates a 14 Myr PARSEC isochrone, corrected for the median extinction of the stars in this region and the LMC distance modulus.}
        \label{fig:TrainingSet_Selection}
    \end{figure*}
    
    \begin{figure*}
        \centering
        \includegraphics[width = \linewidth]{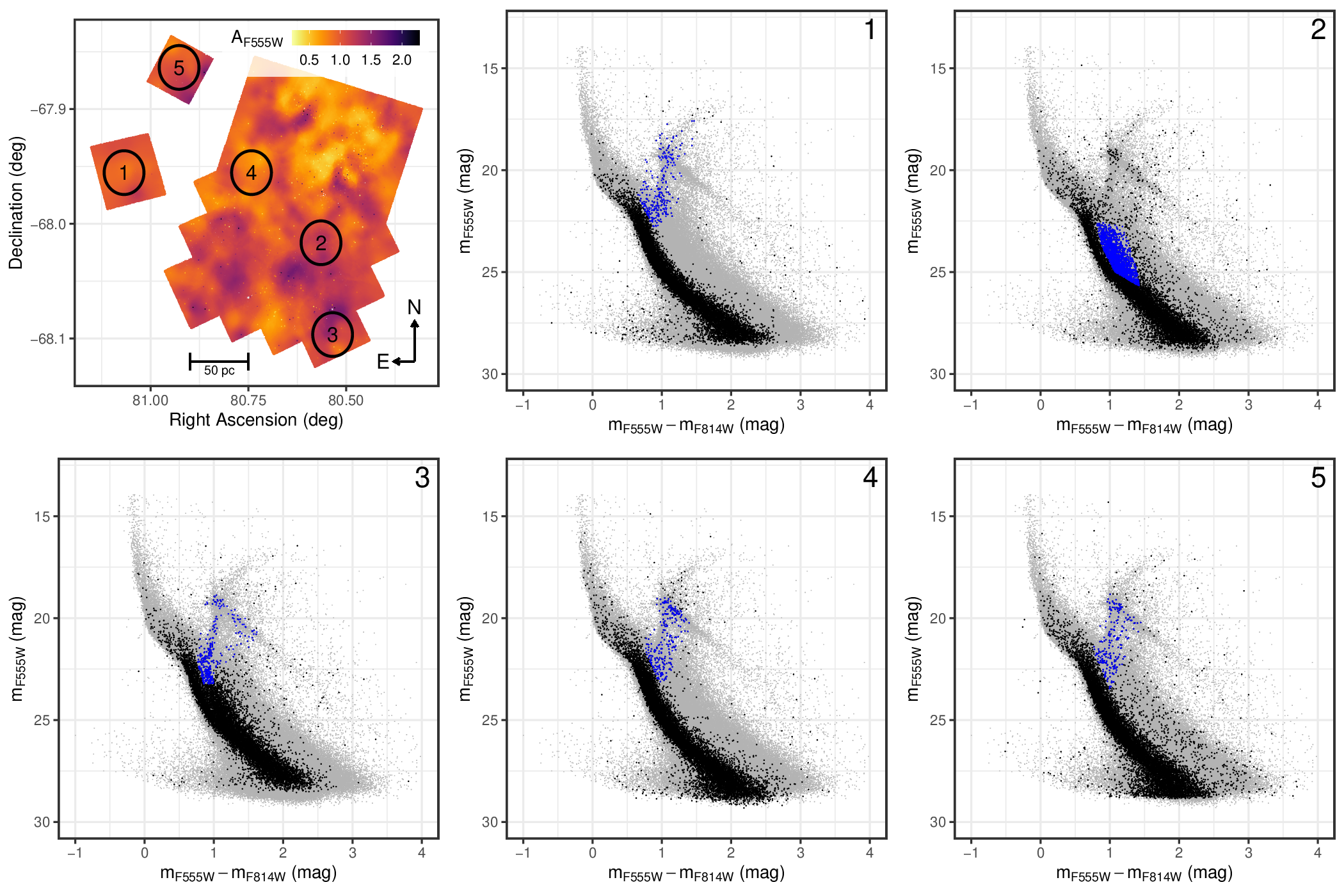}
        \caption{Extinction Map of the MYSST photometric catalog (top left). The black numbered circles indicate regions that are identified to be devoid of PMS stars and used to add examples of RGB stars into the training set. The size of the circular selections is chosen to match the surface area of the training set contour in the center panel of Figure \ref{fig:TrainingSet_Selection}. The remaining five panels show the CMDs of the corresponding circles in black in comparison to the total CMD of the MYSST data (grey). Highlighted in blue are the respective non-PMS examples added to the training set. Note that we do not select RGB samples in the top right panel but rather an emergent feature that resembles a highly extincted main sequence.}
        \label{fig:TrainingSet_RGB}
    \end{figure*}
    
    \begin{figure}
        \centering
        \includegraphics[width = \linewidth]{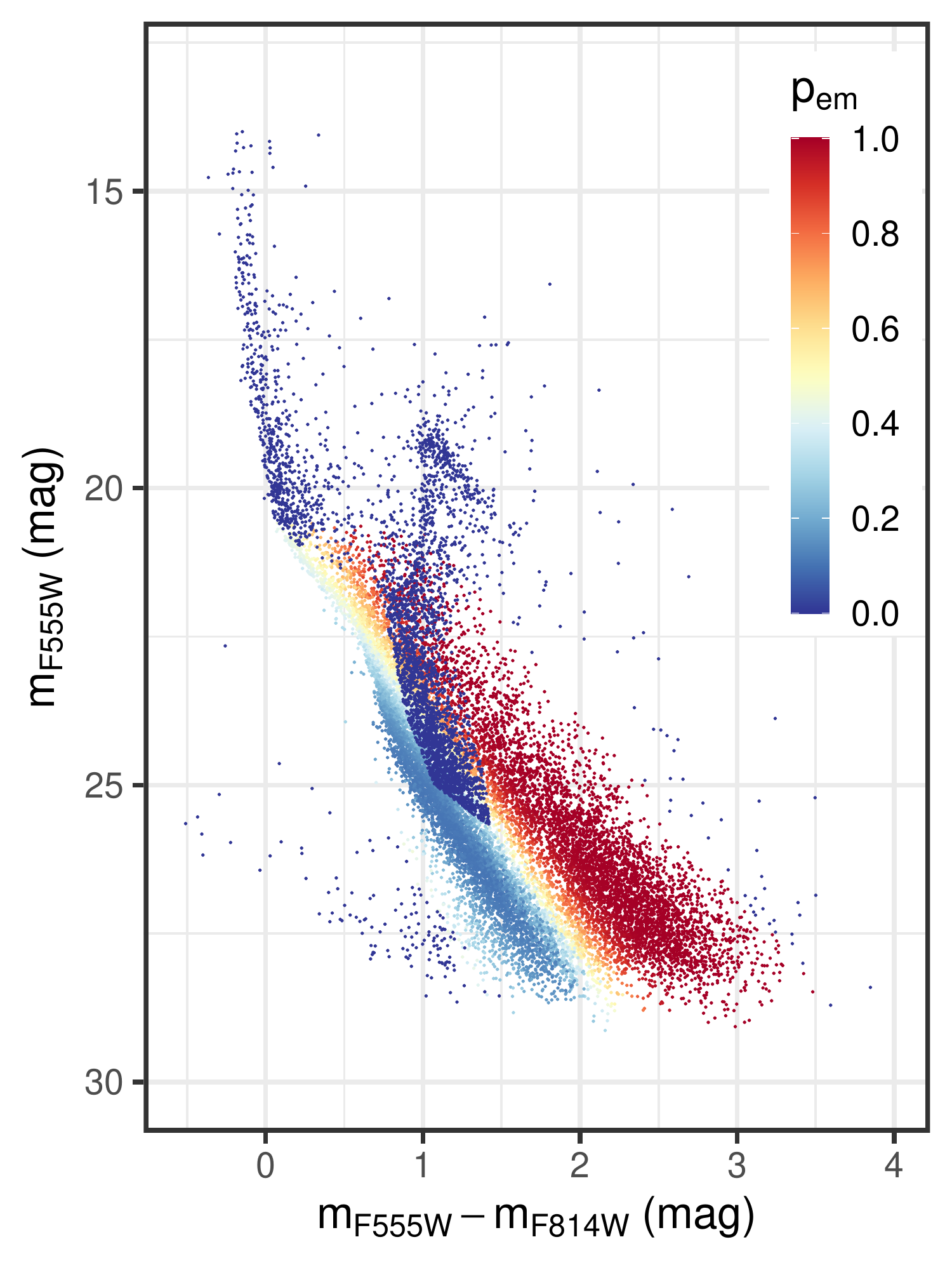}
        \caption{Optical CMD of the final training set selection. Each star is color coded according to the derived probability that it belongs to the PMS population in the CMD. Note that the UMS and additional RGB examples are included with a fixed probability of 0.}
        \label{fig:TrainingSetFinal}
    \end{figure}
    
    \cite{Ksoll2018} establish a machine learning approach for the identification of PMS candidate stars based on HST photometry, which here we apply to the MYSST data. The method entails the careful selection of a training set from the observational data, in which a distinction between examples of PMS and non-PMS stars can be made easily. With this labeled training data the classical machine learning techniques called support vector machine (SVM) and random forest (RF) are then trained to distinguish these two classes of stars based on their broad-band photometry and estimated extinction. 
    
    Due to the different filter passbands between the Hubble Tarantula Treasury Project (HTTP) data of \cite{Ksoll2018} and the MYSST survey, one cannot re-use the HTTP training set. The intrinsic differences between the two star-forming regions would in any case justify the creation of a new training set specific to the MYSST data of N44. 
    
    \subsection{PMS Training Set}
    As a base for our training set we select a subset of the MYSST data which is likely to contain a suitable amount of PMS stars as well as lower main sequence (LMS) contaminants. The latter are likely for the most part field constituents, but could also consist of low-mass remnants of earlier star formation episodes in the N44 region. Given that LMS and PMS stars are located closely together in the low brightness regime in the CMD, we require examples from both populations in order for our ML models to learn to properly distinguish PMS from non-PMS stars. To find a region within the MYSST data that contains enough examples of PMS stars, we first make a very rough selection of potential candidates in the CMD using the red polygon in the left panel of Figure \ref{fig:TrainingSet_Selection}. Performing a kernel density estimate (KDE) on the spatial distribution (using a Gaussian kernel and a fixed bandwidth of 300\,pixels, i.e.~$\sim 3$\,pc) of this rough selection we then determine field areas with high densities of PMS star candidates. Since the majority of these are located in the northern half of the FoV we concentrate on this region. Drawing contours at increasing significant density levels, in units of $\sigma$ above the mean surface density, we find that a $2\,\sigma$ density contour, located at the western edge of the N44 super bubble, entails a large enough sample of LMS and PMS stars. This  region is enclosed by the black contour in the center panel of Figure \ref{fig:TrainingSet_Selection}. The corresponding Hess diagram (Figure \ref{fig:TrainingSet_Selection}, right panel) shows a CMD consisting of a prominent main sequence as well as a nicely separated young PMS population, which provides an ideal base for the training set of our ML approach. Note that this region is also subject to significant differential reddening, covering the entire range of the extinction estimates, so that this selection already entails the broad extinction range towards N44.
    
    Since our classification scheme distinguishes between two classes, "PMS" and "Non-PMS", each star of our training set base requires a label indicating to which of the two categories it belongs. Consequently, we need to quantify which of the stars in our data set are part of the PMS and LMS populations in the low brightness regime. To achieve this we have devised a procedure in \cite{Ksoll2018}, where we fit a Gaussian mixture model to a distance metric in the CMD using the Expectation Maximization (EM) algorithm to determine a probability for every star in the low brightness regime to be part of the PMS population. Figure \ref{fig:TrainingSet_EM} in the Appendix shows the selection of the low brightness stars for this fit. Here we have excluded the UMS and red clump sources as well as a few objects, whose nature we could not identify. While the very red objects among the latter could potentially be PMS stars, which are e.g.~variable sources or are undergoing an extreme accretion event, we cannot ascertain this with the MYSST data alone. Therefore, we opt to only find the most secure PMS examples here. Figure \ref{fig:TrainingSet_EM} also highlights the threshold line derived from PARSEC isochrones \citep{Bressan2012}, which is the basis for the CMD distance measure. Note that this selection and the fit are performed on the extinction corrected CMD in order to achieve the best possible separation between PMS and LMS objects. We also ignore the uncertainties of photometry and extinction during the Gaussian mixture model fit, because we aim to perform a classification and not a regression, so that the precise probability values are not of great importance.
    
    Once these probabilities are established we assign our binary labels by selecting a threshold above which we consider a star a true PMS candidate, taking the need for a balanced (ideally 50\% positive and 50\% negative examples) training set into account. Due to the overall lower abundance of PMS stars we cannot reach an optimal balance, but find that selecting a threshold probability of $p_\mathrm{em} \geq 0.85$ achieves a reasonable trade-off between training set balance, strictness in our PMS example choice, and classifier performance. The strictness of the chosen threshold also indirectly accounts for the uncertainties of photometry and extinction, neglected during the fit, as this selection of PMS examples is more conservative than optimistic, already excluding sources in the transition zone that would show the most changes in PMS candidate probability due to measurement uncertainties.  \\
    
    \subsection{RGB Training Set}
    Aside from the field LMS stars, which need to be distinguished from the PMS sources, old stars on the red giant branch (RGB) can also fall into the PMS regions of the CMD due to either distance, extinction or simply the fact that RGB and PMS tracks can partially overlap in the CMD. Like most of the LMS stars these RGB contaminants are either fore- or background stars of the LMC that do not belong to the young star-forming clusters we are trying to identify. As the third panel of Figure \ref{fig:TrainingSet_Selection} indicates our training set basis contains almost no examples of these stars. Consequently, we need to look elsewhere to find additional RGB examples so that our ML models can take these objects into account. To find such examples we use the KDE of the PMS selection again to now identify regions within the survey that are devoid of PMS stars and entail an RGB population. The top left panel in Figure \ref{fig:TrainingSet_RGB} shows five regions we have identified for this purpose, all encircling the same projected area enclosed by the $2\,\sigma$ irregular contour of our training set basis. We select multiple regions to probe different extinction regimes. The remaining five panels show the corresponding CMDs in comparison to the total CMD of the MYSST survey, the blue points representing the RGB examples to add to the training set. We also include a few example red clump stars along with the RGB selection to avoid potential miss-classification on account of the models never having seen any red clump objects during training. Also important to note here is that we do not select RGB examples in region 2 but rather constituents of a feature that looks akin to a heavily reddened main sequence. This feature does not completely disappear when we correct for extinction. Given that this region appears to be more severely extinguished in the UMS extinction measurements, this feature could potentially be a heavily reddened field population behind N44 for which we are still underestimating the reddening. Since the nature of these objects is unclear and because this region is clearly almost devoid of young PMS stars we decide to include this feature as negative examples so that our ML models can also take it into account. 
    
    We add these RGB examples with a fixed PMS probability of $p_\mathrm{em} = 0$ before applying the previously mentioned labeling threshold to the data.
    
    \subsection{Final Training Set}
    Figure \ref{fig:TrainingSetFinal} shows our final training set before application of the label threshold. In early training attempts of our ML models we realized that the prediction benefits from including the UMS (examples located at about $m_{F555W} < 21$ and $m_{F555W} - m_{F814W} < 0.5$) as additional negative examples, something that was not necessary in our previous study \citep{Ksoll2018}. Similarly to the RGB stars, we add them with zero probability. Note that this decision will likely exclude the detection of more massive, brighter PMS stars that are close to joining the MS, like e.g.~Ae sources. We also re-add the low brightness objects of unclear nature, that were excluded during the EM fit, as negative examples (i.e.~with $p_\mathrm{em}=0$). For the most part these are located roughly at $m_{F555W} > 25$ and $m_{F555W} - m_{F814W} < 1$, as well as around $ 28 > m_{F555W} > 22$ and $1.5 < m_{F555W} - m_{F814W} < 4$.
    
    With that our training set entails 17,942 stars, of which 5,512 ($\sim31\,\%$) are PMS candidate stars with $p_\mathrm{em} \geq 0.85$. Again, the balance between positive and negative examples within the training set is not optimal but with about a third of the data being positive examples we believe our selection is robust enough to not suffer from imbalance issues. At this point it is also important to note that the PMS candidate examples in our training set appear to be mostly younger than $\sim 15$ Myr when compared to PARSEC isochrones (see Figure~\ref{fig:TrainingSet_Selection}, right). As our ML classification approach will find the siblings of the training PMS candidates across all of N44, this means in the following that we will recover only the most recent sites of star formation, younger than $\sim15$ Myrs. Therefore, our method is not sensitive to potential low-mass PMS stars from even earlier star formation events, which are still in the formation process, but very close to joining the main-sequence.
    
    Lastly, we also have to note that we do not account for active galactic nuclei (AGNs) or unresolved (background) galaxies in our training data, because we aim for a MYSST survey intrinsic approach and distinguishing these sources with the available data is not straight forward. Consequently, there may be some minor contamination by these types of sources in our training set.
    
    \subsection{Training and Test Results}
    
    \begin{deluxetable}{lcccc}
        \centering
        \tablecaption{Performance Summary for SVM and RF \label{tab:TrainingPerformance}}
        \tablehead{
            & \multicolumn{4}{c}{Method} \\
            Performance & \multicolumn{2}{c}{SVM} & \multicolumn{2}{c}{Random Forest} \\
            Measure & Train & Test & Train & Test
        }
        \startdata
        Accuracy           & 0.9851 & 0.9807 & 0.9709 & 0.9680 \\
        Balanced Accuracy  & 0.9800 & 0.9737 & 0.9628 & 0.9593 \\
        ROC AUC            & 0.9986 & 0.9976 & 0.9957 & 0.9950 \\
        $F_1$ Score        & 0.9755 & 0.9683 & 0.9521 & 0.9477 \\
        \enddata
        \tablecomments{Both models are trained and tested on the same subsets for comparability.}
    \end{deluxetable}


    Having established the training set we follow the approach of \cite{Ksoll2018}, training a random forest \citep[RF;][]{Breiman2001} and support vector machine \citep[SVM;][]{CortesVapnik1995} to distinguish between the "PMS" and "Non-PMS" classes based on the photometry in F555W and F814W as well as the estimated extinction in the F555W filter $A_\mathrm{F555W}$. Note that within the method framework established in \cite{Ksoll2018} we do not consider photometric uncertainties. They do not contribute further information when considered as features, and in addition, the implementations of SVM and RF do not have mechanisms to treat uncertainty. For training we split the data set established in the previous section 70:30 into a training and held-out test subset. We use the latter to ascertain training success and performance on unknown data (with known labels) by computing the accuracy, balanced accuracy, the area under the receiver operating characteristic (ROC AUC) curve \citep[for a detailed description of these performance measures, see e.g.~the Appendix in][]{Ksoll2018} and $F_1$ score,
    \begin{equation}
        F_1 = \frac{2\mathrm{TP}}{2\mathrm{TP} + \mathrm{FN} + \mathrm{FP}},
    \end{equation}
    where TP, FP and FN denote the number of true positives, false positives and false negatives, respectively. We train both algorithms using a 10-fold cross-validation, repeated five times, on the training subset using the ROC AUC as the performance metric for model selection. For the SVM we employ a Gaussian radial basis kernel and we find the best RF solutions employing 500 trees. As we perform predictions on only three features, the magnitudes in F555W and F814W and $A_\mathrm{F555W}$, each tree will consider all of these for the split decisions during tree construction. Aside from a predicted label, "PMS" or "Non-PMS", we setup the two classifiers such that they also provide a probability for the "PMS" class. For the RF this probability is estimated by the fraction of votes among the 500 trees for the "PMS" class, while we use Platt's posterior probabilities \citep{Platt1999} to perform this estimate for the SVM model. \\
    Table \ref{tab:TrainingPerformance} summarizes the training results and performance of both algorithms on the held-out test set. Overall we find excellent results for both methods. With accuracies, both regular and balanced, above 96\%  and ROC AUC as well as $F_1$ scores close to the optimal value of 1 our ML classification approach shows great success for the given identification task. The almost equal performance results on the training and test subset across both methods further indicates that the trained models do not suffer from over-fitting. Comparing the two algorithms we find that the SVM does slightly better than the RF achieving the highest scores across all measures. However, given the small differences in the performance scores it is safe to say that they exhibit an equal success rate. 
    
\section{Identification of PMS stars}
    \label{sec:Results}

    \begin{figure}
        \centering
        \includegraphics[width = \linewidth]{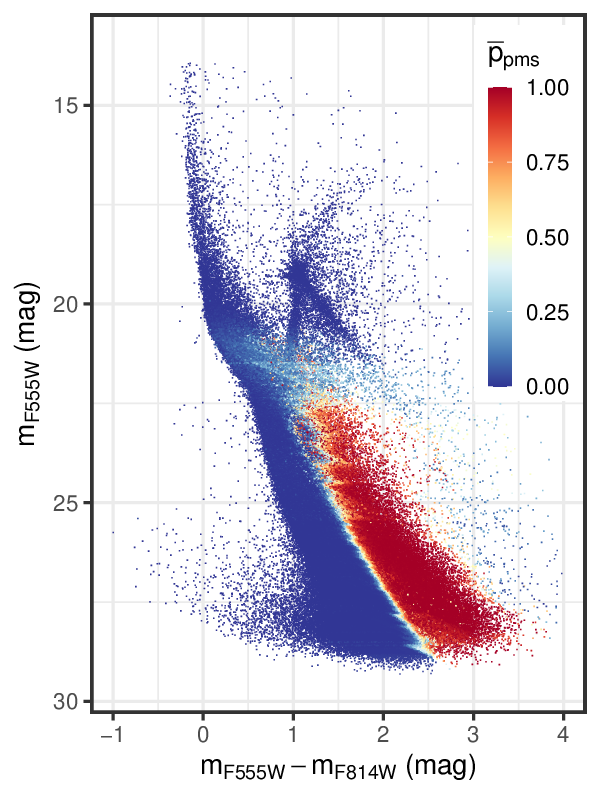}
        \caption{Optical CMD of the MYSST photometric catalog. Each star is color coded according to the mean predicted probability between the SVM and RF model that it belongs to the PMS.}
        \label{fig:PredictionCMD_p_pms_mean}
    \end{figure}
    
    \begin{figure*}
        \centering
        \includegraphics[width = 0.49\linewidth]{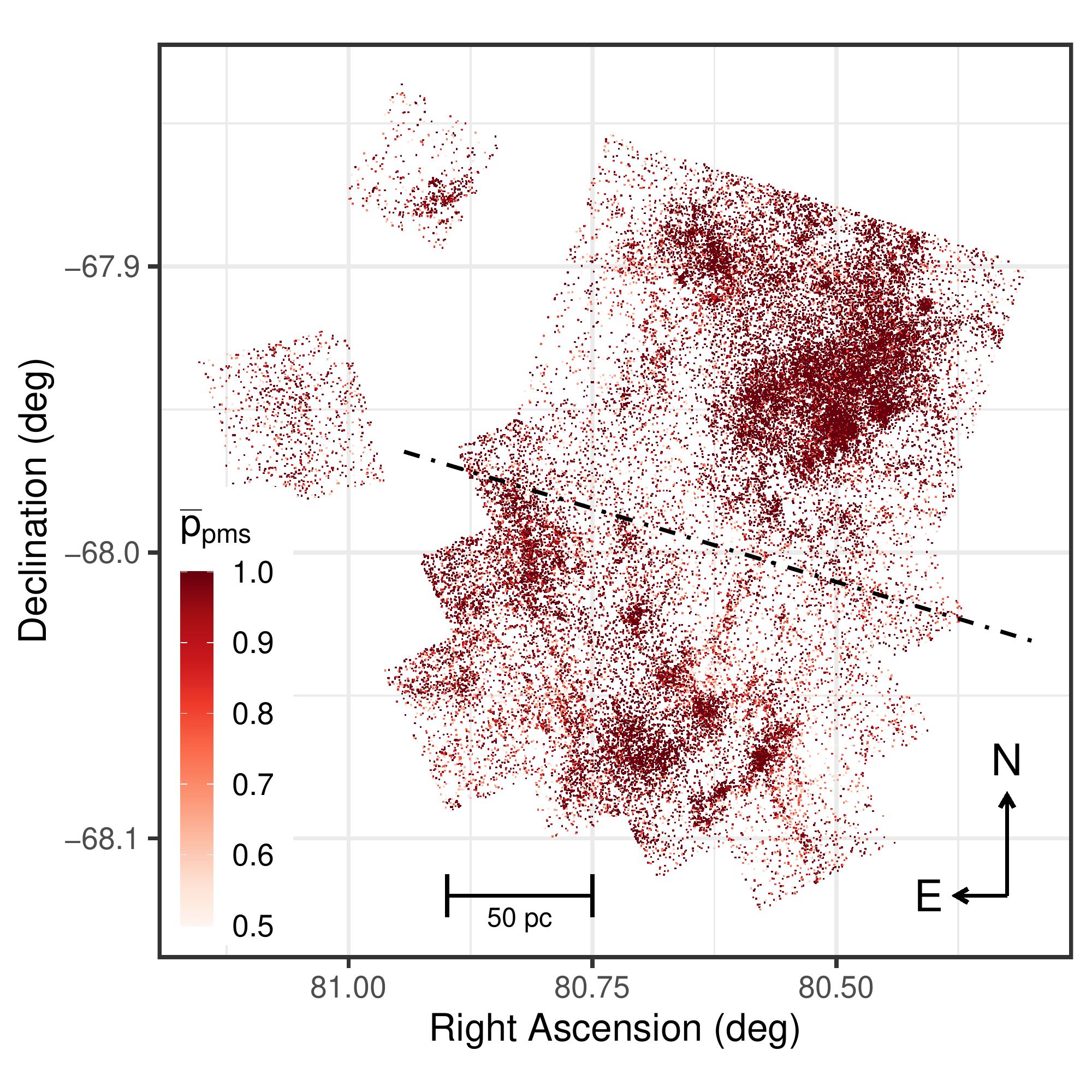}
        \includegraphics[width = 0.49\linewidth]{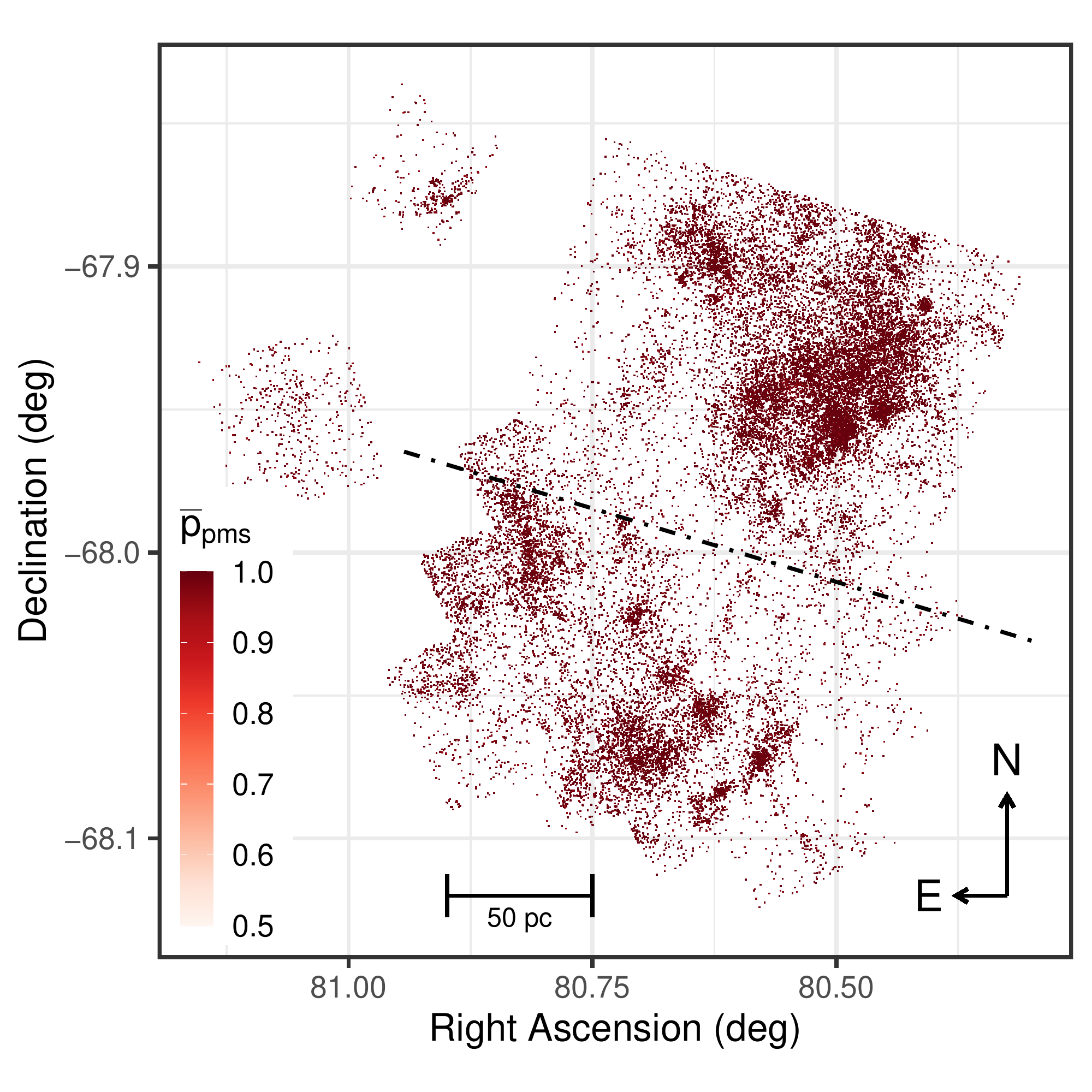}
        \caption{Spatial distribution of the predicted PMS candidate stars. The left panel shows all stars with $\bar{p}_\mathrm{pms} \geq 0.5$, while the right panel indicates the positions of the 28,678 most probable candidate PMS stars with $\bar{p}_\mathrm{pms} \geq 0.95$. The black dot-dashed line in both diagrams indicates our north/south division of the FoV for analysis purposes. }
        \label{fig:PredictionSpatial_p_pms_mean}
    \end{figure*}

        Encouraged by our results on the training and test data we use the trained models to identify the PMS stellar content of the entire MYSST survey by classifying all 461,684 objects. The individual prediction results of the two complementary ML approaches, SVM and RF, in the form of CMDs color coded according to the predicted PMS candidate probabilities, as well as diagrams of the spatial distribution of the most likely PMS candidates, can be found in Figures \ref{fig:PredictionCMD_p_pms_svm_rf_and_diff}, \ref{fig:PredictionSpatial_p_pms_svm} and \ref{fig:PredictionSpatial_p_pms_rf} in the Appendix. Note that the PMS probabilities returned by SVM and RF are a measure of the model's confidence in the prediction of the "PMS" class and \textit{not} the probabilities derived during our Gaussian Mixture Model fit.
        
        In total the SVM identifies 39,818 PMS candidates at a probability of $p_\mathrm{SVM} \geq 0.5$ in the main FoV with a subset of most likely ($p_\mathrm{SVM} \geq 0.95$) candidates consisting of 29,571 stars, while the RF finds 41,909 and 26,610 candidate objects in these two categories, respectively. Therefore, it appears that the SVM is slightly more conservative in the total predicted number of PMS candidates, while the RF seems to put tighter constraints on the most probable PMS constituents. With about 39,000 and 25,500 common predictions in the $p \geq 0.5$ and $p \geq 0.95$ regimes, respectively, both methods nicely agree on the identified PMS population. 
        
        Looking at the predictions more in detail, the SVM exhibits a rather smooth decision boundary in the CMD (Appendix, Figure \ref{fig:PredictionCMD_p_pms_svm_rf_and_diff}, left panel), while the RF entails a more irregular zig-zag shaped class separation, likely an artifact of the underlying partitioning strategy of the RF trees in the low dimensional feature space of our problem. We also see that both classifiers return fairly sharp decision boundaries between the "PMS" and "Non-PMS" classes. From a physical standpoint this may not seem intuitive, because there is source confusion between the LMS and PMS in the low brightness regime and our Gaussian mixture model fit did indeed show a relatively broad transition from one to the other population (c.f.~Figure \ref{fig:TrainingSetFinal}). It is important to emphasize here that this sharp decision boundary is not a physical one, but the one derived by the models to distinguish the two \textit{labels} "PMS" and "Non-PMS" based on the examples in the training set. Since we do not perform a regression on PMS probabilities, but a classification in a low dimensional feature space, the models can, therefore, determine a sharp boundary between our strictly chosen PMS and Non-PMS examples in areas, where the two classes do not overlap significantly.
        
        A direct star by star comparison of the predicted PMS candidate probability (see the right panel of Figure \ref{fig:PredictionCMD_p_pms_svm_rf_and_diff} in the Appendix), shows that the RF tends to make more conservative predictions in the CMD area where PMS and RGB overlap. The SVM, on the other hand, exhibits a more conservative decision boundary between the LMS and PMS in the very low brightness regime. Here we also find that the RF considers several red objects of unclear nature to the right of the PMS as potential candidates, in contrast to the SVM. These very red objects could be young PMS stars that are e.g.~undergoing an extreme accretion event or are variable sources during an event of heightened activity. Since we cannot establish the nature of these objects with the MYSST data alone, we consider the latter RF predictions to be debatable, concluding that the SVM returns more robust results here. On the other hand, there are also some SVM PMS predictions fairly close to and to the left of the RGB, which are likely miss-predictions and are not considered as candidates by the RF. 
        
        Overall, we come to the same conclusion as in our previous study \citep{Ksoll2018}, that a combination of the two classification outcomes provides the most robust prediction result for the PMS stellar content of N44. Figure \ref{fig:PredictionCMD_p_pms_mean} exhibits the classification results if we average the predicted PMS probabilities between SVM and RF as the color code of every star in the CMD. Excluding the two reference fields of the survey, this approach returns a total of 40,509  PMS candidates with $\bar{p}_\mathrm{pms} \geq 0.5$ within the main FoV and a most probable subset consisting of 26,686 stars with $\bar{p}_\mathrm{pms} \geq 0.95$. Figure \ref{fig:PredictionSpatial_p_pms_mean} shows the spatial distributions of these PMS candidates across the area of N44. Notable here is that among the most probable set a majority of 16,976  PMS candidates is located in the northern half of the survey, in and around the massive super bubble of N44, while only 9,710 prospective PMS stars are distributed in the southern region. The black dot-dashed line in Figure \ref{fig:PredictionSpatial_p_pms_mean} indicates our north/south division for the purpose of this discussion. Within the northern part we can see that the PMS stars are mainly concentrated towards the rims of N44's bubble, especially so the western and north-western edge but excluding the south-eastern corner.  
        
        We also recover a number of PMS candidates in the two reference fields of the survey, i.e.~a total of 646 at $\bar{p}_\mathrm{pms} \geq 0.5$ and 346 at $\bar{p}_\mathrm{pms} \geq 0.95$ for the northern field, while the southern one hosts 987 and 439 sources in these two confidence regimes, respectively. The candidate stars in the northern field are concentrated almost entirely at the south-western corner, forming a distinct clump, whereas in the southern field they are more evenly distributed without any apparent structures. Regarding the candidates in the southern field it should be noted that in Paper I we find only very few UMS candidate stars there for the approximation of individual extinction. Additionally, the UMS status of the selected stars remains unclear, so that we believe the estimated extinction values in the southern field to be the most uncertain. Consequently, we recommend to treat the identified PMS candidates in this field with caution.  
    
    \begin{figure*}
        \centering
        \includegraphics[width = \linewidth]{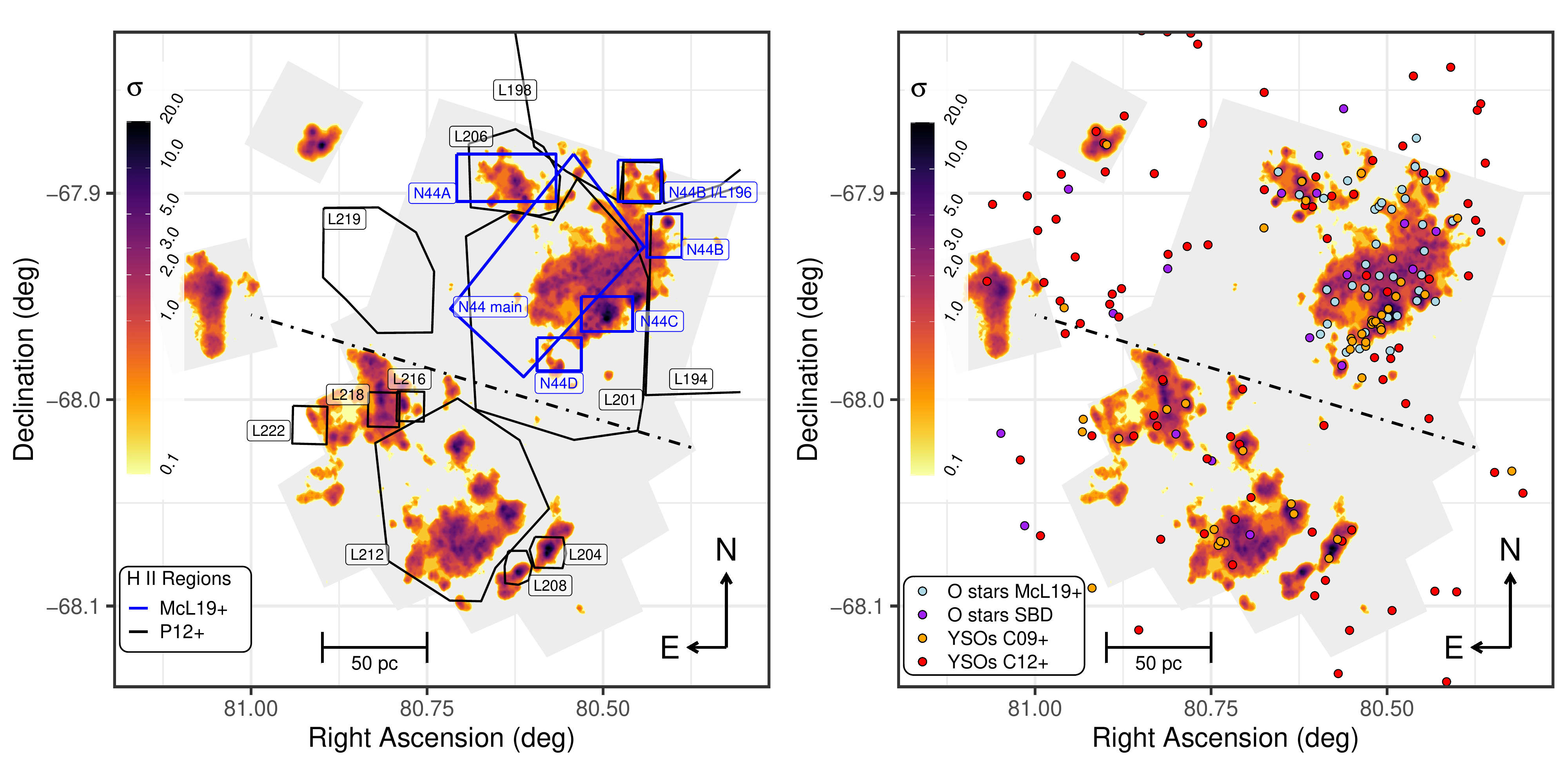}
        \caption{Spatial nearest neighbor contour density diagram of the most probable candidate PMS stars of the MYSST survey (left). The color coding represents the nearest neighbor density in steps of $\sigma$ above the mean density. The black dot-dashed line indicates the north/south-separation for the NNDE. The polygons mark known H II regions around N44, in black as determined by \cite{Pellegrini2012ApJ} and in blue as defined in \cite{McLeod2019}. Note that the boundaries "N44B I" from \cite{McLeod2019} and "L196" from \cite{Pellegrini2012ApJ} coincide. The underlying gray shaded regions (in both panels) indicate the MYSST coverage for comparison. 
        Right: Same diagram as in the left panel, now overlaid with the positions of O stars (light-blue points) as identified by MUSE observations \citep{McLeod2019}. Note that \cite{McLeod2019} only covered the northern half of the MYSST FoV. The purple points signify other known O stars in the SIMBAD database (Appendix, Table \ref{tab:SIMBAD_Ostars}) that are not covered by \cite{McLeod2019}. The orange points indicate massive YSOs identified from Spitzer observations of N44 as found by \cite{Chen2009}. Lastly, the red points mark additional YSOs discovered by \cite{Carlson2012}, excluding matches within 1~arcsec with the \cite{Chen2009} list. Note that 14 YSOs from \cite{Chen2009} and 31 from \cite{Carlson2012} fall outside the shown region.}
        \label{fig:PMS_density_Ostars}
    \end{figure*}
    
    \begin{figure*}
        \centering
        \includegraphics[width = 0.49\linewidth]{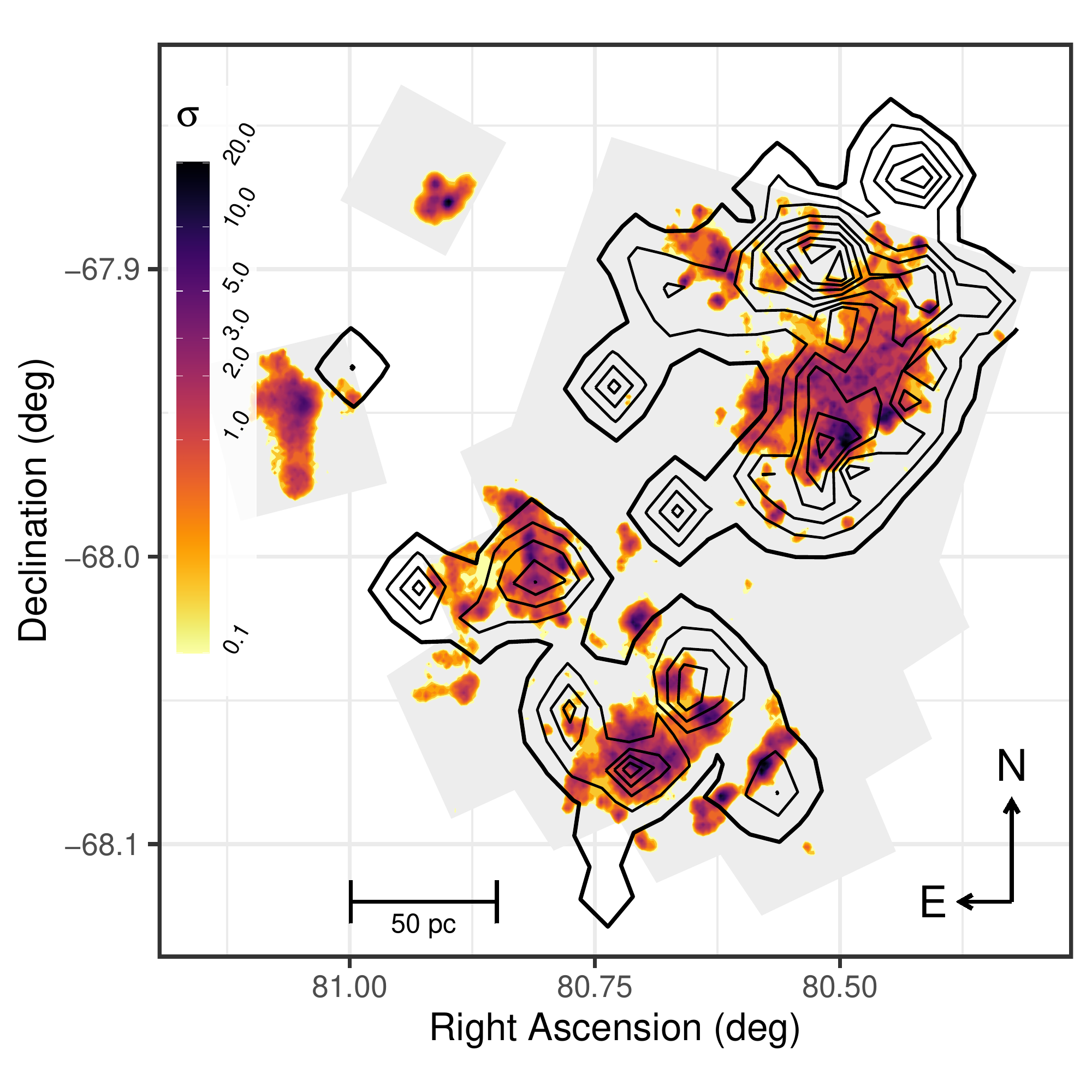}
        \includegraphics[width = 0.49\linewidth]{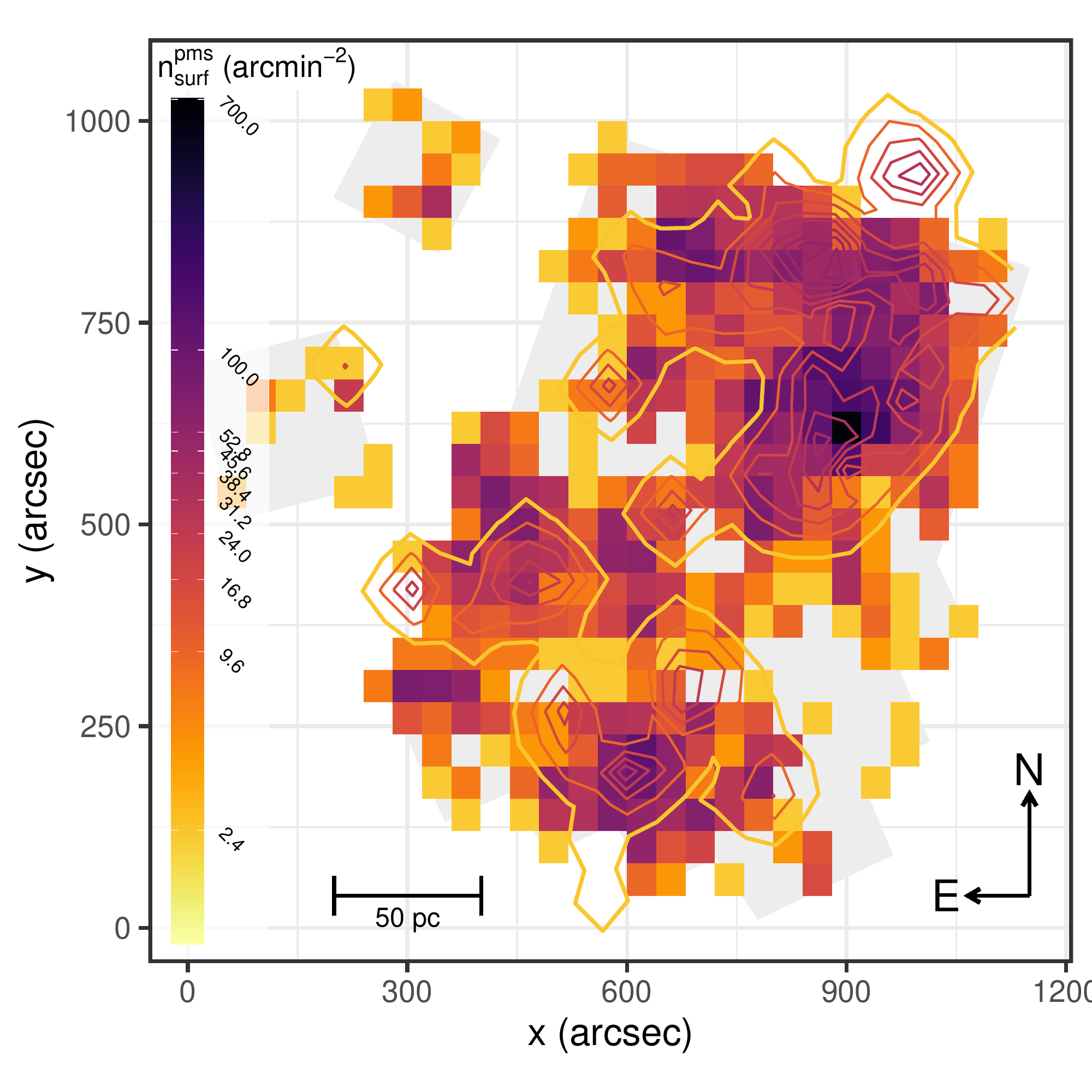}
        \caption{Left: Same PMS nearest neighbor density diagram as in Figure \ref{fig:PMS_density_Ostars}. Overlaid in black are the PMS density contours derived from VMC observations of N44 by \cite{Zivkov2018}. The outermost black contour indicates a number density of $2.4~\mathrm{arcmin^{-2}}$, while each consecutive inner contour indicates an increase by $3 \times 2.4~\mathrm{arcmin^{-2}}$. For comparison the gray shaded region marks the coverage of the MYSST survey in both panels. Right: 2D binned surface density diagram with $40'' \times 40''$ bins of a subset of our PMS catalog matching the PMS mass completeness limit of the VMC survey used in \cite{Zivkov2018}. Overlaid with the same color scheme as the 2d density map are the PMS surface density contours from \cite{Zivkov2018} to allow for quick comparison.}
        \label{fig:PMS_dens_MYSST_vs_Zivkov18}
    \end{figure*}

    \begin{figure}
        \centering
        \includegraphics[width = \linewidth]{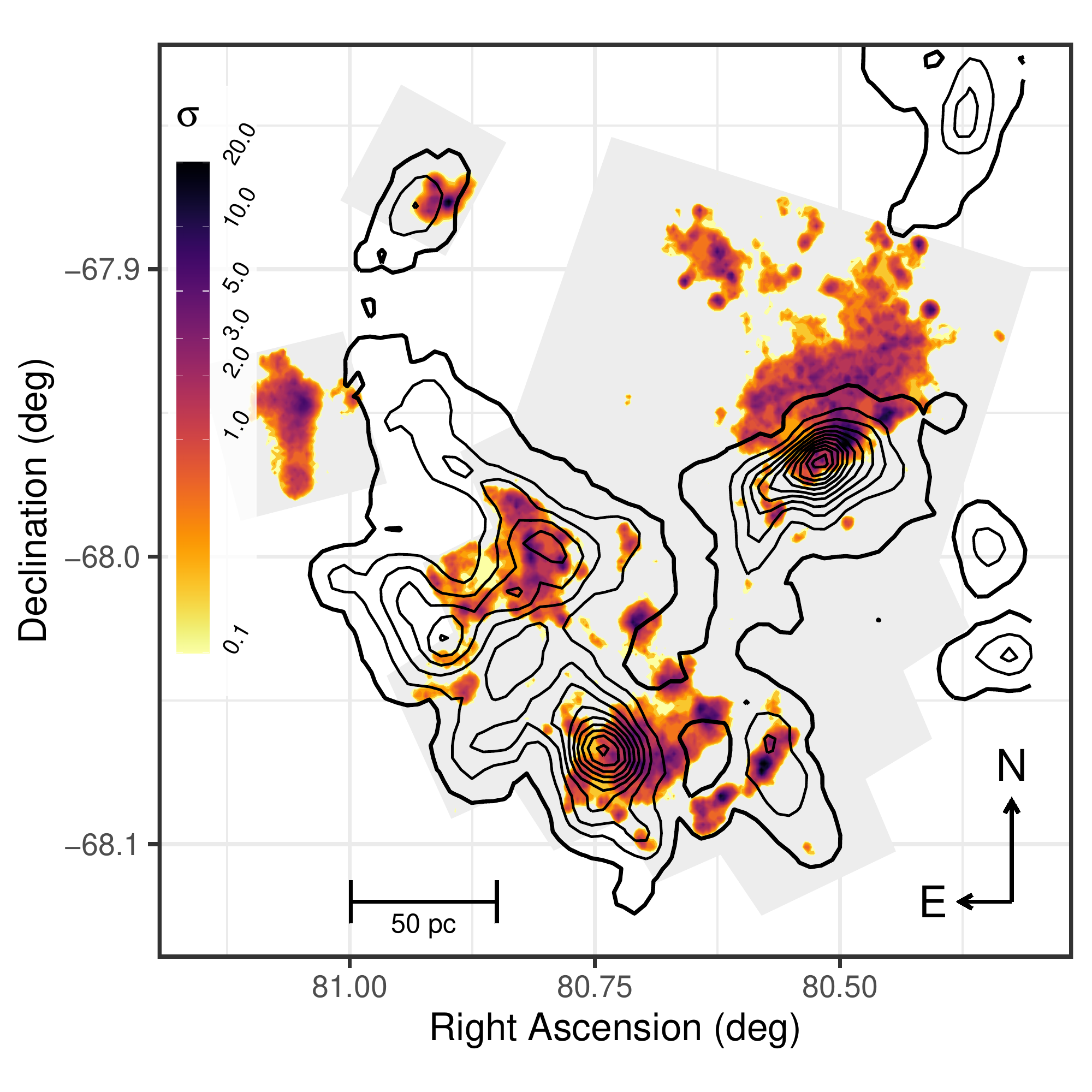}
        \caption{Same PMS nearest neighbor density diagram as in Figure \ref{fig:PMS_density_Ostars}. Overlaid in black are intensity contours of CO emission as observed by the MAGMA survey \citep{Wong2011, Wong2017}. The outermost contour marks the mean CO intensity in the FoV at $1.7~\mathrm{K~kms^{-1}}$, each consecutive inner contour marks an increase in intensity by $1\sigma = 3.5~\mathrm{K~kms^{-1}}$ up to the $10\sigma$ level for the innermost line. For comparison the gray shaded region marks the coverage of the MYSST survey.}
        \label{fig:MYSST_PMS_vs_CO}
    \end{figure}
    
    \begin{figure}
        \centering
        \includegraphics[width = \linewidth]{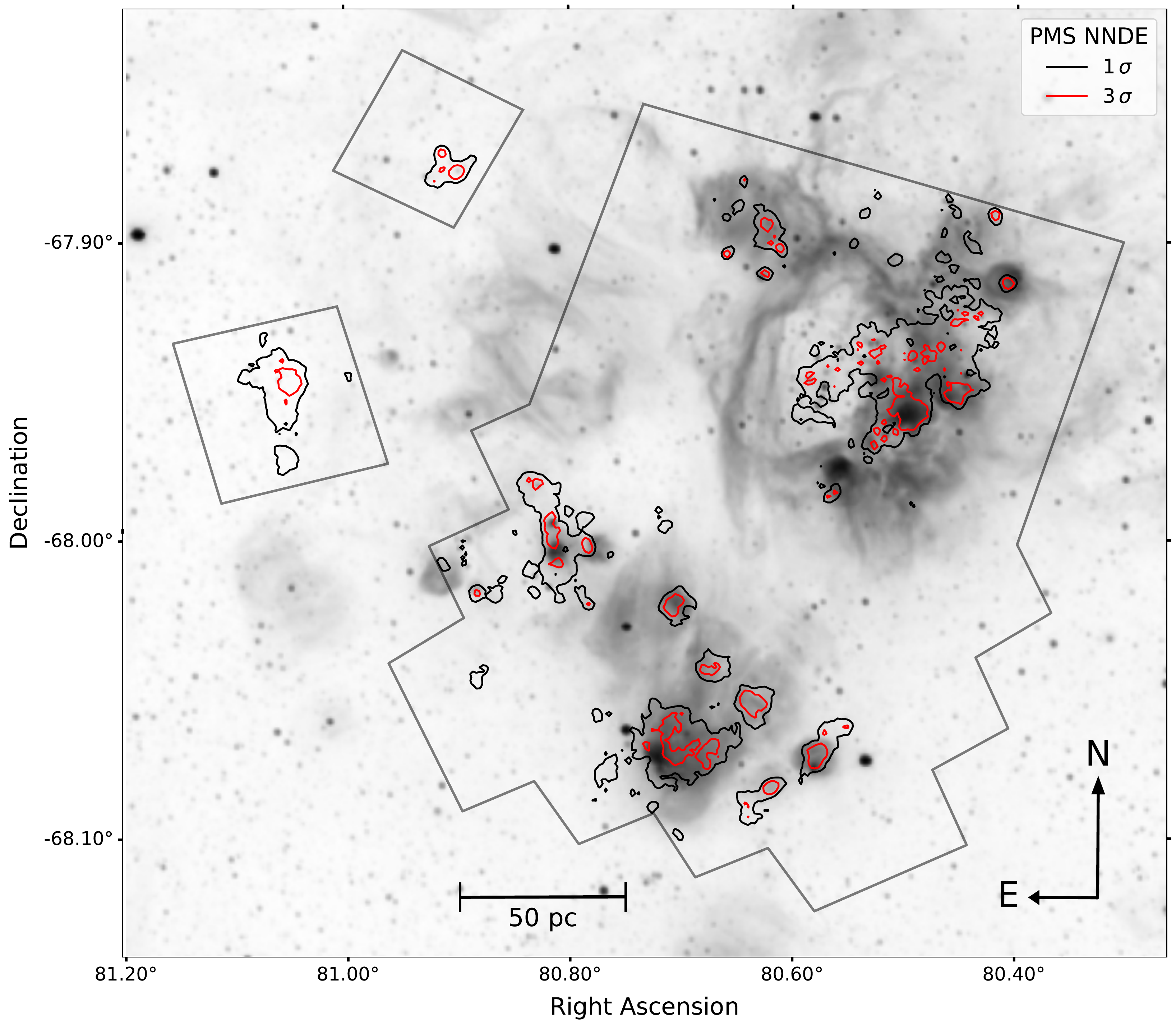}
        \caption{Inverted gray scale image of $\mathrm{H}_\alpha$ emission in N44 as captured by the MCELS survey \citep{Pellegrini2012ApJ}. Overlaid in black and red are our 1 and $3\sigma$ PMS nearest neighbor density contours (c.f. Figure \ref{fig:PMS_density_Ostars}) for comparison. The gray outlines indicate the coverage of the MYSST survey.}
        \label{fig:MYSST_PMS_vs_Halpha}
    \end{figure}
    
    \begin{figure*}
        \centering
        \includegraphics[width = 0.49\linewidth]{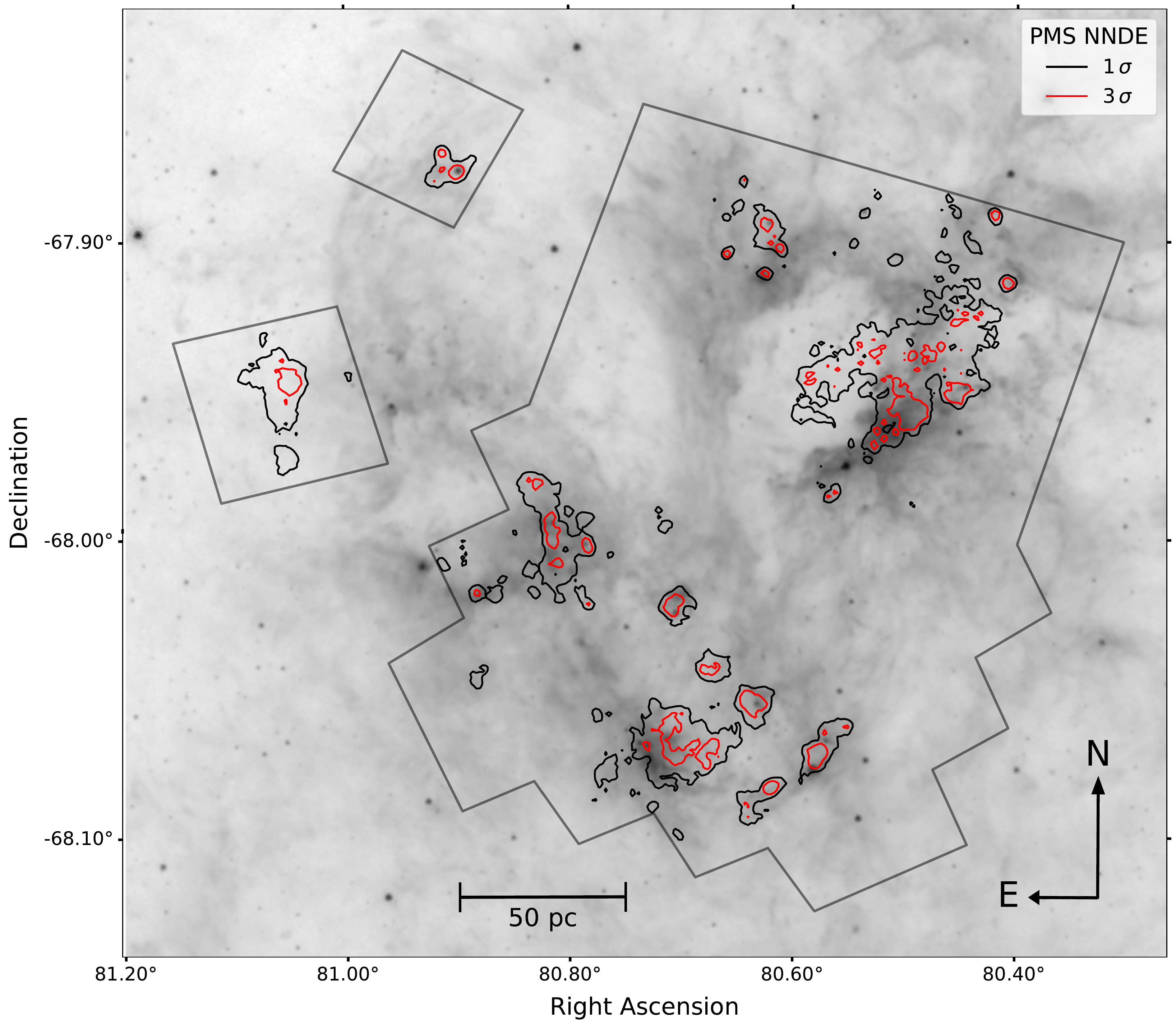}
        \includegraphics[width = 0.49\linewidth]{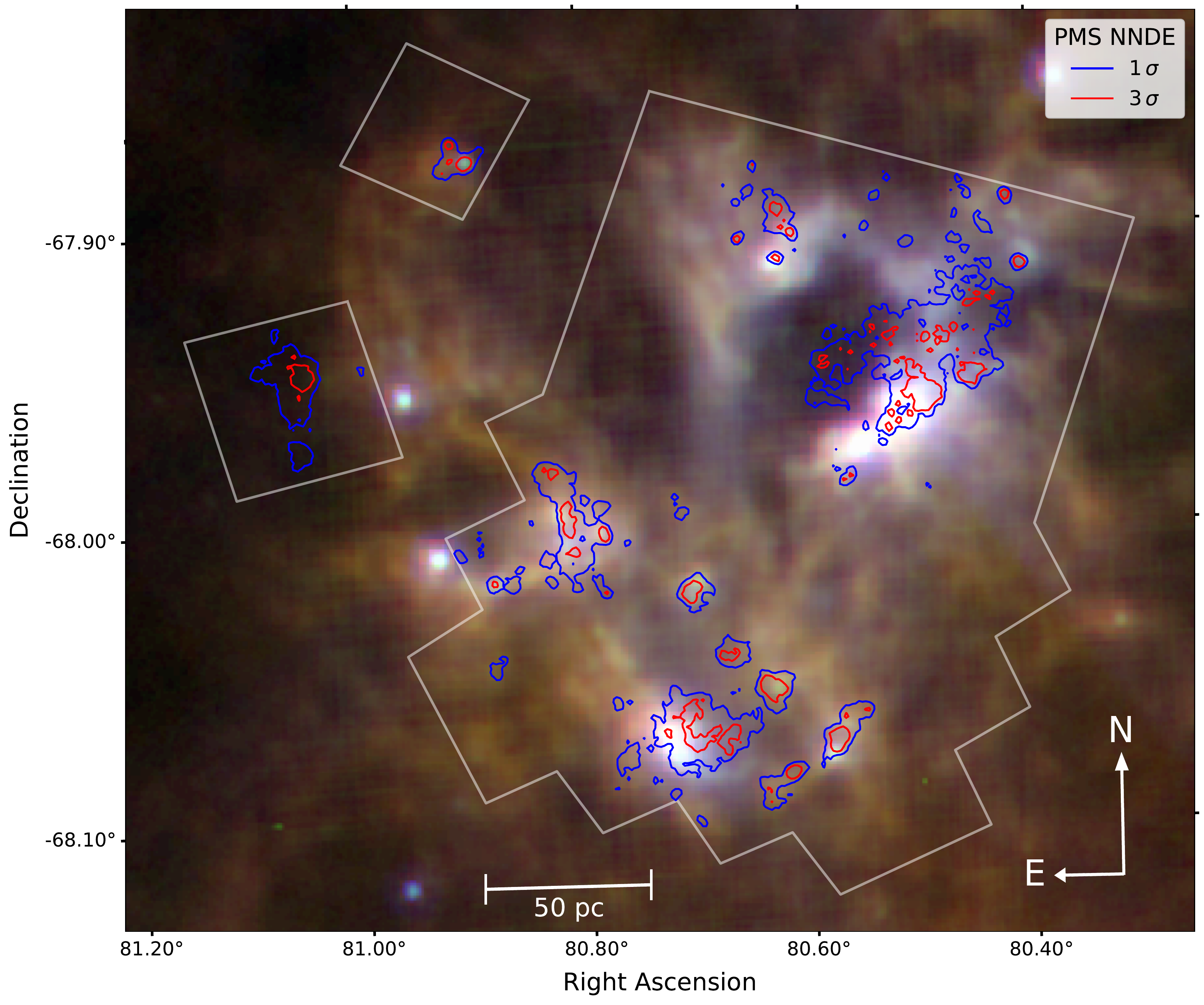}
        \caption{Inverted gray scale image of dust emission at $8~\mu\mathrm{m}$ in N44 from the SAGE survey \citep{Meixner2006} of the LMC (left). Overlaid in black and red are the 1 and 3~$\sigma$ nearest neighbor density contours of the most likely MYSST PMS candidates (as in Figure \ref{fig:PMS_density_Ostars}). They gray outline indicates the coverage of the MYSST survey for comparison. Right: Color composite dust emission image of N44 combining Spitzer $70~\mu\mathrm{m}$ in blue, Herschel $160~\mu\mathrm{m}$ in green and Herschel $350~\mu\mathrm{m}$ in red. Overlaid in blue and red are the 1 and 3~$\sigma$ PMS density contours and the MYSST coverage is indicated in gray. The Herschel observations were taken as part of the HERITAGE survey \citep{Meixner2013}.}
        \label{fig:PMS_density_SpitzerHerschel}
    \end{figure*}
    
    \section{Spatial Distribution of PMS Stars}
    \label{sec:Clustering}

    To better understand the star-formation processes in N44, we investigate the spatial distribution of the PMS candidate stars in more detail. We employ a nearest neighbor search to determine the surface density of PMS candidates and characterize their clustering properties. We also look at the correlation of the PMS candidate stars with other star formation indicators, specifically we compare with the positions of the known O, B stars and YSOs in the region, as well as CO, $\mathrm{H}_\alpha$ and dust emission observations. Additionally, we evaluate how well our spatial PMS candidate distribution matches the one derived by \cite{Zivkov2018} from VMC observations of N44.
    
    \subsection{Location of PMS Stars}
        \label{subse:PMS-clusters}
        To further ascertain the validity of our PMS identification and to study the spatial distribution of these stars we perform a nearest neighbor density estimation (NNDE). We compute the local source density $n_j$, first introduced in astronomy by \cite{Casertano1985}, as 
        \begin{equation}
            n_j = \frac{j-1}{\pi r_j^2},
            \label{eq:NNDE}
        \end{equation}
        where $r_j$ denotes the distance to the $j$th nearest neighbor, on a regular grid within the MYSST main FoV. Note that we modify the density estimate to a surface number density here, instead of the mass density in \cite{Casertano1985}. Similar to our previous application of a KDE, we then compute surface density contours in terms of significance $\sigma$ above the mean estimated density. We find that employing the distance to the 20th nearest neighbor, corresponding to  $j=20$ in Eq. \ref{eq:NNDE}, offers a reasonable compromise between resolution and statistical significance, and allows us to highlight the structures of the identified PMS clusters. Due to difference in number and spatial distribution of the identified PMS candidates between the northern and southern half of the survey (c.f. black dot-dashed line in Figure \ref{fig:PredictionSpatial_p_pms_mean}) we perform the NNDE separately on both regions to better quantify the clustering properties of PMS stars. For the same reasons we also treat the two reference fields individually. Both panels of Figure \ref{fig:PMS_density_Ostars} shows the corresponding nearest neighbor surface density contours. Due to the individual treatment of the four regions the nearest neighbor density differs for the same $\sigma$-significance level between regions. For instance at $1~\sigma$ the nearest neighbor densities are at $2.62~\mathrm{pc^{-2}}$ and $1.49~\mathrm{pc^{-2}}$ in the northern and southern halves of the main FoV, while it reaches only $0.84~\mathrm{pc^{-2}}$ and $0.27~\mathrm{pc^{-2}}$ in the northern and southern reference fields, respectively. As the overall nearest neighbor density in the southern reference field is fairly low, barely reaching $0.5~\mathrm{pc^{-2}}$ even at a $3~\sigma$ significance, it is obvious that the structures here are not entirely comparable to those found in the main FoV.  
        
        For comparison the right panel of Figure \ref{fig:PMS_density_Ostars} also provides the location of O stars derived from MUSE observations \citep[note that this survey only covered the northern half of the MYSST FoV]{McLeod2019}, additional known O type sources in the SIMBAD database (see Appendix, Table \ref{tab:SIMBAD_Ostars}), as well as massive YSOs identified from Spitzer observations \citep{Chen2009} and Spitzer data combined with optical and near infrared photometry \citep{Carlson2012}. Additionally, the left panel indicates the prominent H II regions of N44, as determined by \cite{Pellegrini2012ApJ} and defined in \cite{McLeod2019}. This diagram confirms that the PMS stars identified by our ML classification are primarily located within the H~II regions of N44. The only notable exception here is the H~II region L219, where we do not find a prominent overdensity of PMS candidate stars. Since a large part of this region falls outside of the MYSST FoV, similar to L198 and L194, it is not unlikely that we are simply missing most of the associated PMS clusters. Note also that \cite{Pellegrini2012ApJ} does not find HII regions associated with the structures of PMS candidates we identify in the two reference fields.
        
        In Paper I we use a selection of $\sim1,300$ UMS stars to derive the extinction toward the region. As previously mentioned, the MYSST survey misses the most massive stars of N44 due to saturation effects. Therefore, this selection consists primarily of late O and early B type UMS stars. Comparing this population of young massive stars to our PMS density maps (see Appendix, Figure \ref{fig:MYSST_PMS_vs_UMS}) we also find evidence that they are preferably located in correspondence of the PMS clusters, as more than 35\% (62\%) of them fall into the $1\,\sigma$ ($0\,\sigma$) PMS density contours. For comparison, in a uniform random distribution (averaged over 100 random realizations) only $9.6 \pm 0.5 \%$ ($30.0 \pm 0.5\%$) of objects would fall within the same contours. This provides additional confirmation that the PMS we identify tend to be located in the vicinity of more massive young UMS stars.
        
        Using Hess diagrams to identify PMS regions as density excesses over local field populations, \cite{Zivkov2018} recently provided a PMS surface density map of N44 based on data by the VMC survey. In the left panel of Figure \ref{fig:PMS_dens_MYSST_vs_Zivkov18} we show their PMS surface density contours in comparison to our PMS nearest neighbor density map. Aside from two of their distinct density peaks which fall outside of the MYSST coverage, we find a good match to our nearest neighbor density map within the southern half of the main MYSST FoV. In the northern half we also have a decent agreement along the western edge of the main bubble. However, we identify three density peaks from \cite{Zivkov2018} as well, that do not have a significant counterpart in our PMS nearest neighbor density map. These \cite{Zivkov2018} density peaks are located at the northern bubble rim ($\mathrm{R.A.}=80.53^\circ$; $\mathrm{Dec.}=-67.90^\circ$), the eastern bubble edge ($\mathrm{R.A.}=80.73^\circ$; $\mathrm{Dec.}=-67.94^\circ$) and just south of the bubble ($\mathrm{R.A.} = 80.67^\circ$; $\mathrm{Dec.}=-67.98^\circ$), respectively. The discrepancy in these three regions could be an effect of both the angular resolution and completeness differences between the VMC and MYSST surveys. Employing the VISTA telescope, the VMC project achieves an angular resolution on the order of $0.34''$, a value that is almost ten times larger than the $0.04''$ resolution obtained with the HST in the MYSST observations. Additionally, \cite{Zivkov2018} state that the $5\sigma$ magnitude limit of their photometry catalog corresponds to the brightness of 1 Myr old PMS stars with $0.7~M_\sun$ (reddening corrected), while the MYSST survey reaches down to $0.09~M_\sun$ (albeit unreddened) for stars of that age (Paper I).
        
        To test if the completeness (and resolution) differences between the MYSST and VMC survey can indeed explain the missing density peaks in our PMS distribution in the three identified regions, we select a subset of our PMS candidate catalog that matches the VMC PMS mass limit of $0.7~M_\sun$. Using the 1 Myr PARSEC isochrone and accounting for the average extinction measured in Paper I, the $0.7~M_\sun$ cutoff translates to a limiting magnitude of $24.86~$mag in F555W. Selecting only PMS candidates brighter than this limit reduces our catalog of most likely PMS sources from 27,471 to only 4,002 across the entire MYSST FoV, including the two reference fields. Missing more than 85\% of our identified PMS candidates from this limit alone, it is not unlikely that the PMS density map derived from the VMC data overestimates the significance of these three regions compared to the rest. In fact, \cite{Zivkov2018} only find about $1000\pm38$ PMS stars (as a lower limit) in N44 based on the VMC data.
        
        To approximate the spatial resolution of the \cite{Zivkov2018} approach for identifying PMS regions -- they use a grid of overlapping circular elements with a radius of $40''$-- we compute a 2D binned surface density map with $40'' \times 40''$ bins from the reduced PMS candidate catalog. The right panel of Figure \ref{fig:PMS_dens_MYSST_vs_Zivkov18} shows this map in comparison to the \cite{Zivkov2018} PMS density contours, where bins and contours share the color scheme to easily highlight matching number density levels (in $\mathrm{arcmin}^{-2}$). While our low resolution 2D number density map generally tends to larger values, in particular at the western edge of the bubble, we find that the surface densities in the three regions of question actually match up reasonably well. It is also interesting to note that the small density peak found by \cite{Zivkov2018} close to our southern reference field is matched fairly well in our low resolution density map, even though half of it is actually outside the MYSST FoV. Given these results, the missing density peaks in our full resolution nearest neighbor density map appear to be well explained as a result of the lower completeness and spatial resolution of the VMC data. Therefore, we conclude that our results agree well with the \cite{Zivkov2018} study and provide a significant extension towards very low-mass PMS stars at a higher spatial resolution.
        
        Lastly, we also compare our spatial PMS distribution with other star formation tracers, such as gas and dust emission. In Figure \ref{fig:MYSST_PMS_vs_CO} our PMS nearest neighbor density distribution is shown in comparison to contours of CO emission derived from the Magellanic Mopra Assessment \citep[MAGMA;][]{Wong2011, Wong2017} survey. Here the outermost contour signifies the mean CO intensity in the FoV of $1.7~\mathrm{K~kms^{-1}}$, while each subsequent contour marks an increase by $1\sigma = 3.5~\mathrm{K~kms^{-1}}$ up to a maximum of $10\sigma$. We find a clear correlation of enhanced CO emission to regions of high PMS density in the southern half of the main FoV. In the northern half there is also a very prominent peak in the CO emission that partially coincides with the highest PMS nearest neighbor density at the western edge of the bubble. Quite notable is the absence of CO emission along the northern bubble rim and inside of the bubble, where we still find notable structures of PMS sources. As the very massive stars have cleared out the gas and dust in the bubble, the absence of CO emission there is not surprising. Interesting as well is a small peak of CO emission in the northern reference field, coinciding with the PMS over-density we have identified there. In contrast, the southern field does not exhibit any CO emission.
        
\begin{figure*}
        \centering
        \includegraphics[width = 0.7\linewidth]{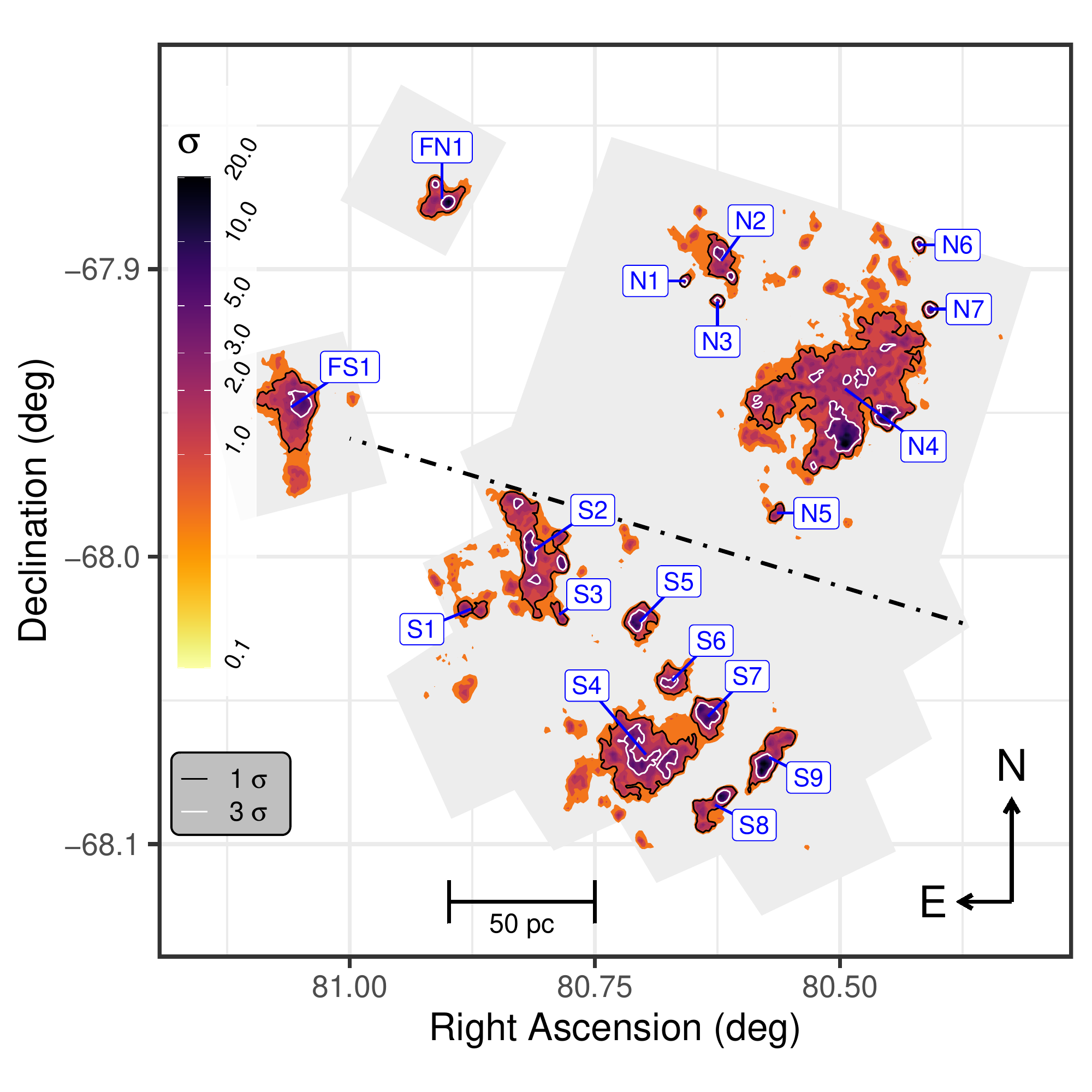}
        \caption{NNDE contour density diagram of the most likely PMS population of N44. The black dot-dashed line indicates the north/south - separation chosen for the NNDE. The gray shaded regions mark the MYSST coverage for comparison. The black contours indicate density structures at $1\sigma$ significance which harbor a minimum of 100 stars and have substructures that persist up to $3\sigma$ significance in density. The white contours mark the $3\sigma$ subclusters which entail at least 50 stars. The labels give an ID of the $1\sigma$ structures for easier reference. Properties of these $1\sigma$ structures and $3\sigma$ subclusters are summarized in Tables \ref{tab:PMS_1sigma_Structures} and \ref{tab:PMS_3sigma_Structures}. }
        \label{fig:MYSST_densityPersistentStructures}
     \end{figure*}

        Figure \ref{fig:MYSST_PMS_vs_Halpha} shows an inverted gray scale image of $\mathrm{H}_\alpha$ emission in N44 as captured by the Magellanic Cloud Emission-Line Survey \citep[MCELS;][]{Pellegrini2012ApJ} in comparison to the $1\sigma$ (black) and $3\sigma$ (red) contours derived from our PMS nearest neighbor density. Overall this Figure demonstrates that the most significant structures of our PMS candidate stars appear correlated with enhanced $\mathrm{H}_\alpha$ emission. As $\mathrm{H}_\alpha$ traces regions of ionized hydrogen this falls inline with our previous assessment that our PMS candidates correlate with the identified H II regions in N44 (which is not entirely surprising as the MCELS $\mathrm{H}_\alpha$ images contributed to the definition of the H II region boundaries in \cite{Pellegrini2012ApJ} to begin with). Notable exceptions to this correlation with enhanced $\mathrm{H}_\alpha$ emission is the part of the large $1\sigma$ contour at the western bubble edge (N4, c.f.~Fig.~\ref{fig:MYSST_densityPersistentStructures}, Section \ref{sec:PMS_Cluster_Identification}) that extends into the bubble interior, and both structures found in the two reference fields. For the bubble interior this is, again, consistent with the fact that the very massive stars located here have driven out most of the gas and dust of their natal environment. 
        
        In the left panel of Figure \ref{fig:PMS_density_SpitzerHerschel} we overlay our $1\sigma$ (black) and $3\sigma$ (red) PMS density contours on an inverted gray scale image of dust emission at $8~\mathrm{\mu m}$, i.e.~emission from polycyclic aromatic hydrocarbon (PAH), as observed by the SAGE survey \citep{Meixner2006} with the Spitzer Space Telescope. In the right panel we show the same comparison with a color composite image of dust emission, combining $70~\mathrm{\mu m}$ emission Spitzer observations from SAGE with $160~\mathrm{\mu m}$ and $350~\mathrm{\mu m}$ Herschel images from the HERschel Inventory of The Agents of Galaxy Evolution project \citep[HERITAGE,][]{Meixner2013}. Visual inspection of both dust maps reveals that many of the structures at $1$ and $3\sigma$ of our PMS nearest neighbor density distribution coincide with areas of increased dust emission, although the $3\sigma$ density peaks are often slightly offset from the maxima of dust surface brightness (e.g.~in Region S4, c.f. Figure \ref{fig:MYSST_densityPersistentStructures}, Section \ref{sec:PMS_Cluster_Identification}). This finding is consistent with the hypothesis that in large concentrations of young stars the irradiation of the dusty remnants of the stellar birth environments leads to bright dust emission in the FIR as the dust re-emits the incoming stellar radiation at longer wavelengths. The large $1\sigma$ structure (N4, c.f.~Fig.~\ref{fig:MYSST_densityPersistentStructures}, Section \ref{sec:PMS_Cluster_Identification}) that partially extends into the bubble is, as for the $\mathrm{H}_\alpha$ emission, again one of the notable exceptions here, explained of course by the feedback of the very massive stars in the bubble interior having cleared out gas and dust. There are also three more prominent structures in the southern half of the main FoV (S6, S7, S8, c.f. Fig.~\ref{fig:MYSST_densityPersistentStructures}, Section \ref{sec:PMS_Cluster_Identification}) that do not appear particularly bright in the dust emission. In the reference fields we find again slightly enhanced emission for the structure found in the northern one, but almost no dust emission in the southern field.
    
    \subsection{Identifying PMS Clusters} 
       \label{sec:PMS_Cluster_Identification}
       The spatial distribution of PMS candidates in N44, e.g.~as indicated in Figure \ref{fig:PredictionSpatial_p_pms_mean}, clearly shows that these stars are distributed in a hierarchical and highly clustered fashion. To identify the PMS clusters, we utilize the nearest neighbor density map (Figure \ref{fig:PMS_density_Ostars}) to first find all density contours at a $1\sigma$ significance level. These contours define our preliminary PMS cluster candidates. We then down select the most prominent PMS clusters if they fulfill a \textit{persistence} criterion of exhibiting substructures at $3\sigma$ density significance. Preliminary we remove all candidate contours that contain less than 100 stars in total (PMS and non-PMS) as they are likely an outcome of noise fluctuations at the $1\sigma$ level and would therefore never fulfill the persistence criterion in the first place. The limit of 100 corresponds to approximately the square root of the number of all PMS sources located in the $1\,\sigma$ contours.  
       
        Applying this contour density based clustering approach, we identify seven prominent PMS structures at $1\sigma$ significance in the northern half of the FoV, nine distinct clusters of PMS candidates in the south and one each in the two reference fields. Again, for this step we use the individual NNDEs of the northern and southern half of the main FoV as well as the reference fields to be more sensitive to local density structures by avoiding the large difference in stellar numbers between the individual regions. Figure \ref{fig:MYSST_densityPersistentStructures} indicates the spatial positions of these eighteen prominent PMS structures. Note that we only show the $1\sigma$ contours (black) of the structures that pass our persistence criteria here. In this figure we also highlight the subclusters at $3\sigma$ significance (white) of the prominent $1\sigma$ structures, excluding, however, those that do not contain at least 50 stars in total. Again, this serves to decrease statistical noise, this time at the $3\sigma$ level, with 50 being approximately the root of the number of all PMS sources inside the $3\,\sigma$ contours. 
    
       Comparison with Figure \ref{fig:PMS_density_Ostars} indicates that almost all of the structures which we identify as PMS clusters are close to or harbor one or more massive O star/YSO. Star formation theory predicts, and recent studies confirm \citep{Cignoni2015, Stephens2017}, that PMS clusters are located primarily in the vicinity of young massive stars. Therefore, our approach identifies clusters of PMS stars where one would expect to encounter them. Combined with the fact that we also find them within the H II regions, which are the remnants of recent star formation events, the comparison with the MUSE \citep{McLeod2019}, SIMBAD and YSO \citep{Chen2009, Carlson2012} data provides an independent confirmation of the validity of our ML classification approach. There is one possible exception, namely the H II region N44D, where we only find a small amount of PMS stars which do not immediately coincide with the two MUSE O stars and three YSOs located there, but only with one O type source in the SIMBAD database. This particular region suffers from a large amount of incompleteness in the MYSST survey due to saturation effects likely caused by the massive O star/YSOs at its center (c.f. Paper I). We note that the small offset between our identified PMS grouping and the other O star/YSOs within this H II region is consistent with the hypothesis that we are simply missing most of the PMS stars around these massive objects.

    \begin{deluxetable*}{lcccccccccccccccc}
        \centerwidetable
        \tablecaption{Properties of the $1\sigma$ PMS Density Structures \label{tab:PMS_1sigma_Structures}}
        \tablehead{
        \colhead{ID} & \colhead{$\mathrm{RA_{cent}}$} & \colhead{$\mathrm{Dec_{cent}}$} & \colhead{$A_\mathrm{surf}$} & \colhead{$R_\mathrm{eff}$} & \colhead{$N_*$} & \colhead{$n_\mathrm{surf}^{total}$} & \colhead{$N_\mathrm{PMS}$} & \colhead{$n_\mathrm{surf}^{pms}$} &
        \colhead{$N_\mathrm{O}^{M19}$} & \colhead{$N_\mathrm{O}^{SD}$} & \colhead{$N_\mathrm{YSO}^{C09}$} & \colhead{$N_\mathrm{YSO}^{C12}$} & \colhead{$N_\mathrm{sub}^{2\sigma}$} & \colhead{$N_\mathrm{sub}^{3\sigma}$} & \colhead{$Q$} & \colhead{$\sigma_Q$} \\
             & (deg) & (deg) & $(\mathrm{pc^2})$ & (pc) & & $(\mathrm{pc^{-2}})$ & & $(\mathrm{pc^{-2}})$ & & & & & & & & 
        }
        \startdata
        N1 & 80.6585    & -67.9041 & 10.3   & 1.8   & 164   & 15.9  & 50    & 4.9 & 0 & 0 & 0 & 0   & 1   & 1(1)     & 0.76 & 0.5 \\
        N2 & 80.6212    & -67.8971 & 95.9   & 5.5   & 1516  & 15.8  & 434   & 4.5 & 1 & 0 & 2 & 1   & 1   & 4(2)     & 0.68 & 0.47 \\
        N3 & 80.6251    & -67.9112 & 12.8   & 2     & 197   & 15.4  & 66    & 5.2 & 0 & 0 & 0 & 0   & 1   & 1(1)     & 0.77 & 0.52 \\
        N4 & 80.4950    & -67.9416 & 1365.3 & 20.8  & 23164 & 17    & 6397  & 4.7 & 16 & 2 & 15 & 5 (2) & 28  & 29(10)    & 0.64 & 0.43  \\
        N5 & 80.5641    & -67.9849 & 18.2   & 2.4   & 252   & 13.9  & 75    & 4.1 & 0 & 1 & 0 & 0   & 1   & 2(0)     & 0.72 & 0.5 \\
        N6 & 80.4202    & -67.8916 & 15.6   & 2.2   & 239   & 15.3  & 85    & 5.4 & 0 & 0 & 1 & 0   & 1   & 1(1)     & 0.78 & 0.5  \\
        N7 & 80.4087    & -67.9139 & 19.6   & 2.5   & 370   & 18.9  & 126   & 6.4 & 1 & 0 & 0 & 0   & 1   & 1(1)     & 0.8 & 0.54 \\
        \hline
        S1 & 80.8765 & -68.0182 & 37.1    & 3.4   & 367   & 9.9   & 86    & 2.3 & 0 & 0 & 1 & 0 & 2   & 1(0)     & 0.60 & 0.43 \\
        S2 & 80.8122 & -67.9978 & 367.9   & 10.8  & 4312  & 11.7  & 902   & 2.5 & 0 & 0 & 2 & 2 & 8   & 5(4)     & 0.55 & 0.39 \\
        S3 & 80.7864 & -68.0201 & 19.5    & 2.5   & 257   & 13.2  & 39    & 2.0 & 0 & 0 & 0 & 0 & 1   & 1(0)     & 0.58 & 0.43 \\
        S4 & 80.6981 & -68.0688 & 504.4   & 12.7  & 6500  & 12.9  & 1308  & 2.6 & 0 & 1 & 3 & 3 (1) & 10  & 4(2)     & 0.71 & 0.46 \\
        S5 & 80.7043 & -68.0226 & 75.8    & 4.9   & 1045  & 13.8  & 233   & 3.1 & 0 & 0 & 1 & 1 & 1   & 1(1)     & 0.73 & 0.47 \\
        S6 & 80.6714 & -68.0432 & 68.8    & 4.7   & 1112  & 16.2  & 181   & 2.6 & 0 & 0 & 0 & 0 & 1   & 1(1)     & 0.75 & 0.46 \\
        S7 & 80.6347 & -68.0556 & 103.8   & 5.7   & 1524  & 14.7  & 336   & 3.2 & 0 & 0 & 2 & 1 (1) & 1   & 1(1)     & 0.76 & 0.48 \\
        S8 & 80.6275 & -68.0863 & 88.3    & 5.3   & 1019  & 11.5  & 255   & 2.9 & 0 & 0 & 0 & 0 & 3   & 2(1)     & 0.54 & 0.41 \\
        S9 & 80.5722 & -68.0698 & 131.4   & 6.5   & 1644  & 12.5  & 454   & 3.5 & 0 & 0 & 2 & 3 (1) & 1   & 3(1)     & 0.63 & 0.48 \\
        \hline
        FN1 & 80.9058 & -67.8756 & 102.7 & 5.7 & 1286 & 12.5 & 179 & 1.7 & 0 & 0 & 1 & 2 & 2 & 3(2) & 0.65 & 0.44 \\
        FS1 & 81.0598 & -67.9480 & 272.5 & 9.3 & 3051 & 11.2 & 116 & 0.4 & 0 & 0 & 0 & 1 & 3 & 3(1) & 0.68 & 0.45 \\
        \enddata
        \tablecomments{
        Properties of the $1\sigma$ PMS Density Structures which persist with substructures up to $3\sigma$ in density and consist of at least 100 stars. Listed are the structure ID as in Figure \ref{fig:MYSST_densityPersistentStructures}, the right ascension $\mathrm{RA_{cent}}$ and declination $\mathrm{Dec_{cent}}$ of the structure center, the surface area $A_\mathrm{surf}$ enclosed by the given density contour, an effective radius $R_\mathrm{eff}$ derived from the surface area, the total number $N_*$ of MYSST catalog stars within the structure, the total surface stellar number density $n_\mathrm{surf}^{total}$, the number of identified most likely PMS stars $N_\mathrm{PMS}$ inside the contour, the corresponding surface number density of PMS sources $n_\mathrm{surf}^{pms}$, the number of enclosed \cite{McLeod2019} O stars $N_\mathrm{O}^{M19}$, SIMBAD O stars (c.f.~Table \ref{tab:SIMBAD_Ostars}) $N_\mathrm{O}^{SD}$, \cite{Chen2009} YSOs $N_\mathrm{YSO}^{C09}$, \cite{Carlson2012} YSOs $N_\mathrm{YSO}^{C12}$ (the number in parenthesis indicates matches in $N_\mathrm{YSO}^{C09}$), and the number of substructures $N_\mathrm{sub}^{2\sigma}$, $N_\mathrm{sub}^{3\sigma}$ at a density significance of 2 and $3\sigma$, respectively. The value in parenthesis in the $N_\mathrm{sub}^{3\sigma}$ column indicates the number of subclusters at $3\sigma$ with at least 50 stars, corresponding to the white contours in Figure \ref{fig:MYSST_densityPersistentStructures}. Lastly, we also provide the \cite{Cartwright2004} $Q$ parameter and its uncertainty as an indicator of cluster 'clumpiness'
        }
    \end{deluxetable*}
 
    \subsection{Properties of the PMS Clusters in N44}
        \label{sec:ClusterProperties}
        To further characterize the properties of the PMS clusters we first determine their center-of-mass position on the sky. This is simply obtained as the average of the position of all cluster members, because we do not have reliable estimates of the physical masses of the PMS candidate stars at this moment\footnote{This situation will be improved with the application of more advanced machine learning techniques \citep[INN]{Ksoll2020} in a future study.}. We also compute the surface area $A_\mathrm{surf}$ encompassed by the corresponding density contour, an effective radius derived as $R_\mathrm{eff} = \sqrt{A_\mathrm{surf}/\pi}$, the total number of MYSST stars $N_*$ inside the structure, as well as the number of most likely PMS candidate stars $N_\mathrm{PMS}$, and corresponding surface number densities for total $n_\mathrm{surf}^{total}$ and PMS candidates $n_\mathrm{surf}^{PMS}$. Additionally, we count the enclosed O stars and YSOs from \cite{McLeod2019}, the SIMBAD database (Table \ref{tab:SIMBAD_Ostars}), \cite{Chen2009} and \cite{Carlson2012}. For the prominent $1\sigma$ structures we determine the number of substructures at 2 and 3$\sigma$ significance in density, $N_\mathrm{sub}^{2\sigma}$ and $N_\mathrm{sub}^{3\sigma}$, based on the dendrogram decomposition \citep{Rosolowsky2008} of the spatial distribution of the PMS candidate stars. Furthermore, we also compute the subclustering parameter $Q$ as defined by \cite{Cartwright2004} and its uncertainty $\sigma_Q$ (more details on the $Q$-parameter follow at the end of this section).
        
        A summary of these properties can be found in Table \ref{tab:PMS_1sigma_Structures} for the $1\sigma$ structures and in Table \ref{tab:PMS_3sigma_Structures} in the Appendix for their $3\sigma$ subclusters. Note that the IDs of the $3\sigma$ substructures in Table~\ref{tab:PMS_3sigma_Structures} indicate the $1\sigma$ structures that they belong to, i.e.~N1.1 is inside N1, N2.1 in N2 etc (see also Figure~\ref{fig:MYSST_persistenStructures_3sigma_labelled} in the Appendix for indicators of their spatial position). Additionally, Figures \ref{fig:DendrogramNorth} and \ref{fig:DendrogramSouth} in the Appendix provide dendrograms of the NNDE density structures in the main FoV up to the $5\sigma$ significance level for a more in depth visualization of the hierarchical clustering structure that we encounter here. 
        
        We find that the PMS clusters in N44 cover a wide range of mass and size, with clearly the most prominent structure being the one denoted as N4. With a surface area of more than 1,300$\,\mathrm{pc}^2$ and effective radius of over 20~pc, it is a very large structure of PMS candidates that traces the western ridge of N44's super bubble and extends into the bubble itself. It stretches across two of the H II regions, namely 'N44 main' and 'N44C', and contains almost 6,400 candidate PMS stars. Given the size of this structure it is unlikely to be a single massive cluster (we are after all working only on a 2D projection).
        
        The radial velocities of the sixteen O stars located within this contour, as measured in \cite{McLeod2019}, do not exhibit a noticeable trend in comparison to the remaining O stars.
        It appears that this cluster formed ``in situ'' in a region of higher gas density as the shell of the expanding H II bubble expands into the ambient medium. Additionally, N4 encloses a total of eighteen YSOs, ranging from $6.5~M_\sun$ to $22.1~M_\sun$ \citep{Chen2009, Carlson2012}, 
        and contains ample amounts of substructure (see Figure \ref{fig:DendrogramNorth}). At the $3\sigma$ density significance level this structure still contains ten subclusters with at least 50 constituents, one of which, the subcluster N4.5 (c.f.~Appendix, Figure~\ref{fig:MYSST_persistenStructures_3sigma_labelled}), entails more than 1,100 PMS candidates. With a PMS surface number density of about $11.1\,\mathrm{pc}^{-2}$ N4.5 is the most prominent star-forming center that we identify in N44. It also harbors three O stars 
        \citep[O5 III, O8 V and O9.5 V;][]{McLeod2019} and one $9.2\,M_\sun$ YSO \citep{Chen2009}. The second largest $3\sigma$ structure, hosting about 400 PMS candidates, is N4.9, which is likely another active star-forming cluster given its PMS surface number density above $11.6\,\mathrm{pc}^{-2}$. It comprises one O5 V star \citep{McLeod2019} and a massive $17.4\,M_\sun$ YSO \citep{Chen2009} as well.
        
        In the south we do not find any structures with $1\sigma$ density significance as large as N4. The most prominent ones are S4 and S2 hosting 1,308 and 902 PMS candidate stars, respectively. Additionally, S4 entails five YSOs \citep[$4.8\,M_\sun$--$15.6\,M_\sun$,][]{Chen2009,Carlson2012} and the O9 II giant Sk-67 82a (c.f.~Table \ref{tab:SIMBAD_Ostars}), whereas S2 hosts four YSOs \citep[$6.9\,M_\sun$--$16.5\,M_\sun$,][]{Chen2009, Carlson2012}. Both clusters exhibit notable sub-structuring with two and four clusters at $3\sigma$ density significance (see also Figure \ref{fig:DendrogramSouth}). Overall, the southern PMS structures appear to be less dense in their PMS stellar content as the PMS surface number density lies on average around $4.9\,\mathrm{pc}^{-2}$ in the subclusters at $3\sigma$ significance, which is only about half the average density of the corresponding structures in the northern part. This lower average surface density of PMS candidate sources could indicate less star-forming activity in the regions south of the main bubble, due to e.g.~less available gas, resulting in fewer present PMS sources. Alternatively, most of the potential PMS sources could be older than 15 Myr, which is the maximum age our classification approach is sensitive to. The most 'active' star-forming subclusters here are S4.1 and S9.1 (c.f.~Appendix, Figure~\ref{fig:MYSST_persistenStructures_3sigma_labelled}) with 312 and 245 PMS candidates, respectively. 
        
        The one structure in the northern reference field, FN1, appears similar in spatial extend to S5 -- S9, but exhibits a notably lower surface density of PMS candidate stars at only $1.7~\mathrm{pc^{-2}}$, which is closer to but still below the smallest $1\sigma$ structures S1 and S3 in the southern main FoV. FN1 also exhibits substructure, with two subclusters at $3\sigma$, and hosts three YSOs \citep[$3.8~M_\sun$--$16~M_\sun$;][]{Chen2009, Carlson2012}. 
        FN1's two $3\sigma$ substructures share PMS candidate surface densities comparable with the $3\sigma$ subclusters in the southern main FoV, with values of $3.7~\mathrm{pc^{-2}}$ and $4.5~\mathrm{pc^{-2}}$ for FN1.1 and FN1.2 (c.f.~Appendix, Figure~\ref{fig:MYSST_persistenStructures_3sigma_labelled}), respectively. This structure appears as a valid cluster candidate along with those identified in the main FoV, although it is located in one of the reference fields, which were supposed to only capture the LMC field population.
        
        The $1\sigma$ structure identified in the southern reference field is among the largest (in area), comparable to S2 and S4. Hosting only 116 PMS candidates, however, it has by far the lowest PMS candidate surface density with $0.4~\mathrm{pc^{-2}}$. This value is by factors of 12.6 and 6.8 smaller than the average PMS surface density of the $1\sigma$ structures in the northern and southern main FoV. Together with the very low PMS nearest neighbor density that defines this $1\sigma$ structure and the uncertainty of the extinction estimate for this field, we believe that it is unclear if FS1 actually traces a star-forming center. There is, however, one YSO \citep[$6.2~M_\sun$,][]{Carlson2012} in FS1, providing some evidence for recent star formation in this structure.
        
        Instead of using the number of substructures identified in the dendrogram analysis as indication of the `clumpiness' or hierarchical nature of the PMS clusters, we can also look at the $Q$ parameter introduced by \cite{Cartwright2004}. It is defined as the ratio of the mean edge length $\bar{m}$ in a minimum spanning tree \citep{Prim1957} constructed from the cluster stars and the mean stellar separation $\bar{s}$, both normalized to the effective cluster radius $R_\mathrm{eff}$. Values of $Q < 0.8$ are indicative of a high degree of substructure, whereas larger values of $Q$ are found in clusters which have a well-defined power-law radial density profile \citep[][]{Cartwright2004, Schmeja2006, Allison2009}. For an application to the structure of young stars in other clusters, see e.g.\ \citet{Schmeja2009} and \citet{Gennaro2017}. The numbers in Table \ref{tab:PMS_1sigma_Structures} indeed indicate that the $Q$ values are lowest in clusters with well defined subclusters ($N_\mathrm{sub}^{2\sigma}$ and $N_\mathrm{sub}^{3\sigma}$ above one) with the one exception possibly being cluster S4. We note that all clusters identified in N44 have $Q < 0.8$ as expected for hierarchically structured or fractal systems. This is confirmed by visual inspection of Figure \ref{fig:PMS_clusters_spatial} which shows the spatial distribution of the PMS candidates in the eighteen clusters N1 -- N7 in the north, S1 -- S9 in the south, as well as FN1 and FS1 in the reference fields, none of which exhibits a clear power-law density fall-off.
        
\section{Summary}
    \label{sec:Summary}
        In this study we present the identification of the pre-main-sequence (PMS) stellar population of the star-forming complex N44 in the Large Magellanic Cloud based on the photometric catalog of the deep HST survey MYSST. For this purpose we apply a machine learning classification approach, which we have previously established \citep{Ksoll2018}, 
        to distinguish the observed sources into the two classes 'PMS' and 'Non-PMS' based on their photometry in the F555W and F814W filters, as well as an estimate of individual stellar extinction.
        
        To apply our classification scheme to the observations of N44 we first construct a suitable training set by selecting 
        a region of N44 which exhibits a high density in PMS sources (as determined by a kernel density estimate on a rough selection of PMS candidate stars). This region provides both a distinct PMS and lower main-sequence (LMS) population, which we distinguish using a Gaussian mixture model approach described 
        in \cite{Ksoll2018}. As stars on the red giant branch (RGB) can also contaminate the CMD region usually occupied by PMS stars we extend our training set through addition of RGB 'Non-PMS' examples selected from a series of LMC field regions within the observed FoV. Our final training set consists of 17,942 stars of which 5,512 are PMS examples.
        
        In the following, we train a support vector machine (SVM) and a random forest (RF) classifier to distinguish the two classes 'PMS' and 'Non-PMS' using the magnitudes in F555W and F814W, as well as the estimated stellar extinction as the feature space. To evaluate training success we hold out a randomly selected subset (30\% of the total training data) as a test set and compute a series of standard performance measures, i.e.~the normal and balanced accuracy, the area under the receiver operating characteristic curve (ROC AUC) and the $F_1$ score. We find that both models achieve excellent results on both the training and test sub-sets with accuracies exceeding 96\%, as well as ROC AUCs above 0.99 and $F_1$ scores beyond 0.94.
        
        Classifying the remaining data of the MYSST survey, 
        we determine that an average of the predicted probability for the 'PMS' class between the SVM and RF methods provides the most robust outcome. \textit{With that we find 40,509 potential PMS candidates satisfying $\bar{p}_\mathrm{pms} \geq 0.5$ and a most likely subset with $\bar{p}_\mathrm{pms} \geq 0.95$ consisting of 26,686 sources across N44.} Adopting the latter criterion, a majority of 16,976 PMS candidate stars are identified in and around N44's massive super bubble, located in the northern half of the MYSST FoV, while only 9,710 candidate PMS sources are found in the region south of the bubble.
        
        We then perform a nearest neighbor density estimate \citep[NNDE;][]{Casertano1985} on the set of most likely PMS candidates to characterize their spatial distribution and clustering structures. 
        Comparing with previous studies of the H II regions of N44 \citep{McLeod2019, Pellegrini2012ApJ}, we confirm that the majority of the dominant groupings of PMS candidate stars revealed by our ML classification approach coincide with N44's known H II regions. Further comparison with MUSE observations \citep{McLeod2019} of the most massive young O star population of N44's bubble reveals that, at least within the FoV overlap of the two studies, almost all of our PMS clusters harbor one or more of the young high-mass stars. We find a similar result comparing with the positions of massive YSOs identified in N44 \citep{Carlson2012, Chen2009}. Therefore, we conclude that our classification approach identifies PMS sources exactly where one would expect to find them, i.e.~within N44's gas reservoirs and in the vicinity of its massive young population. \textit{This supports the hypothesis that stars tend to form in clusters} \citep[see also][]{LadaLada2003, Klessen1998, Bonnell1998}.
        
        Additionally, we perform a comparison of our spatial PMS candidate distribution with the \cite{Zivkov2018} study, which has previously established a lower limit of $1000\pm38$ for the number of PMS stars in N44 and derived a PMS surface density map for the region, based on the VMC survey. We find an overall decent agreement with their results, in particular when we account for the completeness and resolution limits of the VMC survey, and conclude that our study provides an excellent extension of their results to much lower brightness and higher spatial resolution.
        
        We also compare the spatial distribution of our PMS candidate stars to other tracers of star formation, i.e.~images of CO, $\mathrm{H}_\alpha$ and dust (at $7~\mathrm{\mu m}$, $70~\mathrm{\mu m}$, $160~\mathrm{\mu m}$ and $350~\mathrm{\mu m}$) emission. Here we find that \textit{most of the prominent structures of PMS candidates appear correlated with areas of enhanced gas and dust emission}, with the most prominent exception being the interior of N44's super bubble, where massive stellar feedback has cleared out most of the material.
        
        To assess the prominent PMS structures across N44, we use the NNDE to identify dominant groupings as density contours at $1\sigma$ significance (above the mean estimated density) which entail at least 100 stars in total and have sub-structures that persist up to the $3\sigma$ level. Here we perform separate NNDEs for the northern and southern half of the main FoV, as well as the two reference fields, to account for the difference in number of PMS candidates between the four regions and be more sensitive to the local clustering structures. \textit{This procedure reveals eighteen dominant PMS structures at $1\sigma$ in total, seven located in the north, nine in the south and one each in the two fields.} For all of these we derive several properties, i.e., the center coordinates, surface area, effective radius, numbers and surface number densities of total/PMS stars, as well as the \cite{Cartwright2004} $Q$-parameter for cluster 'clumpiness'. In the north the most dominant structure we find is a very large grouping of more than 6,500 PMS stars that stretches along the western edge of the super bubble and extends into the bubble itself. While this structure is too large in size to be a single PMS cluster it appears as a common envelope connecting the numerous star-forming centers at $3\sigma$ significance that fall within it. In the south we find more but slightly smaller PMS groupings which appear overall less densely populated in terms of PMS sources, i.e.~they exhibit PMS surface number densities that are on average only half as large as in the north. We suspect that this \textit{hints at a reduced star-forming activity in the south compared to the north.} On top of that the identified dominant PMS groupings in both the north and south exhibit ample hierarchical substructures. 

        Following the outcomes of this study there are a few open questions, which we plan to address in a future investigation. First and foremost is the physical characterization of the identified PMS candidates by estimating their most fundamental properties, age and mass. We aim to achieve this through further development of an invertible neural network based regression approach which we have recently presented in a pilot study \citep{Ksoll2020} with very promising results on the test cases of Westerlund 2 and NGC6397. Establishing these physical properties of the PMS stars of N44 will allow us to quantify the star formation history of this complex and investigate if there is, e.g.~an age difference between the clustering structures we have identified in the northern and southern part of the MYSST FoV. Furthermore, we plan to re-evaluate our clustering analysis with regards to the predicted physical properties of the PMS stars to establish a comprehensive picture of the spatial distribution of star formation in this star-forming complex. 
        
\section*{Acknowledgments}

We thank the anonymous referee for their timely and thorough review of our manuscript, helping us greatly to provide a more complete and concise version of this study.\\
We also thank Viktor Zivkov for providing access to the results of the \cite{Zivkov2018} study, facilitating the comparison with our data. \\
VFK was funded by the Heidelberg Graduate School of Mathematical and Computational Methods for the Sciences (HGS MathComp), founded by DFG grant GSC 220 in the German Universities Excellence Initiative. VFK also acknowledges support from the International Max Planck Research School for Astronomy and Cosmic Physics at the University of Heidelberg (IMPRS-HD).\\
RSK acknowledges financial support from the German Research Foundation (DFG) via the collaborative research center (SFB 881, Project-ID 138713538) “The Milky Way System” (subprojects A1, B1, B2, and B8). He also thanks for funding from the Heidelberg Cluster of Excellence ``STRUCTURES'' in the framework of Germany’s Excellence Strategy (grant EXC-2181/1, Project-ID 390900948) and for funding from the European Research Council via the ERC Synergy Grant ``ECOGAL'' (grant 855130) and the ERC Advanced Grant ``STARLIGHT'' (grant 339177). \\
The project ``MYSST: "Mapping Young Stars in Space and Time'' is supported by the German Ministry for Education and Research (BMBF) through grant 50OR1801. \\
Based on observations with the NASA/ESA Hubble Space Telescope obtained from the Mikulski Archive for Space Telescopes at the Space Telescope Science Institute, which is operated by the Association of Universities for Research in Astronomy, Incorporated, under NASA contract NAS5-26555. Support for program number GO-14689 was provided through a grant from the STScI under NASA contract NAS5-26555. \\
Based on photographic data obtained using The UK Schmidt Telescope. The UK Schmidt Telescope was operated by the Royal Observatory Edinburgh, with funding from the UK Science and Engineering Research Council, until 1988 June, and thereafter by the Anglo-Australian Observatory. Original plate material is copyright (c) of the Royal Observatory Edinburgh and the Anglo-Australian Observatory. The plates were processed into the present compressed digital form with their permission. The Digitized Sky Survey was produced at the Space Telescope Science Institute under US Government grant NAG W-2166.


\bibliography{paper2.bib}{}

\begin{thebibliography}{}
\expandafter\ifx\csname natexlab\endcsname\relax\def\natexlab#1{#1}\fi
\providecommand{\url}[1]{\href{#1}{#1}}
\providecommand{\dodoi}[1]{doi:~\href{http://doi.org/#1}{\nolinkurl{#1}}}
\providecommand{\doeprint}[1]{\href{http://ascl.net/#1}{\nolinkurl{http://ascl.net/#1}}}
\providecommand{\doarXiv}[1]{\href{https://arxiv.org/abs/#1}{\nolinkurl{https://arxiv.org/abs/#1}}}

\bibitem[{{Allison} {et~al.}(2009){Allison}, {Goodwin}, {Parker}, {Portegies
  Zwart}, {De Grijs}, \& {Kouwenhoven}}]{Allison2009}
{Allison}, R.~J., {Goodwin}, S.~P., {Parker}, R.~J., {et~al.} 2009, \mnras,
  395, 1449, \dodoi{10.1111/j.1365-2966.2009.14508.x}

\bibitem[{{Baron}(2019)}]{Baron2019}
{Baron}, D. 2019, arXiv e-prints, arXiv:1904.07248.
\newblock \doarXiv{1904.07248}

\bibitem[{{Bohannan} \& {Walborn}(1989)}]{Bohannan1989}
{Bohannan}, B., \& {Walborn}, N.~R. 1989, \pasp, 101, 520,
  \dodoi{10.1086/132463}

\bibitem[{{Bonnell} {et~al.}(1998){Bonnell}, {Bate}, \&
  {Zinnecker}}]{Bonnell1998}
{Bonnell}, I.~A., {Bate}, M.~R., \& {Zinnecker}, H. 1998, \mnras, 298, 93,
  \dodoi{10.1046/j.1365-8711.1998.01590.x}

\bibitem[{Breiman(2001)}]{Breiman2001}
Breiman, L. 2001, Machine Learning, 45, 5, \dodoi{10.1023/A:1010933404324}

\bibitem[{{Bressan} {et~al.}(2012){Bressan}, {Marigo}, {Girardi}, {Salasnich},
  {Dal Cero}, {Rubele}, \& {Nanni}}]{Bressan2012}
{Bressan}, A., {Marigo}, P., {Girardi}, L., {et~al.} 2012, \mnras, 427, 127,
  \dodoi{10.1111/j.1365-2966.2012.21948.x}

\bibitem[{{Brunet} {et~al.}(1975){Brunet}, {Imbert}, {Martin}, {Mianes},
  {Pr{\'e}vot}, {Rebeirot}, \& {Rousseau}}]{Brunet1975}
{Brunet}, J.~P., {Imbert}, M., {Martin}, N., {et~al.} 1975, \aaps, 21, 109

\bibitem[{{Cannon} \& {Pickering}(1993)}]{Cannon1993}
{Cannon}, A.~J., \& {Pickering}, E.~C. 1993, VizieR Online Data Catalog,
  III/135A

\bibitem[{{Carlson} {et~al.}(2012){Carlson}, {Sewi{\l}o}, {Meixner}, {Romita},
  \& {Lawton}}]{Carlson2012}
{Carlson}, L.~R., {Sewi{\l}o}, M., {Meixner}, M., {Romita}, K.~A., \& {Lawton},
  B. 2012, \aap, 542, A66, \dodoi{10.1051/0004-6361/201118627}

\bibitem[{{Cartwright} \& {Whitworth}(2004)}]{Cartwright2004}
{Cartwright}, A., \& {Whitworth}, A.~P. 2004, \mnras, 348, 589,
  \dodoi{10.1111/j.1365-2966.2004.07360.x}

\bibitem[{{Casertano} \& {Hut}(1985)}]{Casertano1985}
{Casertano}, S., \& {Hut}, P. 1985, \apj, 298, 80, \dodoi{10.1086/163589}

\bibitem[{{Chabrier}(2003)}]{Chabrier03}
{Chabrier}, G. 2003, \pasp, 115, 763, \dodoi{10.1086/376392}

\bibitem[{{Chen} {et~al.}(2009){Chen}, {Chu}, {Gruendl}, {Gordon}, \&
  {Heitsch}}]{Chen2009}
{Chen}, C. H.~R., {Chu}, Y.-H., {Gruendl}, R.~A., {Gordon}, K.~D., \&
  {Heitsch}, F. 2009, \apj, 695, 511, \dodoi{10.1088/0004-637X/695/1/511}

\bibitem[{{Chu} {et~al.}(1993){Chu}, {Mac Low}, {Garcia-Segura}, {Wakker}, \&
  {Kennicutt}}]{Chu1993}
{Chu}, Y.-H., {Mac Low}, M.-M., {Garcia-Segura}, G., {Wakker}, B., \&
  {Kennicutt}, Robert~C., J. 1993, \apj, 414, 213, \dodoi{10.1086/173069}

\bibitem[{{Cignoni} {et~al.}(2015){Cignoni}, {Sabbi}, {van der Marel}, {Tosi},
  {Zaritsky}, {Anderson}, {Lennon}, {Aloisi}, {de Marchi}, {Gouliermis},
  {Grebel}, {Smith}, \& {Zeidler}}]{Cignoni2015}
{Cignoni}, M., {Sabbi}, E., {van der Marel}, R.~P., {et~al.} 2015, \apj, 811,
  76, \dodoi{10.1088/0004-637X/811/2/76}

\bibitem[{{Cioni} {et~al.}(2011){Cioni}, {Clementini}, {Girardi}, {Guandalini},
  {Gullieuszik}, {Miszalski}, {Moretti}, {Ripepi}, {Rubele}, {Bagheri},
  {Bekki}, {Cross}, {de Blok}, {de Grijs}, {Emerson}, {Evans}, {Gibson},
  {Gonzales-Solares}, {Groenewegen}, {Irwin}, {Ivanov}, {Lewis}, {Marconi},
  {Marquette}, {Mastropietro}, {Moore}, {Napiwotzki}, {Naylor}, {Oliveira},
  {Read}, {Sutorius}, {van Loon}, {Wilkinson}, \& {Wood}}]{Cioni2011}
{Cioni}, M. R.~L., {Clementini}, G., {Girardi}, L., {et~al.} 2011, \aap, 527,
  A116, \dodoi{10.1051/0004-6361/201016137}

\bibitem[{{Conti} {et~al.}(1986){Conti}, {Garmany}, \& {Massey}}]{Conti1986}
{Conti}, P.~S., {Garmany}, C.~D., \& {Massey}, P. 1986, \aj, 92, 48,
  \dodoi{10.1086/114133}

\bibitem[{Cortes \& Vapnik(1995)}]{CortesVapnik1995}
Cortes, C., \& Vapnik, V. 1995, Machine Learning, 20, 273,
  \dodoi{10.1007/BF00994018}

\bibitem[{{Da Rio} {et~al.}(2010){Da Rio}, {Gouliermis}, \&
  {Gennaro}}]{DaRio2010}
{Da Rio}, N., {Gouliermis}, D.~A., \& {Gennaro}, M. 2010, \apj, 723, 166,
  \dodoi{10.1088/0004-637X/723/1/166}

\bibitem[{{Da Rio} {et~al.}(2012){Da Rio}, {Gouliermis}, {Rochau}, {Pasquali},
  {Setiawan}, \& {De Marchi}}]{DaRio2012}
{Da Rio}, N., {Gouliermis}, D.~A., {Rochau}, B., {et~al.} 2012, \mnras, 422,
  3356, \dodoi{10.1111/j.1365-2966.2012.20851.x}

\bibitem[{{De Marchi} {et~al.}(2010){De Marchi}, {Panagia}, \&
  {Romaniello}}]{DeMarchi2010}
{De Marchi}, G., {Panagia}, N., \& {Romaniello}, M. 2010, \apj, 715, 1,
  \dodoi{10.1088/0004-637X/715/1/1}

\bibitem[{{De Marchi} {et~al.}(2016){De Marchi}, {Panagia}, {Sabbi}, {Lennon},
  {Anderson}, {van der Marel}, {Cignoni}, {Grebel}, {Larsen}, {Zaritsky},
  {Zeidler}, {Gouliermis}, \& {Aloisi}}]{DeMarchi2016}
{De Marchi}, G., {Panagia}, N., {Sabbi}, E., {et~al.} 2016, \mnras, 455, 4373,
  \dodoi{10.1093/mnras/stv2528}

\bibitem[{{Elmegreen}(2011)}]{Elmegreen2011}
{Elmegreen}, B.~G. 2011, in EAS Publications Series, Vol.~51, EAS Publications
  Series, ed. C.~{Charbonnel} \& T.~{Montmerle}, 45--58,
  \dodoi{10.1051/eas/1151004}

\bibitem[{{Fluke} \& {Jacobs}(2020)}]{Fluke2020}
{Fluke}, C.~J., \& {Jacobs}, C. 2020, WIREs Data Mining and Knowledge
  Discovery, 10, e1349, \dodoi{10.1002/widm.1349}

\bibitem[{{Gennaro} {et~al.}(2017){Gennaro}, {Goodwin}, {Parker}, {Allison}, \&
  {Brandner}}]{Gennaro2017}
{Gennaro}, M., {Goodwin}, S.~P., {Parker}, R.~J., {Allison}, R.~J., \&
  {Brandner}, W. 2017, \mnras, 472, 1760, \dodoi{10.1093/mnras/stx2098}

\bibitem[{{Gouliermis} {et~al.}(2006){Gouliermis}, {Brandner}, \&
  {Henning}}]{Gouliermis2006}
{Gouliermis}, D., {Brandner}, W., \& {Henning}, T. 2006, \apj, 641, 838,
  \dodoi{10.1086/500500}

\bibitem[{{Gouliermis}(2012)}]{Gouliermis2012}
{Gouliermis}, D.~A. 2012, \ssr, 169, 1, \dodoi{10.1007/s11214-012-9868-2}

\bibitem[{{Gouliermis} {et~al.}(2012){Gouliermis}, {Schmeja}, {Dolphin},
  {Gennaro}, {Tognelli}, \& {Prada Moroni}}]{Gouliermis2012b}
{Gouliermis}, D.~A., {Schmeja}, S., {Dolphin}, A.~E., {et~al.} 2012, \apj, 748,
  64, \dodoi{10.1088/0004-637X/748/1/64}

\bibitem[{Hartmann {et~al.}(2016)Hartmann, Herczeg, \& Calvet}]{Hartmann16}
Hartmann, L., Herczeg, G., \& Calvet, N. 2016, \araa, 54, 135,
  \dodoi{10.1146/annurev-astro-081915-023347}

\bibitem[{{Henize}(1956)}]{Henize1956}
{Henize}, K.~G. 1956, \apjs, 2, 315, \dodoi{10.1086/190025}

\bibitem[{{Jaskot} {et~al.}(2011){Jaskot}, {Strickland}, {Oey}, {Chu}, \&
  {Garc{\'\i}a-Segura}}]{Jaskot2011}
{Jaskot}, A.~E., {Strickland}, D.~K., {Oey}, M.~S., {Chu}, Y.~H., \&
  {Garc{\'\i}a-Segura}, G. 2011, \apj, 729, 28,
  \dodoi{10.1088/0004-637X/729/1/28}

\bibitem[{{Kato} {et~al.}(2007){Kato}, {Nagashima}, {Nagayama}, {Kurita},
  {Koerwer}, {Kawai}, {Yamamuro}, {Zenno}, {Nishiyama}, {Baba}, {Kadowaki},
  {Haba}, {Hatano}, {Shimizu}, {Nishimura}, {Nagata}, {Sato}, {Murai},
  {Kawazu}, {Nakajima}, {Nakaya}, {Kandori}, {Kusakabe}, {Ishihara},
  {Kaneyasu}, {Hashimoto}, {Tamura}, {Tanab{\'e}}, {Ita}, {Matsunaga},
  {Nakada}, {Sugitani}, {Wakamatsu}, {Glass}, {Feast}, {Menzies}, {Whitelock},
  {Fourie}, {Stoffels}, {Evans}, \& {Hasegawa}}]{Kato2007}
{Kato}, D., {Nagashima}, C., {Nagayama}, T., {et~al.} 2007, \pasj, 59, 615,
  \dodoi{10.1093/pasj/59.3.615}

\bibitem[{{Kennicutt} \& {Hodge}(1986)}]{KennicutHodge1986}
{Kennicutt}, R.~C., J., \& {Hodge}, P.~W. 1986, \apj, 306, 130,
  \dodoi{10.1086/164326}

\bibitem[{{Klessen} {et~al.}(1998){Klessen}, {Burkert}, \&
  {Bate}}]{Klessen1998}
{Klessen}, R.~S., {Burkert}, A., \& {Bate}, M.~R. 1998, \apjl, 501, L205,
  \dodoi{10.1086/311471}

\bibitem[{{Klessen} \& {Glover}(2016)}]{Klessen16}
{Klessen}, R.~S., \& {Glover}, S.~C.~O. 2016, Star Formation in Galaxy
  Evolution: Connecting Numerical Models to Reality, Saas-Fee Advanced Course,
  Volume 43.~ISBN 978-3-662-47889-9.~Springer-Verlag Berlin Heidelberg, 2016,
  p.~85, 43, 85, \dodoi{10.1007/978-3-662-47890-5_2}

\bibitem[{{Kroupa}(2002)}]{Kroupa02}
{Kroupa}, P. 2002, Science, 295, 82, \dodoi{10.1126/science.1067524}

\bibitem[{{Ksoll} {et~al.}(2018){Ksoll}, {Gouliermis}, {Klessen}, {Grebel},
  {Sabbi}, {Anderson}, {Lennon}, {Cignoni}, {de Marchi}, {Smith}, {Tosi}, \&
  {van der Marel}}]{Ksoll2018}
{Ksoll}, V.~F., {Gouliermis}, D.~A., {Klessen}, R.~S., {et~al.} 2018, \mnras,
  479, 2389, \dodoi{10.1093/mnras/sty1317}

\bibitem[{{Ksoll} {et~al.}(2020{\natexlab{a}}){Ksoll}, {Gouliermis}, {Sabbi},
  {Ryon}, {Robberto}, {Gennaro}, {Klessen}, {Koethe}, {de Marchi}, {Chen},
  {Cignoni}, \& {Dolphin}}]{Ksoll2020b}
{Ksoll}, V.~F., {Gouliermis}, D., {Sabbi}, E., {et~al.} 2020{\natexlab{a}},
  arXiv e-prints, arXiv:2012.00521.
\newblock \doarXiv{2012.00521}

\bibitem[{{Ksoll} {et~al.}(2020{\natexlab{b}}){Ksoll}, {Ardizzone}, {Klessen},
  {Koethe}, {Sabbi}, {Robberto}, {Gouliermis}, {Rother}, {Zeidler}, \&
  {Gennaro}}]{Ksoll2020}
{Ksoll}, V.~F., {Ardizzone}, L., {Klessen}, R., {et~al.} 2020{\natexlab{b}},
  \mnras, 499, 5447, \dodoi{10.1093/mnras/staa2931}

\bibitem[{{Lada} \& {Lada}(2003)}]{LadaLada2003}
{Lada}, C.~J., \& {Lada}, E.~A. 2003, \araa, 41, 57,
  \dodoi{10.1146/annurev.astro.41.011802.094844}

\bibitem[{{Lasker} {et~al.}(1996){Lasker}, {Doggett}, {McLean}, {Sturch},
  {Djorgovski}, {de Carvalho}, \& {Reid}}]{Lasker1996}
{Lasker}, B.~M., {Doggett}, J., {McLean}, B., {et~al.} 1996, in Astronomical
  Society of the Pacific Conference Series, Vol. 101, Astronomical Data
  Analysis Software and Systems V, ed. G.~H. {Jacoby} \& J.~{Barnes}, 88

\bibitem[{{Lee} \& {Chen}(2007)}]{Lee2007}
{Lee}, H.-T., \& {Chen}, W.~P. 2007, \apj, 657, 884, \dodoi{10.1086/510893}

\bibitem[{{Lucke} \& {Hodge}(1970)}]{LuckeHodge1970}
{Lucke}, P.~B., \& {Hodge}, P.~W. 1970, \aj, 75, 171, \dodoi{10.1086/110959}

\bibitem[{{McLeod} {et~al.}(2019){McLeod}, {Dale}, {Evans}, {Ginsburg},
  {Kruijssen}, {Pellegrini}, {Ramsay}, \& {Testi}}]{McLeod2019}
{McLeod}, A.~F., {Dale}, J.~E., {Evans}, C.~J., {et~al.} 2019, \mnras, 486,
  5263, \dodoi{10.1093/mnras/sty2696}

\bibitem[{{Meixner} {et~al.}(2006){Meixner}, {Gordon}, {Indebetouw}, {Hora},
  {Whitney}, {Blum}, {Reach}, {Bernard}, {Meade}, {Babler}, {Engelbracht},
  {For}, {Misselt}, {Vijh}, {Leitherer}, {Cohen}, {Churchwell}, {Boulanger},
  {Frogel}, {Fukui}, {Gallagher}, {Gorjian}, {Harris}, {Kelly}, {Kawamura},
  {Kim}, {Latter}, {Madden}, {Markwick-Kemper}, {Mizuno}, {Mizuno}, {Mould},
  {Nota}, {Oey}, {Olsen}, {Onishi}, {Paladini}, {Panagia}, {Perez-Gonzalez},
  {Shibai}, {Sato}, {Smith}, {Staveley-Smith}, {Tielens}, {Ueta}, {van Dyk},
  {Volk}, {Werner}, \& {Zaritsky}}]{Meixner2006}
{Meixner}, M., {Gordon}, K.~D., {Indebetouw}, R., {et~al.} 2006, \aj, 132,
  2268, \dodoi{10.1086/508185}

\bibitem[{{Meixner} {et~al.}(2013){Meixner}, {Panuzzo}, {Roman-Duval},
  {Engelbracht}, {Babler}, {Seale}, {Hony}, {Montiel}, {Sauvage}, {Gordon},
  {Misselt}, {Okumura}, {Chanial}, {Beck}, {Bernard}, {Bolatto}, {Bot},
  {Boyer}, {Carlson}, {Clayton}, {Chen}, {Cormier}, {Fukui}, {Galametz},
  {Galliano}, {Hora}, {Hughes}, {Indebetouw}, {Israel}, {Kawamura}, {Kemper},
  {Kim}, {Kwon}, {Lebouteiller}, {Li}, {Long}, {Madden}, {Matsuura}, {Muller},
  {Oliveira}, {Onishi}, {Otsuka}, {Paradis}, {Poglitsch}, {Reach},
  {Robitaille}, {Rubio}, {Sargent}, {Sewi{\l}o}, {Skibba}, {Smith},
  {Srinivasan}, {Tielens}, {van Loon}, \& {Whitney}}]{Meixner2013}
{Meixner}, M., {Panuzzo}, P., {Roman-Duval}, J., {et~al.} 2013, \aj, 146, 62,
  \dodoi{10.1088/0004-6256/146/3/62}

\bibitem[{{Nota} {et~al.}(2006){Nota}, {Sirianni}, {Sabbi}, {Tosi}, {Clampin},
  {Gallagher}, {Meixner}, {Oey}, {Pasquali}, {Smith}, {Walterbos}, \&
  {Mack}}]{Nota2006}
{Nota}, A., {Sirianni}, M., {Sabbi}, E., {et~al.} 2006, \apjl, 640, L29,
  \dodoi{10.1086/503301}

\bibitem[{{Oey} \& {Massey}(1995)}]{Oey1995}
{Oey}, M.~S., \& {Massey}, P. 1995, \apj, 452, 210, \dodoi{10.1086/176292}

\bibitem[{{Panagia} {et~al.}(1991){Panagia}, {Gilmozzi}, {Macchetto}, {Adorf},
  \& {Kirshner}}]{Panagia1991}
{Panagia}, N., {Gilmozzi}, R., {Macchetto}, F., {Adorf}, H.~M., \& {Kirshner},
  R.~P. 1991, \apjl, 380, L23, \dodoi{10.1086/186164}

\bibitem[{{Pellegrini} {et~al.}(2012){Pellegrini}, {Oey}, {Winkler}, {Points},
  {Smith}, {Jaskot}, \& {Zastrow}}]{Pellegrini2012ApJ}
{Pellegrini}, E.~W., {Oey}, M.~S., {Winkler}, P.~F., {et~al.} 2012, \apj, 755,
  40, \dodoi{10.1088/0004-637X/755/1/40}

\bibitem[{Platt(1999)}]{Platt1999}
Platt, J.~C. 1999, in ADVANCES IN LARGE MARGIN CLASSIFIERS (MIT Press), 61--74

\bibitem[{Portegies~Zwart {et~al.}(2010)Portegies~Zwart, McMillan, \&
  Gieles}]{PortegiesZwart2010}
Portegies~Zwart, S.~F., McMillan, S.~L., \& Gieles, M. 2010, \araa, 48, 431,
  \dodoi{10.1146/annurev-astro-081309-130834}

\bibitem[{{Prim}(1957)}]{Prim1957}
{Prim}, R.~C. 1957, The Bell System Technical Journal, 36, 1389,
  \dodoi{10.1002/j.1538-7305.1957.tb01515.x}

\bibitem[{{Rosolowsky} {et~al.}(2008){Rosolowsky}, {Pineda}, {Kauffmann}, \&
  {Goodman}}]{Rosolowsky2008}
{Rosolowsky}, E.~W., {Pineda}, J.~E., {Kauffmann}, J., \& {Goodman}, A.~A.
  2008, \apj, 679, 1338, \dodoi{10.1086/587685}

\bibitem[{{Rousseau} {et~al.}(1978){Rousseau}, {Martin}, {Pr{\'e}vot},
  {Rebeirot}, {Robin}, \& {Brunet}}]{Rousseau1978}
{Rousseau}, J., {Martin}, N., {Pr{\'e}vot}, L., {et~al.} 1978, \aaps, 31, 243

\bibitem[{{Sabbi} {et~al.}(2007){Sabbi}, {Sirianni}, {Nota}, {Tosi},
  {Gallagher}, {Meixner}, {Oey}, {Walterbos}, {Pasquali}, {Smith}, \&
  {Angeretti}}]{Sabbi2007}
{Sabbi}, E., {Sirianni}, M., {Nota}, A., {et~al.} 2007, \aj, 133, 44,
  \dodoi{10.1086/509257}

\bibitem[{{Sabbi} {et~al.}(2016){Sabbi}, {Lennon}, {Anderson}, {Cignoni}, {van
  der Marel}, {Zaritsky}, {De Marchi}, {Panagia}, {Gouliermis}, {Grebel},
  {Gallagher}, {Smith}, {Sana}, {Aloisi}, {Tosi}, {Evans}, {Arab}, {Boyer}, {de
  Mink}, {Gordon}, {Koekemoer}, {Larsen}, {Ryon}, \& {Zeidler}}]{Sabbi2016}
{Sabbi}, E., {Lennon}, D.~J., {Anderson}, J., {et~al.} 2016, \apjs, 222, 11,
  \dodoi{10.3847/0067-0049/222/1/11}

\bibitem[{{Sanduleak}(1970)}]{Sanduleak1970}
{Sanduleak}, N. 1970, Contributions from the Cerro Tololo Inter-American
  Observatory, 89

\bibitem[{{Schmeja} {et~al.}(2009){Schmeja}, {Gouliermis}, \&
  {Klessen}}]{Schmeja2009}
{Schmeja}, S., {Gouliermis}, D.~A., \& {Klessen}, R.~S. 2009, \apj, 694, 367,
  \dodoi{10.1088/0004-637X/694/1/367}

\bibitem[{{Schmeja} \& {Klessen}(2006)}]{Schmeja2006}
{Schmeja}, S., \& {Klessen}, R.~S. 2006, \aap, 449, 151,
  \dodoi{10.1051/0004-6361:20054464}

\bibitem[{{Schulz}(2012)}]{Schulz2012}
{Schulz}, N.~S. 2012, {The Formation and Early Evolution of Stars},
  \dodoi{10.1007/978-3-642-23926-7}

\bibitem[{{Smith Neubig} \& {Bruhweiler}(1999)}]{Smith1999}
{Smith Neubig}, M.~M., \& {Bruhweiler}, F.~C. 1999, \aj, 117, 2856,
  \dodoi{10.1086/300867}

\bibitem[{{Stahler} \& {Palla}(2005)}]{Stahler_Palla2005}
{Stahler}, S.~W., \& {Palla}, F. 2005, {The Formation of Stars}

\bibitem[{{Stasi{\'n}ska} {et~al.}(1986){Stasi{\'n}ska}, {Testor}, \&
  {Heydari-Malayeri}}]{Stasinska1986}
{Stasi{\'n}ska}, G., {Testor}, G., \& {Heydari-Malayeri}, M. 1986, \aap, 170,
  L4

\bibitem[{{Stephens} {et~al.}(2017){Stephens}, {Gouliermis}, {Looney},
  {Gruendl}, {Chu}, {Weisz}, {Seale}, {Chen}, {Wong}, {Hughes}, {Pineda},
  {Ott}, \& {Muller}}]{Stephens2017}
{Stephens}, I.~W., {Gouliermis}, D., {Looney}, L.~W., {et~al.} 2017, \apj, 834,
  94, \dodoi{10.3847/1538-4357/834/1/94}

\bibitem[{{Will} {et~al.}(1997){Will}, {Bomans}, \& {Dieball}}]{Will1997}
{Will}, J.~M., {Bomans}, D.~J., \& {Dieball}, A. 1997, \aaps, 123, 455,
  \dodoi{10.1051/aas:1997169}

\bibitem[{{Wong} {et~al.}(2011){Wong}, {Hughes}, {Ott}, {Muller}, {Pineda},
  {Bernard}, {Chu}, {Fukui}, {Gruendl}, {Henkel}, {Kawamura}, {Klein},
  {Looney}, {Maddison}, {Mizuno}, {Paradis}, {Seale}, \& {Welty}}]{Wong2011}
{Wong}, T., {Hughes}, A., {Ott}, J., {et~al.} 2011, \apjs, 197, 16,
  \dodoi{10.1088/0067-0049/197/2/16}

\bibitem[{{Wong} {et~al.}(2017){Wong}, {Hughes}, {Tokuda}, {Indebetouw},
  {Bernard}, {Onishi}, {Wojciechowski}, {Bandurski}, {Kawamura}, {Roman-Duval},
  {Cao}, {Chen}, {Chu}, {Cui}, {Fukui}, {Montier}, {Muller}, {Ott}, {Paradis},
  {Pineda}, {Rosolowsky}, \& {Sewi{\l}o}}]{Wong2017}
{Wong}, T., {Hughes}, A., {Tokuda}, K., {et~al.} 2017, \apj, 850, 139,
  \dodoi{10.3847/1538-4357/aa9333}

\bibitem[{{Zaritsky} {et~al.}(1997){Zaritsky}, {Harris}, \&
  {Thompson}}]{Zaritsky1997}
{Zaritsky}, D., {Harris}, J., \& {Thompson}, I. 1997, \aj, 114, 1002,
  \dodoi{10.1086/118531}

\bibitem[{{Zinnecker} \& {Yorke}(2007)}]{Zinnecker07}
{Zinnecker}, H., \& {Yorke}, H.~W. 2007, \araa, 45, 481,
  \dodoi{10.1146/annurev.astro.44.051905.092549}

\bibitem[{{Zivkov} {et~al.}(2018){Zivkov}, {Oliveira}, {Petr-Gotzens}, {Cioni},
  {Rubele}, {van Loon}, {Bekki}, {Cusano}, {de Grijs}, {Ivanov}, {Marconi},
  {Niederhofer}, {Ripepi}, \& {Sun}}]{Zivkov2018}
{Zivkov}, V., {Oliveira}, J.~M., {Petr-Gotzens}, M.~G., {et~al.} 2018, \aap,
  620, A143, \dodoi{10.1051/0004-6361/201833951}

\end{thebibliography}
\bibliographystyle{aasjournal}


\appendix
\restartappendixnumbering
\section{Additional Material}
    
    \begin{figure}
        \centering
        \includegraphics[width = \linewidth]{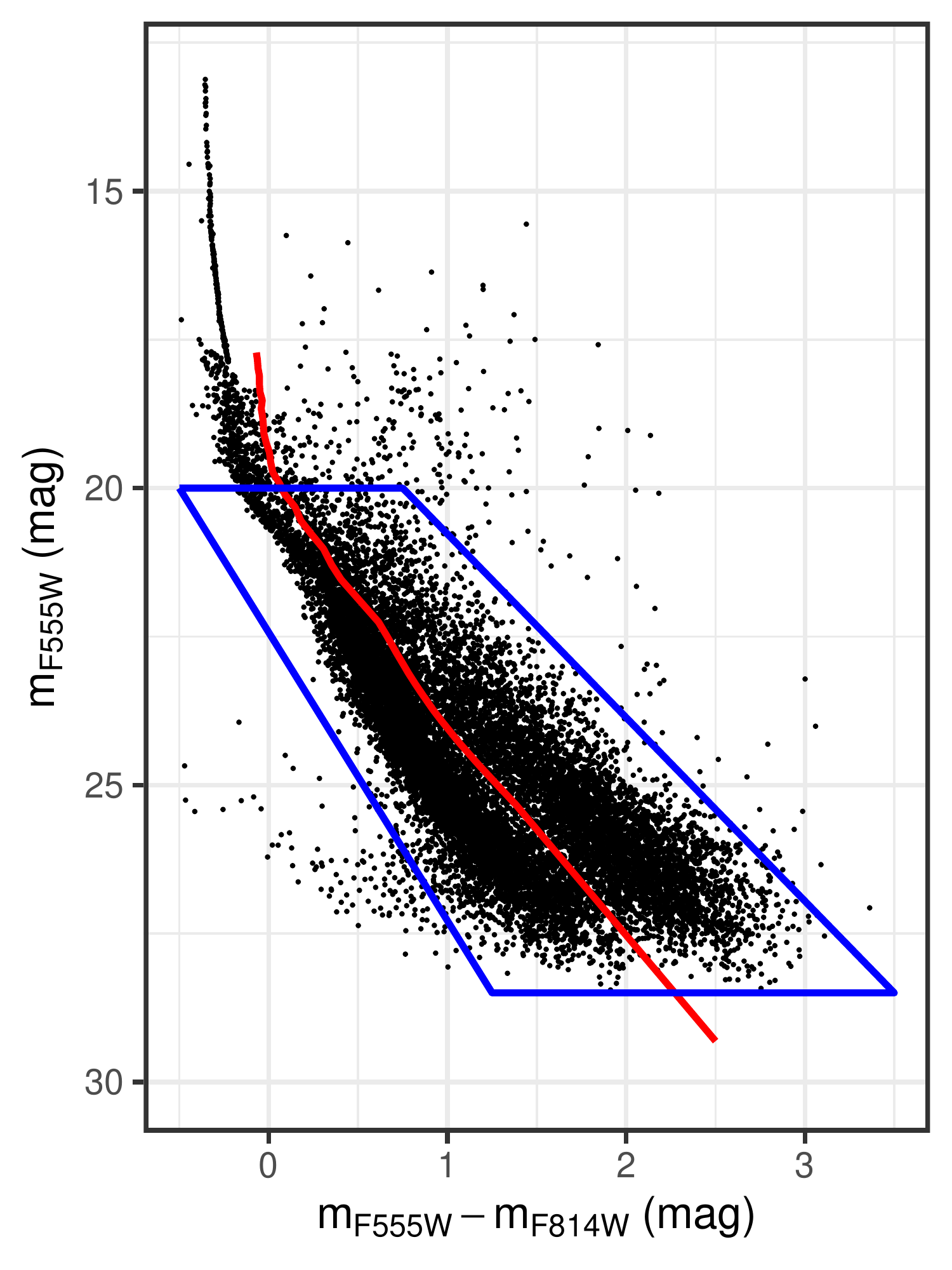}
        \caption{Optical CMD of the region selected as the training set base, corrected for extinction according the $A_\mathrm{F555W}$ measurements. The red line indicates a threshold line, derived from PARSEC isochrones in the age range of 1 to 14 Myr. This threshold is used to generate slopes for our EM approach to quantify the constituents of the two populations. The blue polygon indicates the stars considered for the EM fit, excluding the UMS, red clump and a few objects of unclear nature.}
        \label{fig:TrainingSet_EM}
    \end{figure}
    
    \begin{figure*}
        \centering
        \includegraphics[width = 0.3\linewidth]{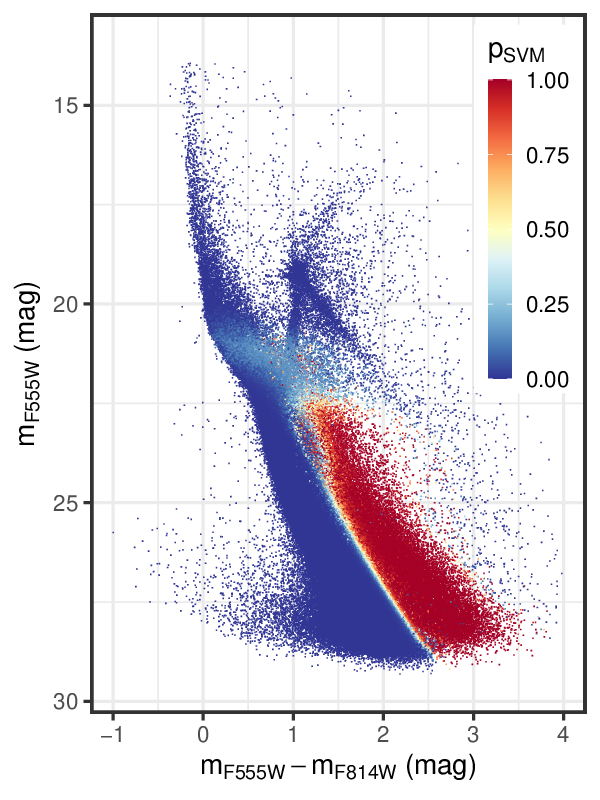}
        \includegraphics[width = 0.3\linewidth]{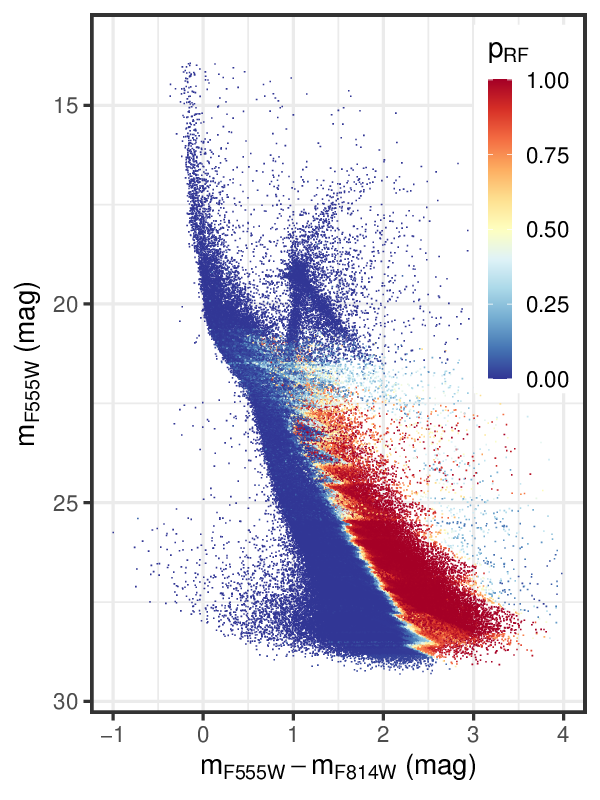}
        \includegraphics[width = 0.3\linewidth]{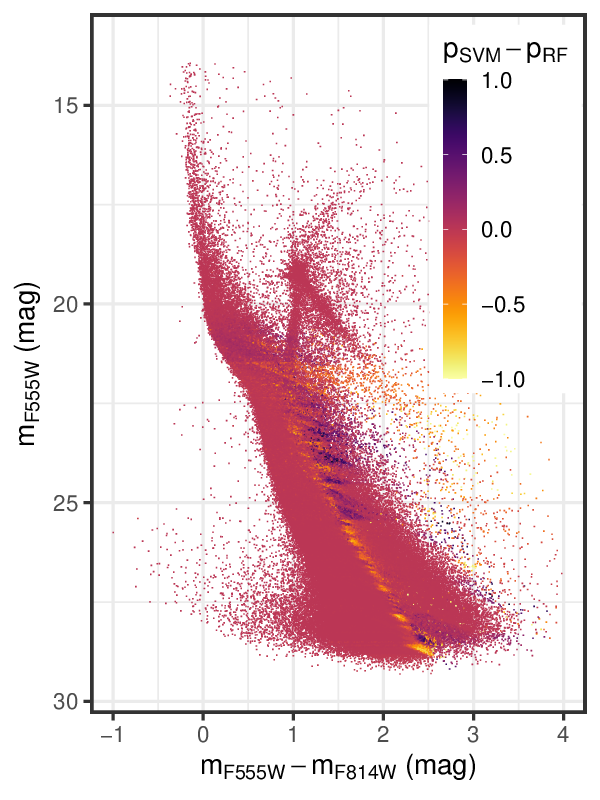}
        \caption{Optical CMD of the entire MYSST photometric catalog. Each star is color coded according to the predicted probability that it belongs to the PMS as given by the SVM model (left), by the RF model (center), and the difference $p_\mathrm{svm} - p_\mathrm{rf}$ of the predicted PMS candidate probabilities (right), respectively.}
        \label{fig:PredictionCMD_p_pms_svm_rf_and_diff}
    \end{figure*}
    
    \begin{figure*}
        \centering
        \includegraphics[width = 0.49\linewidth]{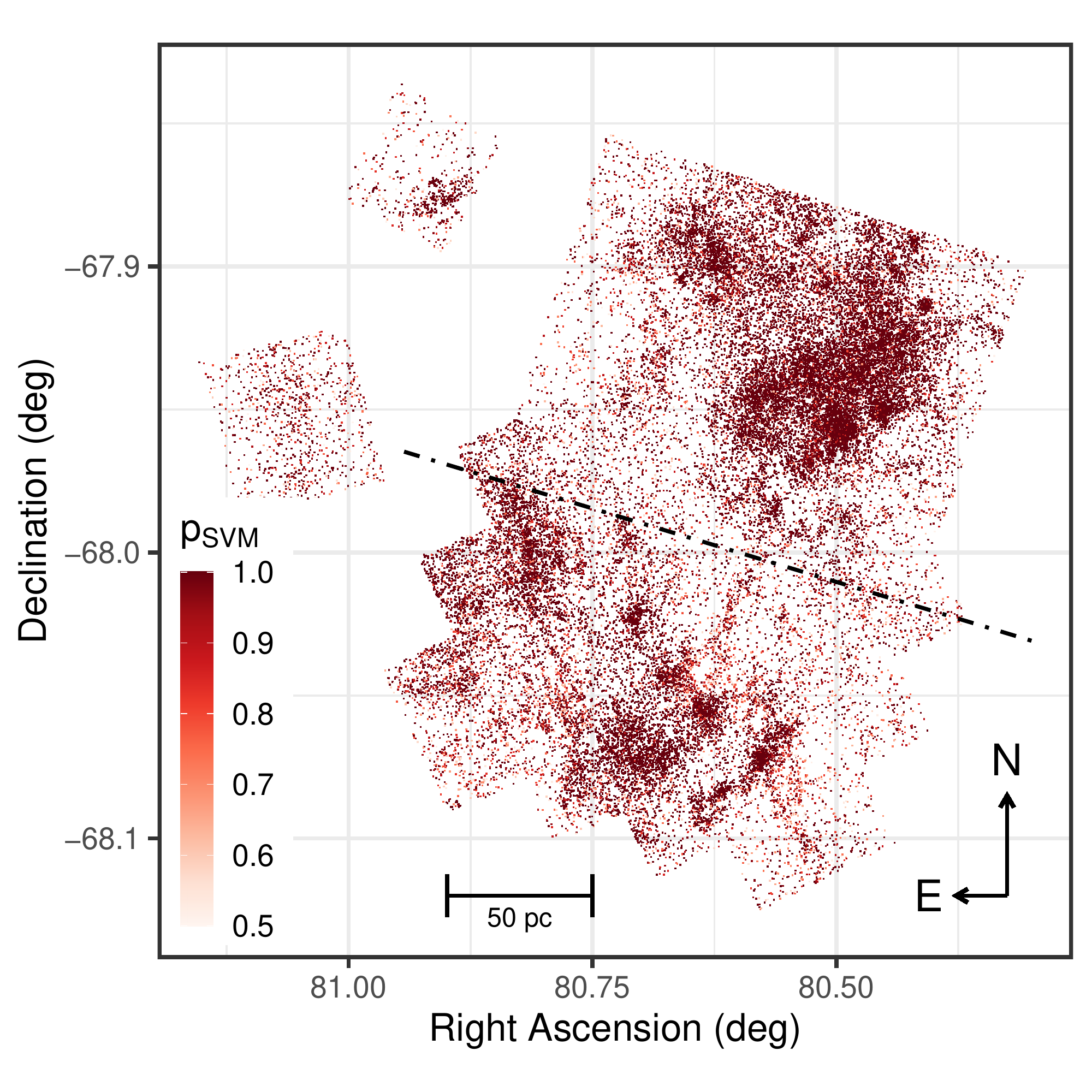}
        \includegraphics[width = 0.49\linewidth]{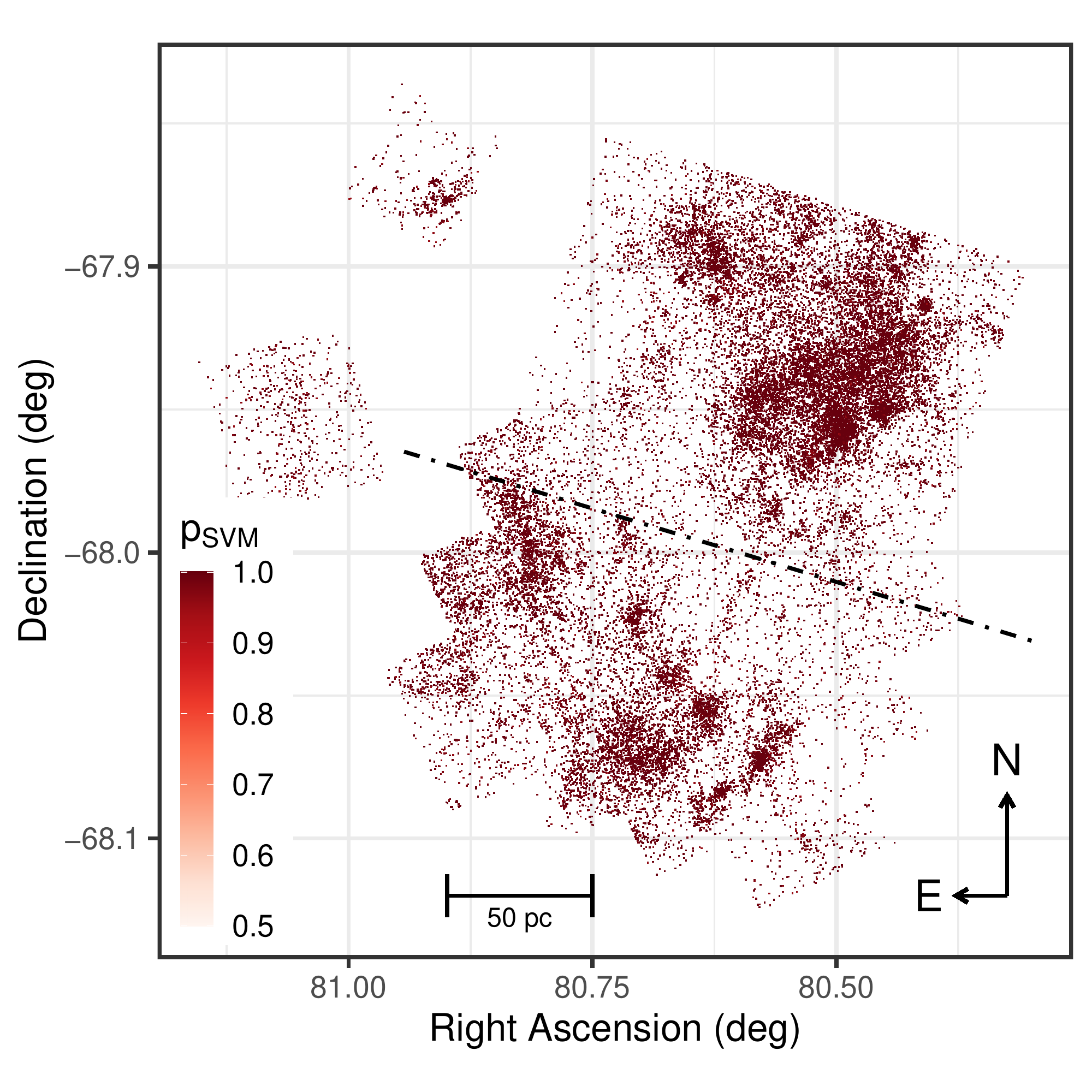}
        \caption{Spatial distribution diagrams of the candidate PMS stars as predicted by the SVM model. The left panel shows all candidate stars with $p_\mathrm{SVM} \geq 0.5$, while the right panel indicates the positions of the most likely PMS stars with $p_\mathrm{SVM} \geq 0.95$.}
        \label{fig:PredictionSpatial_p_pms_svm}
    \end{figure*}

    \begin{figure*}
        \centering
        \includegraphics[width = 0.49\linewidth]{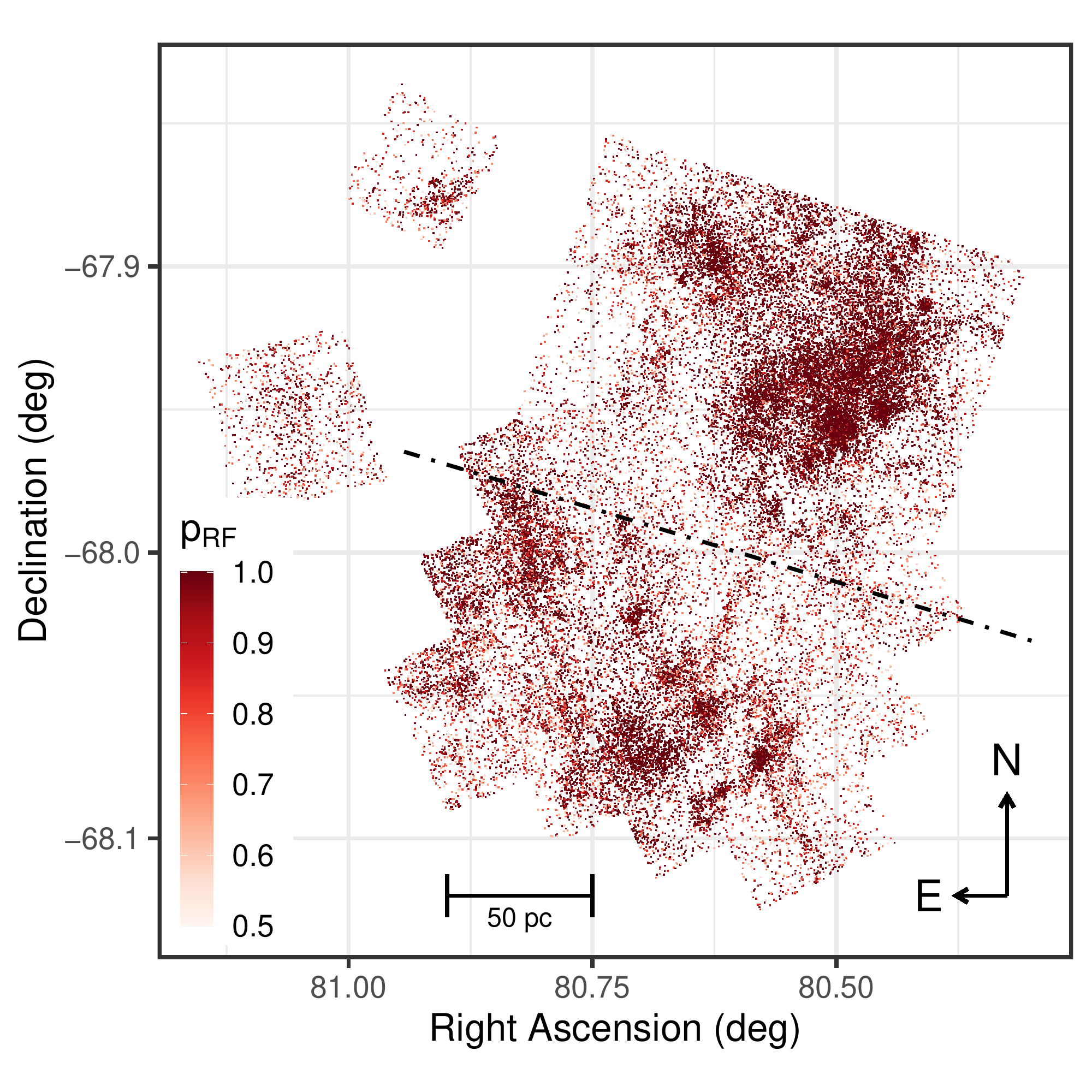}
        \includegraphics[width = 0.49\linewidth]{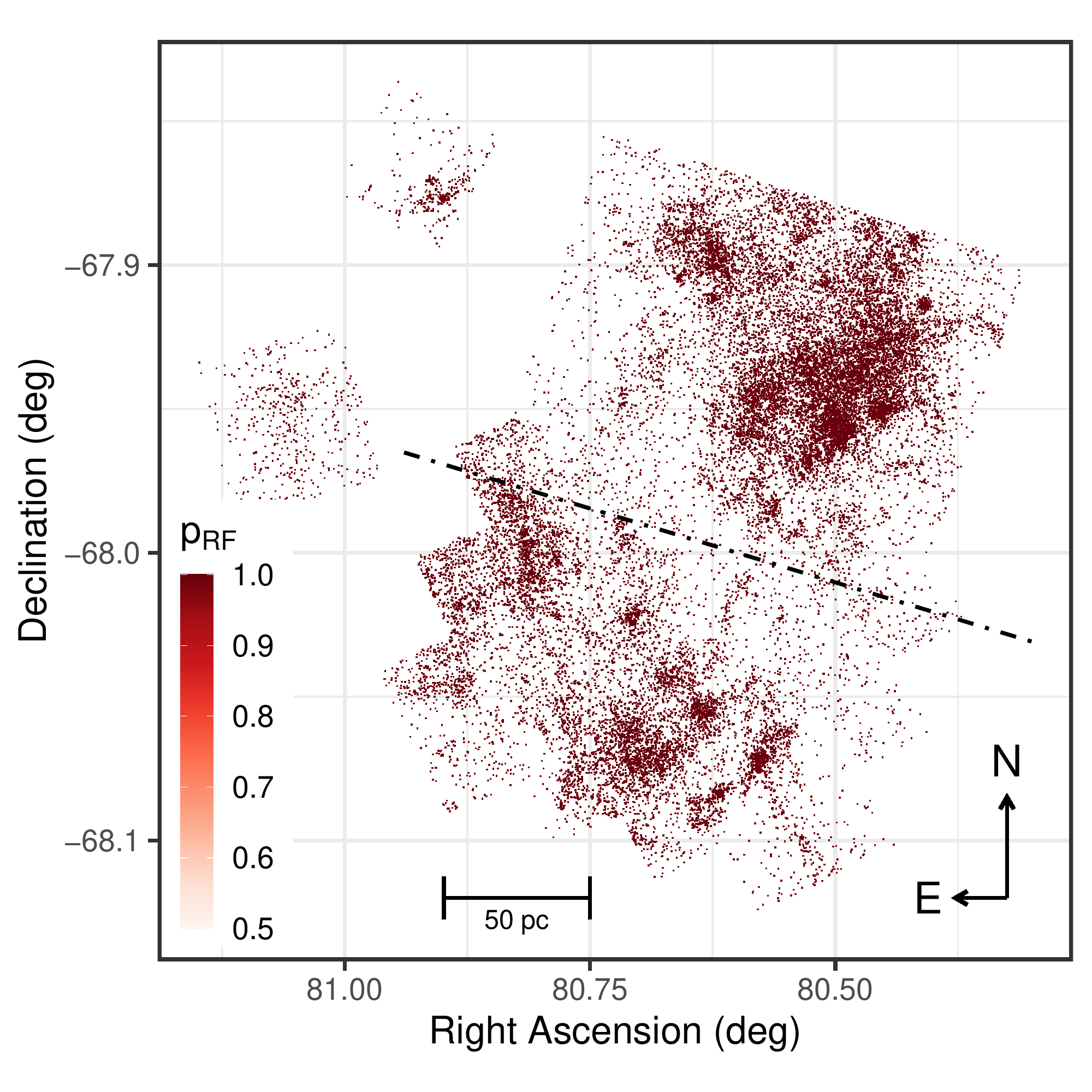}
        \caption{Spatial distribution diagrams of the candidate PMS stars as predicted by the RF model. In the left panel all stars with $p_\mathrm{RF} \geq 0.5$ are show, while the right panel marks the locations of the most likely candidate PMS stars with $p_\mathrm{RF} \geq 0.95$.}
        \label{fig:PredictionSpatial_p_pms_rf}
    \end{figure*}
    
    \begin{deluxetable*}{lllll}
        \centerwidetable
        \tablecaption{List of known O stars within or close to the MYSST FoV \label{tab:SIMBAD_Ostars}}
        \tablehead{
        \colhead{Identifier} & \colhead{RA} & \colhead{Dec} & \colhead{Spectral Type} & \colhead{Reference} \\
            & \colhead{(deg)} & \colhead{(deg)} & &
        }
        \startdata
        SK -67 86   & 80.56181 & -67.85908 & OB & \cite{Sanduleak1970} \\
        HD 269412   & 80.47527 & -67.91469 & OB & \cite{Sanduleak1970} \\
        SK -67 94   & 80.88919 & -67.95822 & OB & \cite{Sanduleak1970} \\
        SK -68 76   & 81.01482 & -68.06106 & OB & \cite{Sanduleak1970} \\
        SK -68 72a  & 80.69472 & -68.06542 & O9II & \cite{Conti1986} \\
        HD 269445   & 80.74911 & -68.02962 & Ofpe/WN9 & \cite{Bohannan1989} \\
        LH 47-355   & 80.56437 & -67.98347 & O9.5V & \cite{Oey1995} \\
        LH 47-335   & 80.55683 & -67.93951 & O9.5V & \cite{Oey1995} \\
        LH 47-14    & 80.43046 & -67.9185 & O9.5V & \cite{Oey1995} \\
        LH 48-122   & 80.59713 & -67.88168 & O9.5V & \cite{Oey1995} \\
        LH 47-84    & 80.46338 & -67.93679 & O9.5V & \cite{Oey1995} \\
        BI 155      & 80.95265 & -67.89803 & O7V & \cite{Smith1999} \\
        BI 159      & 81.04867 & -68.0163 & O/B0 & \cite{Brunet1975} \\
        {[L72]} LH 48-9     & 80.65 & -67.9 & O7III & \cite{Conti1986} \\
        {[L72]} LH 48-21    & 80.6 & -67.9 & O5III & \cite{Conti1986} \\
        HD 269449   & 80.8 & -68.01667 & O & \cite{Cannon1993} \\
        {[STH86]} Star 2    & 80.61 & -67.97 & O & \cite{Stasinska1986} \\
        SK -67 92& 80.81166 & -67.93655 & OB & \cite{Sanduleak1970} \\
        \enddata
        \tablecomments{
        All O stars found in the SIMBAD database that are not captured by the \cite{McLeod2019} MUSE observations. Listed are each stars identifier, right ascension, declination, spectral type and the literature reference for the studies that derive the latter.
        }
    \end{deluxetable*}

    \begin{figure*}
        \centering
        \includegraphics[width = 0.55\linewidth]{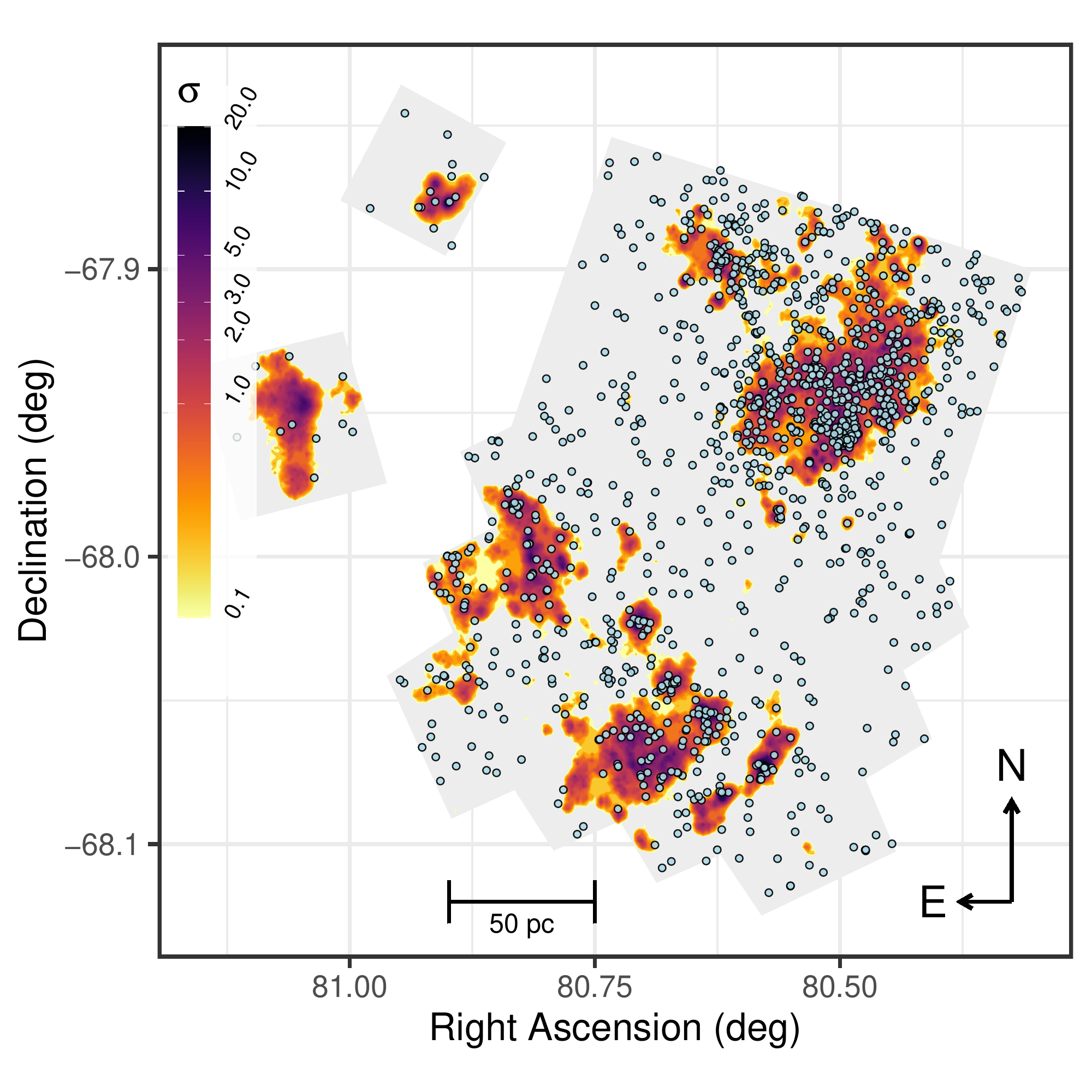}
        \caption{Spatial nearest neighbor contour density diagram of the most probable candidate PMS stars of the MYSST survey (as in Figure \ref{fig:PMS_density_Ostars}). The color coding represents the nearest neighbor density in steps of $\sigma$ above the mean density. The black dot-dashed line indicates the north/south-separation for the NNDE. The light blue points mark the positions of the 1,291 UMS stars selected in Paper I to derive extinction measures for the survey. For comparison the grey shaded regions mark the MYSST coverage.}
        \label{fig:MYSST_PMS_vs_UMS}
    \end{figure*}
    
   \begin{sidewaysfigure}
        \centering
        \includegraphics[width = \linewidth]{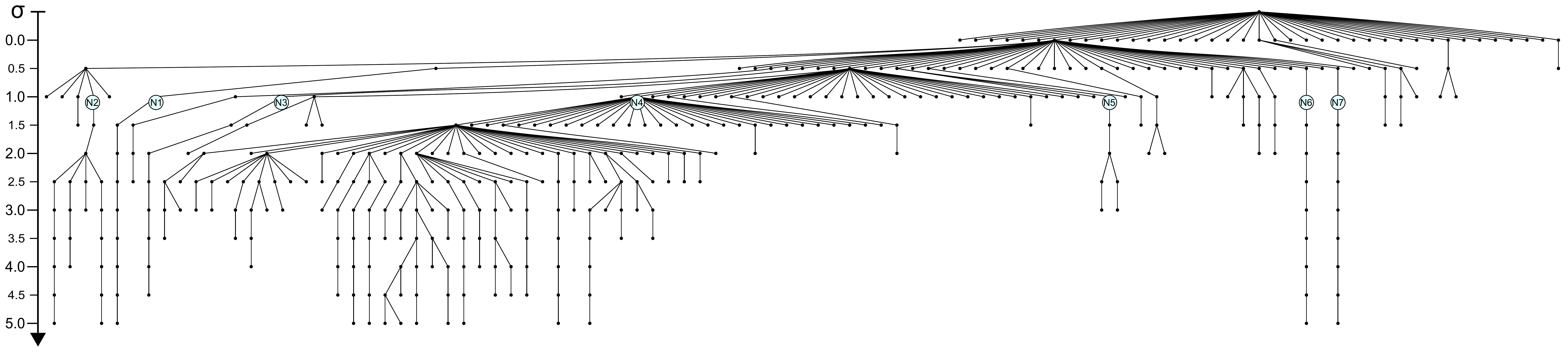}
        \caption{Dendrogram of the NNDE density structures of the most likely PMS candidates located in the northern half of the FoV. The large labeled circles indicate the $1\sigma$ density structures with at least 100 stars that persist up to $3\sigma$ density significance. The labels correspond to the IDs of the regions marked in Figure \ref{fig:MYSST_densityPersistentStructures} and summarized in Table \ref{tab:PMS_1sigma_Structures}.}
        \label{fig:DendrogramNorth}
    \end{sidewaysfigure}
    
    \begin{sidewaysfigure}
        \centering
        \includegraphics[width = 0.9\linewidth]{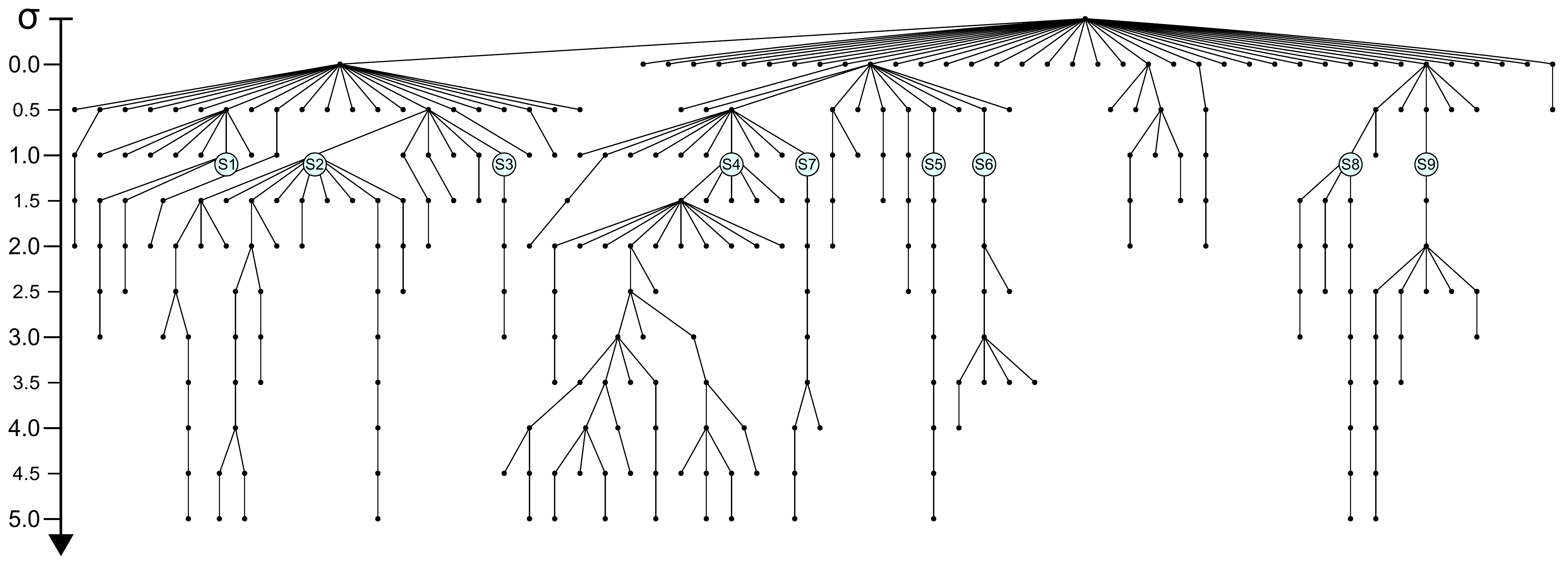}
        \caption{Dendrogram of the NNDE density structures of the most likely PMS candidates located in the southern half of the FoV. The large labeled circles indicate the $1\sigma$ density structures with at least 100 stars that persist up to $3\sigma$ density significance. The labels correspond to the IDs of the regions marked in Figure \ref{fig:MYSST_densityPersistentStructures} and summarized in Table \ref{tab:PMS_1sigma_Structures}.}
        \label{fig:DendrogramSouth}
    \end{sidewaysfigure}

    \begin{deluxetable*}{lccccccccccccccc}
        \centerwidetable
        \tablecaption{Properties of the PMS Subclusters at $3\sigma$ Density \label{tab:PMS_3sigma_Structures}}
        \tablehead{
        \colhead{ID} & \colhead{$\mathrm{RA_{cent}}$} & \colhead{$\mathrm{Dec_{cent}}$} & \colhead{$A_\mathrm{surf}$} & \colhead{$R_\mathrm{eff}$} & \colhead{$N_*$} & \colhead{$n_\mathrm{surf}^{total}$} & \colhead{$N_\mathrm{PMS}$} & \colhead{$n_\mathrm{surf}^{pms}$} & \colhead{$N_\mathrm{O}^{M19}$} & \colhead{$N_\mathrm{YSO}^{C09}$} & \colhead{$N_\mathrm{YSO}^{C12}$} & \colhead{$Q$} & \colhead{$\sigma_Q$} \\
        & (deg) & (deg) & $(\mathrm{pc^2})$ & $(\mathrm{pc})$ & & $(\mathrm{pc^{-2}})$ & & $(\mathrm{pc^{-2}})$ & & & & & 
        }
        \startdata
        N1.1 & 80.6590  & -67.9046  & 2.6   & 0.9   & 50    & 19.3  & 28    & 10.8  & 0 & 0 & 0 & 0.88  & 0.61 \\
        N2.1 & 80.6236  & -67.8946  & 10.3  & 1.8   & 210   & 20.5  & 82    & 8     & 0 & 1 & 1 & 0.73  & 0.48 \\
        N2.2 & 80.6116  & -67.9024  & 4.3   & 1.2   & 94    & 22.1  & 36    & 8.5   & 0 & 0 & 0 & 0.8   & 0.53 \\
        N3.1 & 80.6247  & -67.9110  & 2.8   & 0.9   & 57    & 20.5  & 25    & 9     & 0 & 0 & 0 & 0.76  & 0.53 \\
        N4.1 & 80.5843  & -67.9474  & 3.3   & 1     & 64    & 19.4  & 29    & 8.8   & 0 & 0 & 0 & 0.64  & 0.46 \\
        N4.2 & 80.5852  & -67.9450  & 2.6   & 0.9   & 62    & 23.7  & 17    & 6.5   & 0 & 0 & 0 & 0.73  & 0.47 \\
        N4.3 & 80.5250  & -67.9372  & 8.2   & 1.6   & 163   & 19.8  & 55    & 6.7   & 0 & 0 & 0 & 0.61  & 0.45 \\
        N4.4 & 80.5271  & -67.9684  & 2.9   & 1     & 55    & 19    & 29    & 10    & 0 & 0 & 0 & 0.77  & 0.49 \\
        N4.5 & 80.4973  & -67.9561  & 100   & 5.6   & 2147  & 21.5  & 1109  & 11.1  & 3 & 1 & 0 & 0.69  & 0.48 \\
        N4.6 & 80.4933  & -67.9384  & 5.7   & 1.4   & 120   & 20.9  & 53    & 9.2   & 0 & 0 & 0 & 0.74  & 0.47 \\
        N4.7 & 80.4788  & -67.9380  & 15.3  & 2.2   & 311   & 20.3  & 124   & 8.1   & 0 & 0 & 0 & 0.7   & 0.44 \\
        N4.8 & 80.4682  & -67.9352  & 4.5   & 1.2   & 88    & 19.6  & 30    & 6.7   & 0 & 0 & 0 & 0.75  & 0.47 \\
        N4.9 & 80.4532  & -67.9506  & 32.8  & 3.2   & 750   & 22.9  & 381   & 11.6  & 1 & 1 & 0 & 0.75  & 0.49 \\
        N4.10 & 80.4534 & -67.9271  & 6.2   & 1.4   & 143   & 23.1  & 52    & 8.4   & 0 & 0 & 0 & 0.68  & 0.5 \\
        N6.1 & 80.4203  & -67.8912  & 4.9   & 1.3   & 92    & 18.6  & 45    & 9.1   & 0 & 0 & 0 & 0.8   & 0.5 \\
        N7.1 & 80.4082  & -67.9140  & 7.9   & 1.6   & 200   & 25.3  & 90    & 11.4  & 1 & 0 & 0 & 0.76  & 0.5 \\
        \hline
        S2.1 & 80.8289 & -67.9815   & 6.6   & 1.5   & 101   & 15.3  & 27    & 4.1   & 0 & 0 & 0 & 0.81  & 0.5 \\
        S2.2 & 80.8161 & -67.9974   & 31.3  & 3.2   & 441   & 14.1  & 151   & 4.8   & 0 & 0 & 0 & 0.6   & 0.44 \\
        S2.3 & 80.8113 & -68.0080   & 6.9   & 1.5   & 102   & 14.8  & 27    & 3.9   & 0 & 0 & 0 & 0.74  & 0.49 \\
        S2.4 & 80.7842 & -68.0023   & 9.9   & 1.8   & 139   & 14.1  & 47    & 4.8   & 0 & 1 & 0 & 0.74  & 0.5 \\
        S4.1 & 80.7049 & -68.0678   & 73.5  & 4.8   & 1075  & 14.6  & 312   & 4.2   & 0 & 0 & 0 & 0.51  & 0.36 \\
        S4.2 & 80.6758 & -68.0716   & 31.1  & 3.1   & 544   & 17.5  & 127   & 4.1   & 0 & 0 & 0 & 0.62  & 0.42 \\
        S5.1 & 80.7065 & -68.0223   & 25    & 2.8   & 409   & 16.4  & 138   & 5.5   & 0 & 1 & 1 & 0.74  & 0.48 \\
        S6.1 & 80.6747 & -68.0440   & 12.8  & 2     & 236   & 18.5  & 43    & 3.4   & 0 & 0 & 0 & 0.69  & 0.49 \\
        S7.1 & 80.6359 & -68.0550   & 40    & 3.6   & 677   & 16.9  & 214   & 5.3   & 0 & 1 & 0 & 0.75  & 0.48 \\
        S8.1 & 80.6185 & -68.0834   & 13.6  & 2.1   & 257   & 18.9  & 87    & 6.4   & 0 & 0 & 0 & 0.8   & 0.5 \\
        S9.1 & 80.5774 & -68.0728   & 31.8  & 3.2   & 592   & 18.6  & 245   & 7.7   & 0 & 0 & 0 & 0.79  & 0.51 \\
        \hline
        FN1.1 & 80.9122 & -67.8704  & 3.8   & 1.1   & 55    & 14.5  & 14    & 3.7   & 0 & 0 & 1 & 0.75  & 0.50 \\
        FN1.2 & 80.8991 & -67.8768  & 14.1  & 2.1   & 214   & 15.2  & 64    & 4.5   & 0 & 1 & 1 & 0.77  & 0.51 \\
        FS1.1 & 81.0500 &-67.9466   & 41.7  & 3.6   & 491   & 11.8  & 32    & 0.8   & 0 & 0 & 0 & 0.72  & 0.48 \\
        \enddata
        \tablecomments{
        Properties of the PMS subclusters at $3\sigma$ density within the prominent $1\sigma$ density structures. This list only contains subclusters which entail at least 50 stars, so it corresponds to the solid light blue contours depicted in Figure \ref{fig:MYSST_densityPersistentStructures}. Each object's ID indicates the $1\sigma$ structure it belongs to, e.g.\ N1.1 is within N1, N2.1 in N2 etc. As in Table \ref{tab:PMS_1sigma_Structures} listed are the right ascension $\mathrm{RA_{cent}}$ and declination $\mathrm{Dec_{cent}}$ of the subcluster center, the surface area $A_\mathrm{surf}$ enclosed by the given density contour, an effective radius $R_\mathrm{eff}$ derived from the surface area, the total number $N_*$ of MYSST catalog stars within the structure, the total surface stellar number density $n_\mathrm{surf}^{total}$, the number of identified most likely PMS candidates $N_\mathrm{PMS}$ inside the contour, the corresponding surface number density of PMS sources $n_\mathrm{surf}^{pms}$, the number of enclosed \cite{McLeod2019} O stars $N_\mathrm{O}^{M19}$, \cite{Chen2009} YSOs $N_\mathrm{YSO}^{C09}$, \cite{Carlson2012} YSOs $N_\mathrm{YSO}^{C12}$ and the \cite{Cartwright2004} $Q$ parameter along with its uncertainty $\sigma_Q$ as an indicator of cluster 'clumpiness'. Note that none of the SIMBAD O stars (Table \ref{tab:SIMBAD_Ostars}) fall into any of the $3~\sigma$ contours, so they are not listed here.
        }
    \end{deluxetable*}

    \begin{figure*}
        \centering
        \includegraphics[width = 0.7\linewidth]{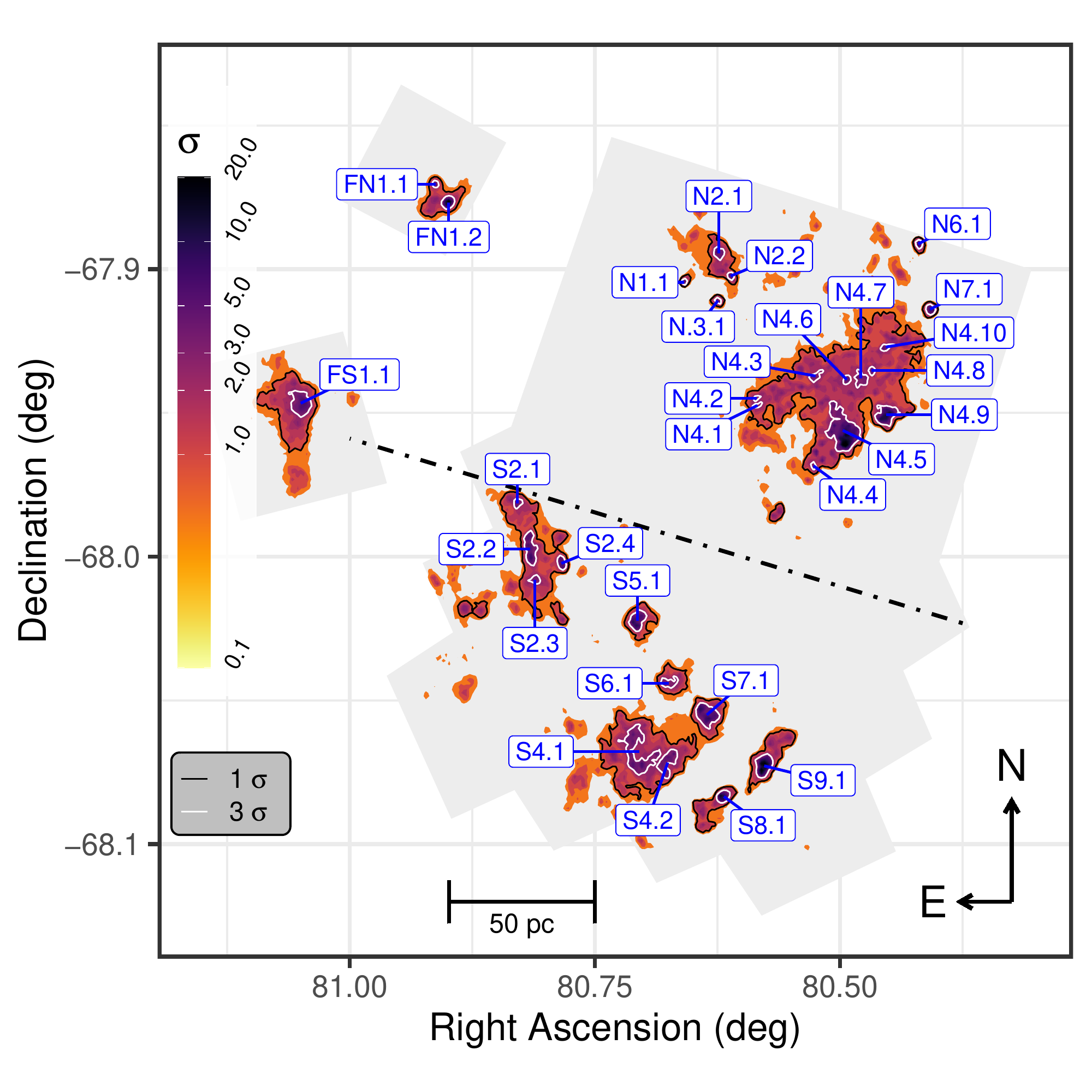}
        \caption{Same as Figure \ref{fig:MYSST_densityPersistentStructures}, but now the ID labels for the $3\sigma$ structures (c.f.~Table~\ref{tab:PMS_3sigma_Structures}) are provided instead of the $1\sigma$ ones.}
        \label{fig:MYSST_persistenStructures_3sigma_labelled}
    \end{figure*}
    
    \begin{figure*}
        \centering
        \includegraphics [width = 0.9\linewidth]{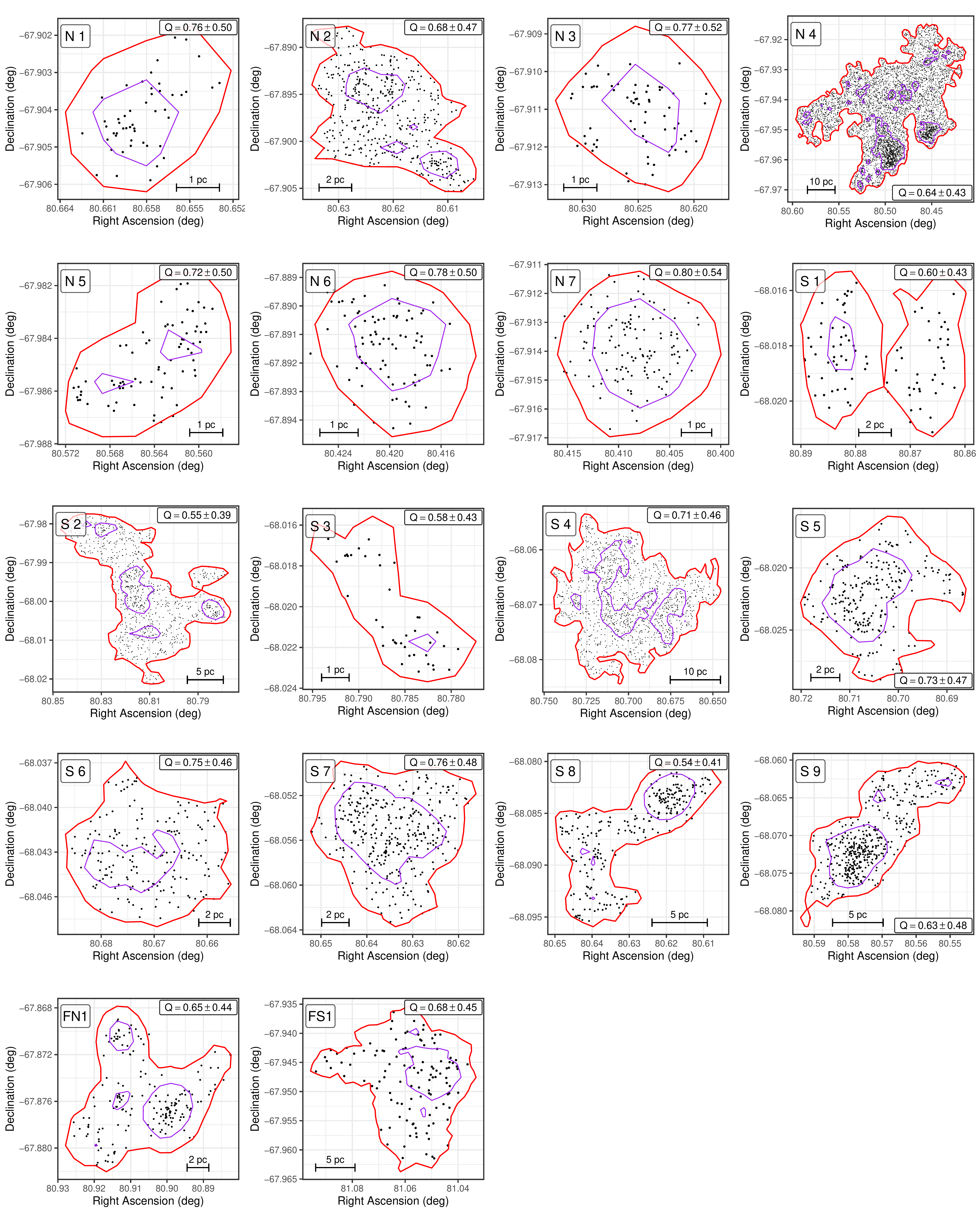}
        \caption{Spatial distribution diagrams of the PMS candidates in the eighteen prominent PMS structures identified across the main MYSST FoV and the two reference fields. The identifiers correspond to the list presented in Table \ref{tab:PMS_1sigma_Structures}. In each diagram the red line indicates the $1\,\sigma$ contour associated with each cluster, while the purple lines mark the substructures at the $3\,\sigma$ density significance level. Note that in a few of these diagrams a small number of stars may fall outside the $1\,\sigma$density contours of their assigned cluster. This is caused by minor inaccuracies in the transformation of the density contours from pixel space, in which they are defined, to the RA-Dec coordinate system presented in this diagram. In pixel space all stars are strictly interior to their respective cluster contours.}
        \label{fig:PMS_clusters_spatial}
    \end{figure*}
    
    This Appendix provides complimentary figures and discussion to the main paper. Figure \ref{fig:TrainingSet_EM} shows the CMD of the $2\,\sigma$ surface density region used as a basis for our training set in Section \ref{sec:TrainingSet}. It highlights in particular the data selection we make for the Gaussian mixture model fit \citep[see][]{Ksoll2018} that we perform in order to distinguish the PMS and LMS populations, as well as the threshold curve that provides the basis for the underlying distance measure of this fit. The threshold curve corresponds to the 14 Myr PARSEC isochrone between 22.3 and 25.3 mag in F555W, extended by isochrones down to 1 Myr above 22.3 mag and follows a combination of isochrones up to 50 Myr below 25.3 mag. Please note that this threshold does not serve as a hard cut between the PMS and the LMS but as a guide for the Gaussian mixture model fit that determines the final population assignments. We refer to \cite{Ksoll2018} for more details on this mixture model approach. \\
    In Figure \ref{fig:PredictionCMD_p_pms_svm_rf_and_diff} we show the individual prediction outcomes on the entire MYSST data set of the trained SVM and RF (left and middle panel) as the color coding of the CMD, as well as a direct star by star comparison of the predicted PMS probabilities (right panel). As previously described in Section \ref{sec:Results} these diagrams demonstrate how well the SVM and RF predictions agree overall and the few regions where they disagree, which lead us to the conclusion that a combination of the outcomes provides the most robust solution. \\
    Complementary to Figure \ref{fig:PredictionSpatial_p_pms_mean} that presents the spatial distribution of identified PMS stars from the combination of the two ML approaches, Figures \ref{fig:PredictionSpatial_p_pms_svm} and \ref{fig:PredictionSpatial_p_pms_rf} provide the corresponding distributions derived from SVM and RF individually. \\
    Analogous to Figure \ref{fig:PMS_density_Ostars} we show the spatial contour density diagram of our NNDE on the most probable candidate PMS stars in Figure \ref{fig:MYSST_PMS_vs_UMS}, here in comparison to the positions of the 1,291 UMS sources (light blue points), likely late O to early B type, that we select in Paper I to derive extinction estimates for N44. As Figure \ref{fig:PMS_density_Ostars} this diagram demonstrates that the prominent PMS groupings we identify tend to be located in the vicinity of the massive young population of N44.\\
    Figures \ref{fig:DendrogramNorth} (north) and \ref{fig:DendrogramSouth} (south) show the corresponding dendrograms of the clustering structures we have identified in Figure \ref{fig:MYSST_densityPersistentStructures} in the main paper. These dendrograms are based on the NNDE we perform in the northern and southern half of the FoV and are iteratively constructed by considering each significance density contour as the root/parent structure of the contours/subclusters located inside of it. Both of these dendrograms highlight the intricate hierarchical substructure of the identified PMS groupings. \\
    Analogous to Table \ref{tab:PMS_1sigma_Structures} we present the characteristic properties of the 34 subclusters at $3\sigma$ surface density significance in Table \ref{tab:PMS_3sigma_Structures}. The properties include center position, surface area, effective radius, numbers and number densities of the total/PMS stars of each $3\sigma$ substructure, as well as the $Q$-parameter \citep{Cartwright2004} as a measure of cluster clumpiness. Figure \ref{fig:MYSST_persistenStructures_3sigma_labelled} indicates the positions of the $3\sigma$ subclusters analogous to Figure \ref{fig:MYSST_densityPersistentStructures}.\\
    To complement the analysis of the cluster substructures in Section \ref{sec:ClusterProperties} Figure \ref{fig:PMS_clusters_spatial} provides the spatial distribution diagrams of the PMS stars within the eighteen prominent PMS structures presented in Table \ref{tab:PMS_1sigma_Structures}. These diagrams provide a visual confirmation of the $Q$-parameter \citep{Cartwright2004} analysis, indicating the overall clumpiness and hierarchical structure of all the prominent PMS clusterings we have identified. 
\end{document}